\documentclass[aps,prd,twocolumn,nofootinbib]{revtex4}
\usepackage{graphicx}
\usepackage{tabularx}
\usepackage{hhline}
\usepackage[mathscr]{euscript}
\usepackage{outlines}
\usepackage{xcolor}
\usepackage{placeins}
\usepackage{amssymb, amsmath}
\newcommand{\rns}{\rho_{\rm sat}}
\newcommand{\Ms}{M_{\odot}}

\newcommand{\Rsmall}{$R_{1.4}\approx11$~km\ }
\newcommand{\Rbig}{$R_{1.4}\approx14$~km\ }
\usepackage{tikz}
\usepackage{pgfplots}
\usepackage{hyperref}

\newcommand{\aap}{Astron.\ and Astrophys.}

\usepackage[normalem]{ulem}

\newcommand{\coderefs}{\cite{Duez:2005sg,Paschalidis:2010dh,Etienne:2011re,Etienne:2015cea}}

\begin{document}

\title{Influence of stellar compactness on finite-temperature effects in neutron star merger simulations}

\author{Carolyn A. Raithel,$^{1,2,3}$ Vasileios Paschalidis,$^{4,5}$}
\affiliation{$^1$School of Natural Sciences, Institute for Advanced Study, 1 Einstein Drive, Princeton, NJ 08540, USA}
\affiliation{$^2$Princeton Center for Theoretical Science, Jadwin Hall, Princeton University, Princeton, NJ 08540, USA}
\affiliation{$^3$Princeton Gravity Initiative, Jadwin Hall, Princeton University, Princeton, NJ 08540, USA}
\affiliation{$^4$Department of Astronomy and Steward Observatory, University of Arizona, 933 N. Cherry Avenue, Tucson, Arizona 85721, USA}
\affiliation{$^5$Department of Physics, University of Arizona, 1118 E. Fourth Street, Arizona 85721, USA}

\begin{abstract}
 Binary neutron star mergers probe the dense-matter equation of state (EoS) across a wide range of densities and temperatures, from the cold conditions of the inspiral to the high-temperature matter of the massive neutron star remnant. In this paper, we explore the sensitivity of neutron star mergers to uncertainties in the finite-temperature part of the EoS with a series of merger simulations performed in full general relativity. We expand on our previous work to explore the interplay between the thermal prescription and the stiffness of the zero-temperature EoS, which determines the compactness of the initial neutron stars. Using a phenomenological model of the particle effective mass, $M^*$, to calculate the finite-temperature part of the EoS, we perform merger simulations for a range of thermal prescriptions, together with two cold EoSs that predict either compact or large-radius initial neutron stars.  We report on how the choice of $M^*$-parameters influences the thermal properties of the post-merger remnant, and how this varies for stars with different initial stellar compactness. We characterize the post-merger gravitational wave signals, and find differences in the peak frequencies of up to 190~Hz depending on the choice of $M^*$-parameters. 
 Finally, we find that the total dynamical ejecta is in general only weakly sensitive to the thermal prescription, but that a particular combination of $M^*$-parameters, together with a soft cold EoS, can lead to significant enhancements in the ejecta.
\end{abstract}

\maketitle

\section{Introduction}

Binary neutron star mergers probe the dense-matter equation of state (EoS) across a wide range of parameter space, from the zero-temperature and $\beta$-equilibrated conditions of the inspiral, to the hot and high-density conditions of the massive neutron star remnant that forms following the merger.

The first detection of gravitational waves (GWs) from the inspiral of a binary neutron star (BNS) merger, event GW170817 \cite{LIGOScientific:2017vwq}, has already provided exciting new constraints on the EoS of cold, dense matter, e.g.,  \cite{LIGOScientific:2018cki,Baiotti:2019sew,GuerraChaves:2019foa,Raithel:2019uzi,Chatziioannou:2020pqz,Annala:2021gom}. After the merger, however, significant shock-heating is expected to raise the temperature of the system to several tens of MeV \cite{Baiotti:2016qnr,Paschalidis:2016vmz}, at which point the thermal pressure can become a significant fraction of the cold pressure. This heating can influence the lifetime and dynamical evolution of the massive neutron star remnant, as well as the launching of the dynamical ejecta during the merger itself \cite[e.g.,][]{Bauswein:2010dn,Paschalidis:2012ff}. Observations of these post-merger properties thus provide insight into the EoS in a new region parameter space: at \textit{finite temperatures}.

In order to connect finite-temperature EoS constraints inferred from post-merger observations to the zero-temperature constraints inferred from the inspiral, one useful approach is to decouple the EoS into a cold and thermal component. For example, one can expand the pressure as 
\begin{equation}
\label{eq:Pth}
P(n,T,Y_e) = P_{\rm cold}(n, T, Y_e) + P_{\rm th}(n, T, Y_e),
\end{equation}
where $P_{\rm cold}$ is the zero-temperature pressure, $P_{\rm th}$ is the thermal contribution to the pressure, and $n$, $T$, and $Y_e$ indicate the baryon number density, temperature, and electron fraction of the matter. Such a decomposition makes it possible to systematically explore the relative sensitivity of a merger to current uncertainties in our understanding of cold dense matter, independently from the uncertainties in the finite-temperature part of the EoS. This decomposition also forms the basis of the so-called ``hybrid approach"  \cite{Janka1993}, in which thermal effects are parametrically calculated according to
\begin{equation}
\label{eq:hybrid}
P_{\rm th} = n E_{\rm th}\left(\Gamma_{\rm th}-1\right),
\end{equation}
where $E_{\rm th}$ is the thermal energy per baryon and $\Gamma_{\rm th}$ is the effective thermal index, which is assumed to constant.

In one early study of thermal effects in BNS mergers using this approach, Ref.~\cite{Bauswein:2010dn} showed that by using different values for $\Gamma_{\rm th}$ -- while keeping the zero-temperature part of the EoS the same -- it is possible to change the amount of dynamical ejecta by a factor of two and the peak frequency of the post-merger GWs by up to 390~Hz (see also \cite{Figura:2020fkj}). However, in the cores of realistic neutron stars, $\Gamma_{\rm th}$ has a strong density-dependence due to the degeneracy of the matter \cite[e.g.,][]{Constantinou:2015mna,Carbone:2019pkr,Huth:2020ozf}, which is neglected in the hybrid approach.

In order to go beyond the simplified assumption of a constant thermal index, Ref.~\cite{Raithel:2019gws} developed a new, phenomenological framework for calculating finite-temperature effects that includes the leading-order effects of degeneracy in the thermal prescription. The framework is based on Landau's Fermi liquid theory, in which the thermal contribution to the pressure and energy of the matter can be written purely in terms of the particle effective mass function $M^*(n)$ \cite{Baym1991,Constantinou:2015zia}. Ref. \cite{Raithel:2019gws}  introduced a new, two-parameter approximation of $M^*(n)$,  which can be used to robustly calculate the thermal pressure in Eq.~\ref{eq:Pth}, and thus to generically extend any cold EoS  to finite-temperatures.

With this ``$M^*$-framework", fully-finite temperature EoS models can be calculated to probe new regions of parameter space. The framework has been shown to recreate the pressure profiles of  a sample of tabulated, microphysical EoSs to within $\lesssim 15$\% accuracy across a range of densities and temperatures \cite{Raithel:2019gws}; and its performance, compared to a tabulated nuclear EoS, was recently validated in the context of numerical merger simulations in \cite{Raithel:2022nab}. 

In \cite{Raithel:2021hye}, we performed the first parameter study of the $M^*$-framework, exploring outcomes of BNS merger simulations for four different sets of effective mass parameters. These parameters were chosen to bracket the range of uncertainty in the finite-temperature part of the EoS spanned by a sample of models that are commonly used in merger simulations. The BNS mergers were evolved using a single cold EoS, onto which the four different thermal prescriptions were attached. It was found that the thermal profile of the remnant neutron stars and the post-merger GW signals indeed depend on the choice of $M^*$-parameters, but that the total ejecta depends only weakly on the finite-temperature part of the EoS.

In this work, we expand on the results of \cite{Raithel:2021hye} (hereafter Paper I) to explore the relative importance of finite-temperature effects on the post-merger properties for two new cold EoSs, which are designed to  span a wide range of stellar compactness. In particular, we perform BNS merger simulations for one soft and one stiff cold EoS, corresponding to models that predict small (11~km) and large (14~km) radii for intermediate-mass neutron stars, respectively. To each of these cold EoSs, we attach one of four different thermal prescriptions, corresponding to different choices of $M^*$-parameters. In general, we find that there is a complex interplay between the choice of $M^*$-parameters, the stellar compactness (or, more generally, the stiffness of the underlying cold EoS), and the thermal structure of the post-merger remnant. We quantify the degree of heating for each merger, including how this affects the structure and post-merger evolution of the remnant neutron star. In addition, we take advantage of recent improvements to our code, in terms of the post-merger convergence and improved modeling of the cold EoS \cite{Raithel:2022san}, to go beyond what was reported in Paper I and provide detailed analyses of the post-merger GW spectra and ejecta.

The outline of the paper is as follows. We start in Sec.~\ref{sec:methods} with an overview of our numerical methods and describe the details of our EoS construction. We present the results of our simulations in Sec.~\ref{sec:results}, starting with an overview of the merger dynamics in Sec.~\ref{sec:overview} and the heating of the system throughout the evolution in Sec.~\ref{sec:thermal_evo}. We describe the thermal structure of the remnant neutron stars in Sec.~\ref{sec:thermal}, the gravitational wave emission in Sec.~\ref{sec:gw}, and the properties of the dynamical ejecta in Sec.~\ref{sec:ejecta}. We summarize and discuss the results in Sec.~\ref{sec:discussion}.

Unless otherwise noted, we use geometric units, where $G=c=1$.

\section{Numerical methods}
\label{sec:methods}
All simulations were performed using the dynamical spacetime, general-relativistic (magneto)-hydrodynamics code with adaptive mesh refinement of \coderefs, as it was recently extended in \cite{Raithel:2021hye,Raithel:2022san}. The code is built within the Cactus/Carpet framework \cite{Allen2001,Schnetter:2003rb,Schnetter:2006pg}. We refer the reader to these works for detailed information on the code; and we here highlight a few key aspects relevant to the present simulations.

\subsection{EoS construction}
In order to investigate the influence of stellar compactness on the role of finite-temperature effects in a BNS merger, we adopt two zero-temperature EoSs that span a wide range of neutron star compactness. For each of these cold EoSs, we add on one of four different thermal prescriptions, for a total of eight simulations.

\begin{figure}[!ht]
\centering
\includegraphics[width=0.45\textwidth]{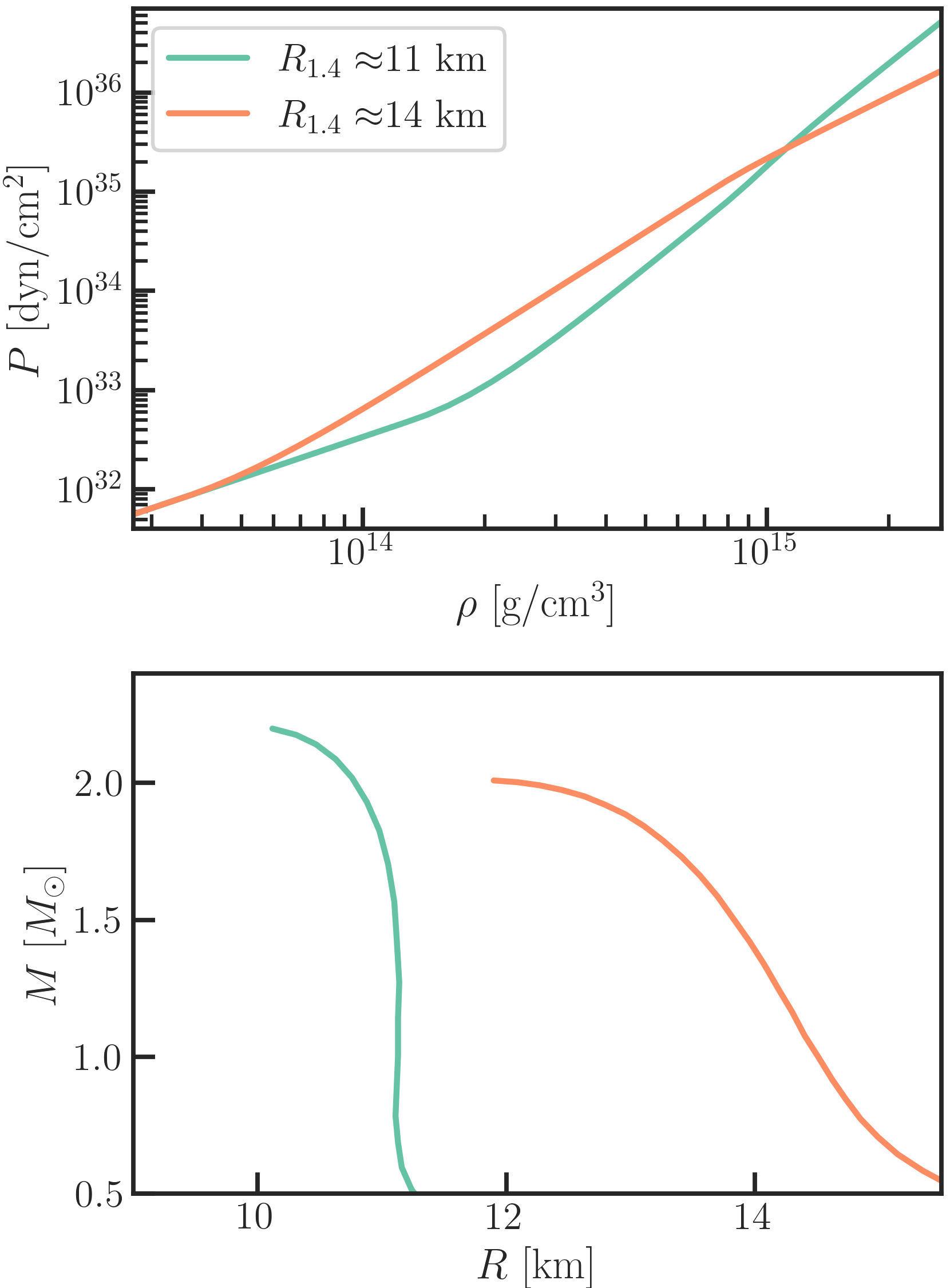} 
\caption{\label{fig:MR} Top: Pressure as a function of density for the zero-temperature, $\beta$-equilibrium slice of each equation of state considered in this work.  Bottom: The corresponding mass-radius relations. The $R_{1.4}\approx11$~km model is shown in green, while the $R_{1.4}\approx14$~km model is shown in orange.}
\end{figure}

For our baseline, zero-temperature models, we use a parametrized representation of the wff2 \cite{Wiringa:1988tp} and H4 \cite{Lackey:2005tk} tabular EoSs. We represent these EoSs with the generalized piecewise polytropic parametrization of \cite{OBoyle:2020qvf}, which we recently studied in the context of numerical merger simulations in \cite{Raithel:2022san}. This parametrization allows us to analytically represent these EoSs with a small number of parameters, while still ensuring that the EoS pressure is smooth and differentiable at all densities. We describe the fits to these EoSs in  Appendix~\ref{sec:appendixGPP}.

The resulting smoothed-polytropic EoSs are shown, together with their mass-radius relations, in Fig.~\ref{fig:MR}. The smoothed-wff2 model predicts the areal radius of a 1.4~$\Ms$ star to be $R_{1.4}=11.1$~km and has a maximum mass of $M_{\rm max}=2.2~\Ms$. In contrast, the smoothed-H4 model corresponds to  $R_{1.4}=13.99$~km and $M_{\rm max}=2.01~\Ms$.  Thus, compared to the exploration of thermal effects in Paper I, which used a zero-temperature EoS with $R_{1.4}=12$~km, the models used here span a broad range of stellar compactness. For convenience, and to focus on the impact of stellar compactness in our simulations, we will refer to these models by their characteristic radii, $R_{1.4}$, throughout this work. 
 
We extend these cold EoSs to finite-temperatures using the $M^*$-framework of \cite{Raithel:2019gws}. This framework is based on a two-parameter model of the particle effective mass function, $M^*(n)$. The free parameters include a density-transition parameter, $n_0$, which describes the density at which the effective mass starts to decrease from the vacuum rest mass, and a  power-law parameter, $\alpha$, which characterizes the rate at which the effective mass function decreases with  density. This model of the effective mass function accounts for the leading order effects of degeneracy, through a Fermi Liquid Theory based model. We use the same four sets of $M^*$-parameters that were studied in Paper I: i.e., $n_0=0.08$ and 0.22~fm$^{-3}$ and $\alpha=0.6$ and 1.3.\footnote{We note that in some figure legends, we will suppress the units on $n_0$ in order to save space. Throughout this paper, the units for $n_0$ are always fm$^{-3}$.} This choice of parameters was found to approximately bracket the range of $M^*$-values for a sample of commonly-used, finite-temperature EoS tables  \cite{Raithel:2019gws} (see also Appendix~\ref{sec:compareM}).

\begin{figure}[!ht]
\centering
\includegraphics[width=0.5\textwidth]{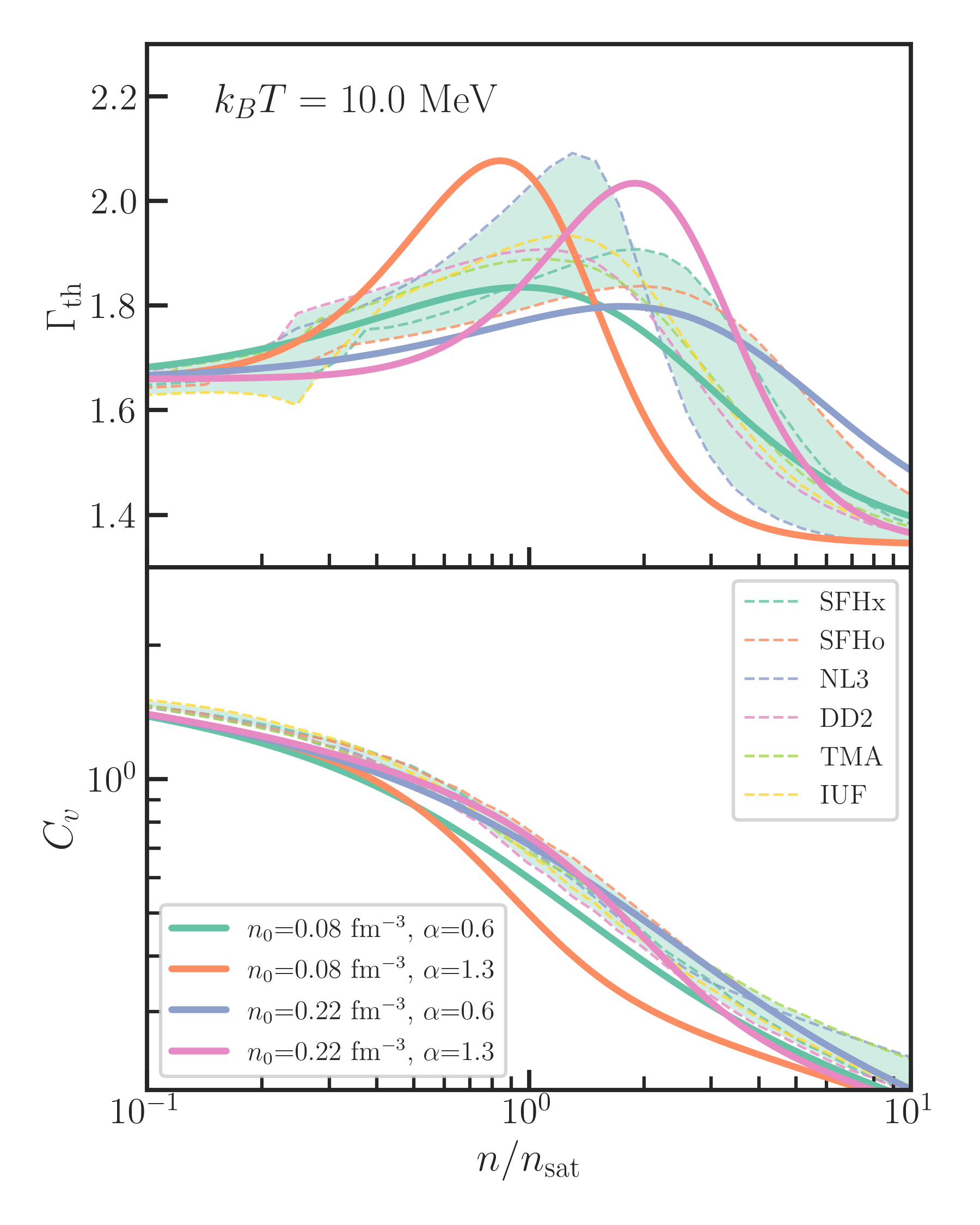} 
\caption{\label{fig:gth} Top: Effective thermal index for the four sets of $M^*$-parameters explored in this work, for a fixed temperature of $k_BT=10$~MeV, and for a composition corresponding to the conditions for cold, neutrinoless, $\beta$-equilibrium. The dotted lines indicate the thermal index spanned by a sample of five tabulated, finite-temperature EoSs, with the green shading added to guide the eye. Bottom: Specific heat for the same models and same conditions. The quantities are plotted as a function of the number density, relative to the nuclear saturation density, $n_{\rm sat}=0.16$~fm$^{-3}$. }
\end{figure}

We compare the predictions of this set of $M^*$-parameters to a subset of these public finite-temperature EoS tables in Fig.~\ref{fig:gth}, which shows the effective thermal index for each model, at a fixed temperature of $T=10$~MeV. From eq.~\ref{eq:hybrid}, we can rearrange to solve for the thermal index according to
\begin{equation}
\label{eq:gammath}
\Gamma_{\rm th} = 1 + \left( \frac{P_{\rm th}(n, T, Y_e)}{n E_{\rm th}(n,T,Y_e)} \right),
\end{equation}
where the thermal pressure and energy are provided by the particular EoS model. Figure~\ref{fig:gth} also shows the specific heat at constant volume, $C_V$, which is defined according to
\begin{equation}
C_V = \frac{\partial E_{\rm th}}{\partial{T} } \biggr \rvert_n.
\end{equation}
The full $M^*$-framework expressions for $P_{\rm th}$ and $E_{\rm th}$, as a function of density, temperature, and electron fraction, are provided in \cite{Raithel:2019gws,Raithel:2021hye}.
 
\begin{figure*}[!ht]
\centering
\includegraphics[width=0.9\textwidth]{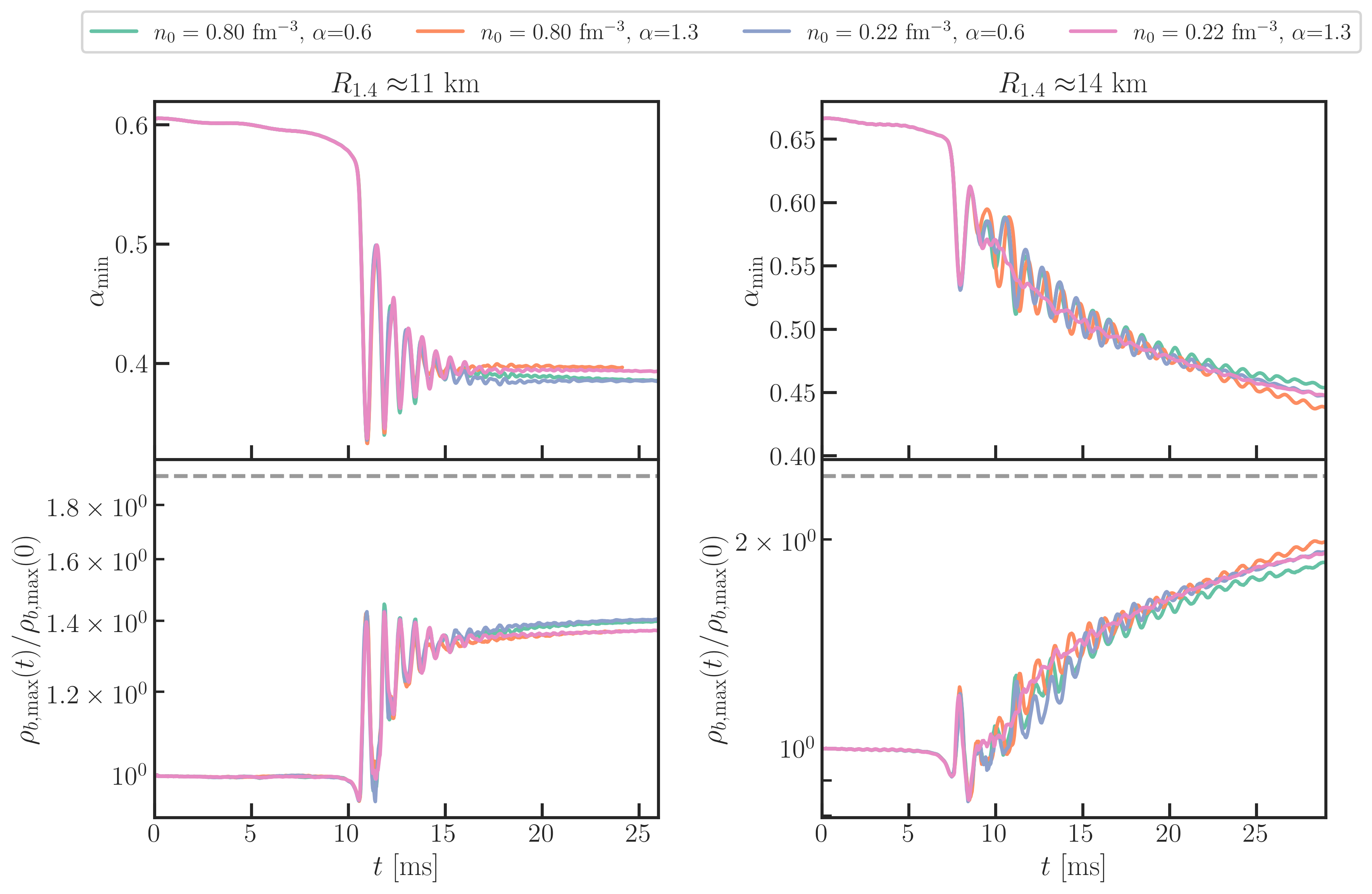} 
\caption{\label{fig:rhob} Top: Evolution of the minimum lapse function, $\alpha_{\rm min}$. Bottom: Evolution of the maximum rest-mass density, $\rho_{\rm b,~max}$, relative to the initial rest-mass density at the start of the evolutions. The evolutions with the \Rsmall cold EoS are shown in the left column, while those with the \Rbig cold EoS are shown in the right column. For comparison, the normalized central density for a star with the maximum mass of the supramassive sequence (i.e., the maximum mass supported by uniform rotation) is shown in the dashed gray lines, for each cold EoS.} 
\end{figure*}

In extending the EoS to finite-temperatures, we also follow Paper I in assuming that the matter maintains its initial $\beta$-equilibrium composition  throughout the evolution. This allows  us to focus our analysis on the impact of thermal effects, thanks  to the separability of the $M^*$-framework, which treats thermal and out-of-equilibrium effects independently \cite{Raithel:2019gws}. The assumption of $\beta$-equilibrium throughout the evolution is further motivated by a recent study, which  showed that Urca processes can act to restore departures  from equilibrium within the post-merger remnant on a millisecond-timescale \cite{Most:2022yhe}.  In order to set the initial composition of the neutron stars, we set the leading-order coefficients of the nuclear symmetry energy to $S_0=32$~MeV for both cold EoSs, and $L=34.8$~MeV and $L=112.1$~MeV for the \Rsmall and \Rbig EoSs, respectively. The value for $S_0$ is chosen in order to be consistent with experimental and theoretical constraints \cite{Li:2021thg}; while, at leading order, the value for $L$ is determined by $S_0$ and  the zero-temperature pressure at $\rns$ \cite{Raithel:2019ejc,Most:2021ktk}. These two parameters, $S_0$ and $L$, together with a free parameter $\gamma$, uniquely determine the initial, $\beta$-equilibrium composition for the neutron stars \cite{Raithel:2019gws}.  We adopt $\gamma=0.6$, which is consistent with values inferred from tabulated EoSs. This initial composition is then held fixed throughout the evolutions. For additional details, see \cite{Raithel:2019gws,Raithel:2021hye}.

\subsection{Initial conditions and numerical set-up}
\label{sec:numerics}

For all simulations, we construct initial data using \texttt{LORENE} \cite{Lorene}. The initial configurations describe two unmagnetized, irrotational, equal-mass neutron stars in a quasi-circular orbit, with an Arnowitt-Deser-Misner (ADM) mass of 2.6~$\Ms$ and an initial separation of 40 km. The neutron stars start at zero temperature and have an initial coordinate equatorial radius of either 8.8~km or 12.0~km, depending on the cold EoS model.

For each binary evolution, we use nine spatial refinement levels which are separated by a 2:1 refinement ratio.  The grid is set up such that the innermost refinement level is approximately 30\% larger than the initial neutron stars. The resulting computational domain extends to 2880~km for the \Rsmall evolutions, and to 3900~km for the \Rbig evolutions. We impose equatorial symmetry to reduce
computational costs.

The finest-level grid spacing covers the coordinate diameter of each initial neutron star with $\sim$125 grid points across  the x-direction. This corresponds to a finest-level grid spacing of $\Delta x=$140~m for the \Rsmall evolutions and $\Delta x=195$~m for the \Rbig evolutions.

Finally, the evolutions implement a continuous parabolic damping of the Hamiltonian constraint, as was introduced in \cite{Raithel:2022san}, to ensure convergence of the Hamiltonian constraint violations. The implementation is identical to what was described in that work, namely, we use the same constant damping coefficient $c_H = 0.0045$~km.

Except where explicitly indicated above, the rest of the details of the evolution are identical to \cite{Raithel:2021hye}.

\section{Results}
\label{sec:results}

We turn now to the results of our simulations. We start with a brief overview of the global properties of the merger.  In all cases, we simulate the final few orbits of the binary,  through the merger itself, and for 12-20~ms post-merger. 

\subsection{Merger overview}
\label{sec:overview}

All binaries start with an initial coordinate separation of 40 km. For the \Rbig (\Rsmall) models, this corresponds to the final $\sim$3  (4.5) orbits prior to merger. In general, we find negligible variations in the time-to-merger,
defined as the time when the gravitational wave strain reaches a peak amplitude (see Sec.~\ref{sec:gw}), for the different choices of $M^*$-parameters.  The minimal impact of the $M^*$-parameters  on the time-to-merger is expected, given the negligible inspiral heating typically found in merger simulations \cite[e.g.,][]{Oechslin:2006uk,Bauswein:2010dn} (see also Sec.~\ref{sec:thermal_evo}).

In all evolutions, a massive neutron star remnant forms after the merger, which does not collapse on the time-scales simulated, as evidenced by the slow growth of the maximum rest-mass density of the remnant, $\rho_{\rm b,~max}$,  and the correspondingly slow decay of the minimum lapse function at late times in our simulations. We show these functions in Fig.~\ref{fig:rhob}.

For the \Rsmall evolutions, the rest mass of the remnant is 2.94-2.95~$\Ms$, which is smaller than the maximum rest mass of the  zero-temperature Kepler sequence of $M_{\rm Kep}=3.15~\Ms$ for this EoS. Thus, we expect these remnants to remain stable to late times. In contrast, for the \Rbig evolutions, the rest mass of the remnant is 2.85-2.86~$\Ms$, which exceeds the Kepler maximum rest mass of   $M_{\rm Kep}=2.68~\Ms$ for this EoS.  This suggests that the remnants evolved with this cold EoS are supported against gravitational collapse by a combination of differential rotation and thermal pressure  \cite{Paschalidis:2012ff};  and that over longer timescales, the remnants will collapse. We do not find evidence of collapse during our simulations. We note, however, that Fig.~\ref{fig:rhob} shows that the remnants for this EoS are still contracting throughout our simulations, with the rate of contraction slowing at late times, suggesting that the remnant may be starting to stabilize. Throughout this work, we will use the late-time properties of the \Rbig evolutions as approximately representative of the final remnant, but we note that the remnant may still experience further contraction.
 
 Finally, we note a weak dependence of the late-time $\rho_{\rm b,~max}$  on the choice of $M^*$-parameters in Fig~\ref{fig:rhob}. For the evolutions with the \Rsmall EoS, we find a 2\% difference in $\rho_{\rm b,~max}$ at the end of our evolutions, with the $\alpha=0.6$ evolutions leading to larger maximum densities. The picture is less clear for the \Rbig remnants due to the continued contraction, but in this case, the
$(n_0=0.08~\text{fm}^{-3}, \alpha=1.3)$ thermal prescription leads to the largest late-time $\rho_{\rm b,~max}$, while the $\alpha=0.6$ prescriptions lead to smaller maximum densities. In summary, the choice
of $M^*$-parameters has a small, but complicated effect on the maximum rest-mass density of the remnant. We revisit this dependence in Sec.~\ref{sec:ejecta}.

\begin{figure}[!ht]
\centering
\includegraphics[width=0.45\textwidth]{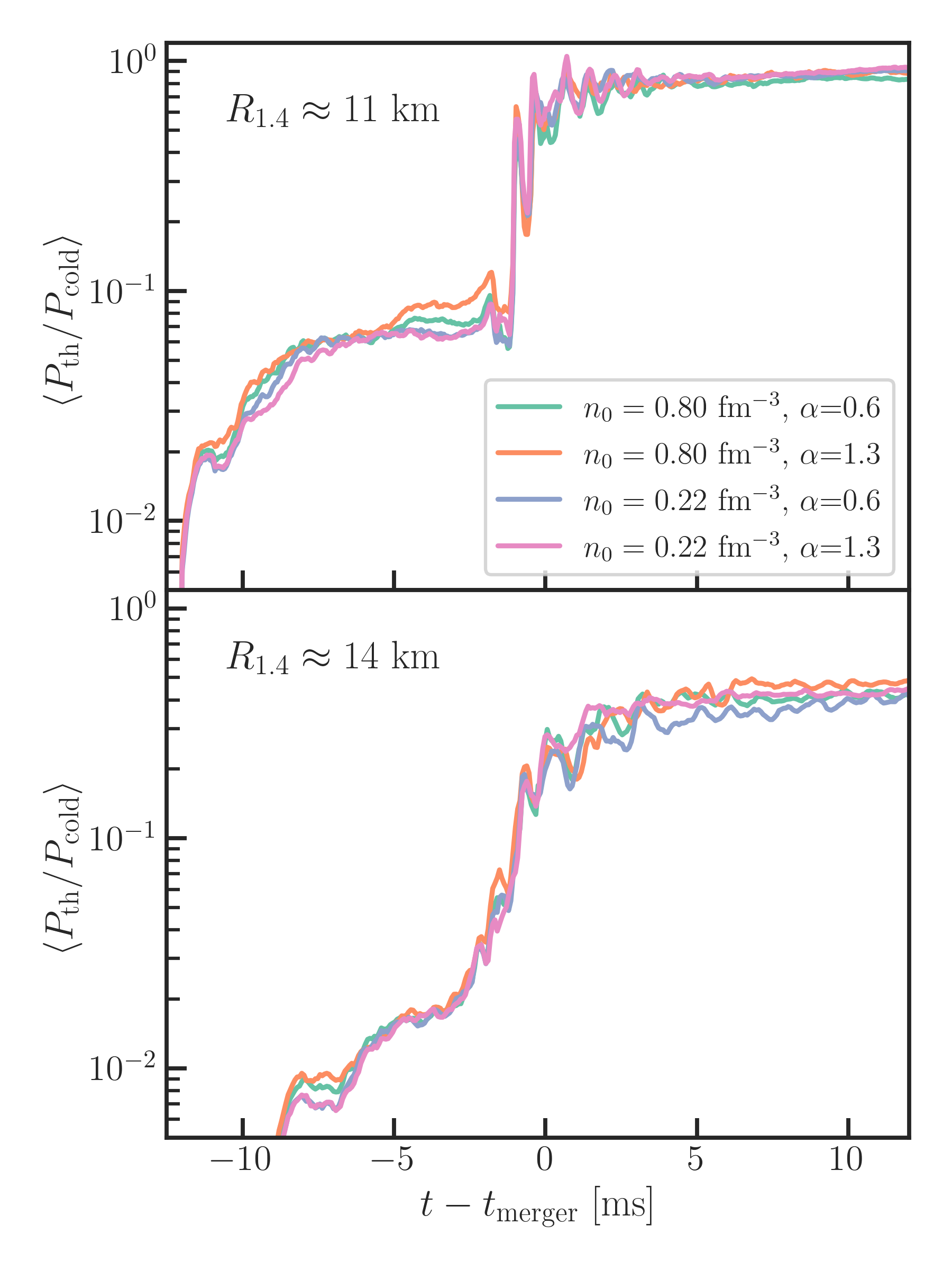}  
\caption{\label{fig:Pth_vs_t}  Evolution of the average thermal pressure, $P_{\rm th}$, relative to the zero-temperature pressure, $P_{\rm cold}$, throughout the simulations. The averages are density-weighted and include all matter with $\rho_b > 0.1 \rns$. The neutron stars stay thermodynamically cold (i.e., the thermal pressure is subdominant to the cold pressure) until merger.}
\end{figure}

\begin{figure*}[!ht]
\centering
\includegraphics[width=0.9\textwidth]{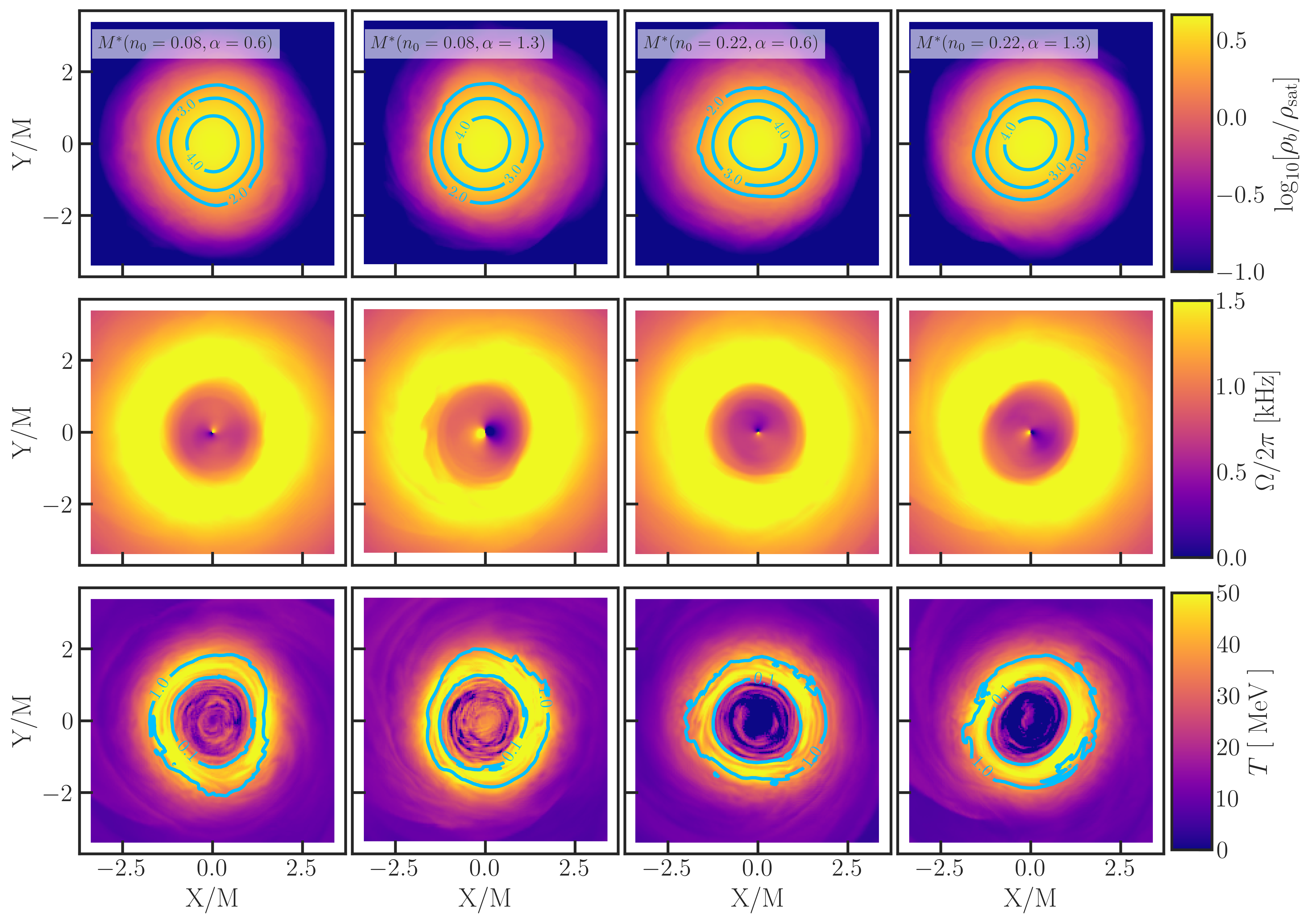} 
\caption{\label{fig:PthT_wff2} Late-time ($t-t_{\rm merger}\approx12$~ms), equatorial snapshots for the \Rsmall evolutions. The top row shows the rest-mass density relative to the nuclear saturation density ($\rho_{\rm sat}=2.7\times10^{14}$g/cm$^3$), with blue lines indicating specific contours of $\rho_b = 2,3$ and $4\rns$. The middle row shows the angular velocity of the fluid in each snapshot. The bottom row  shows the corresponding temperature of the fluid, with contours indicating where the thermal pressure is equal to 100\% and 10\% of the cold pressure. The columns correspond to the different thermal prescriptions  (i.e., the different choices of $M^*$-parameters).}
\end{figure*}

\begin{figure*}[!ht]
\centering
\includegraphics[width=0.9\textwidth]{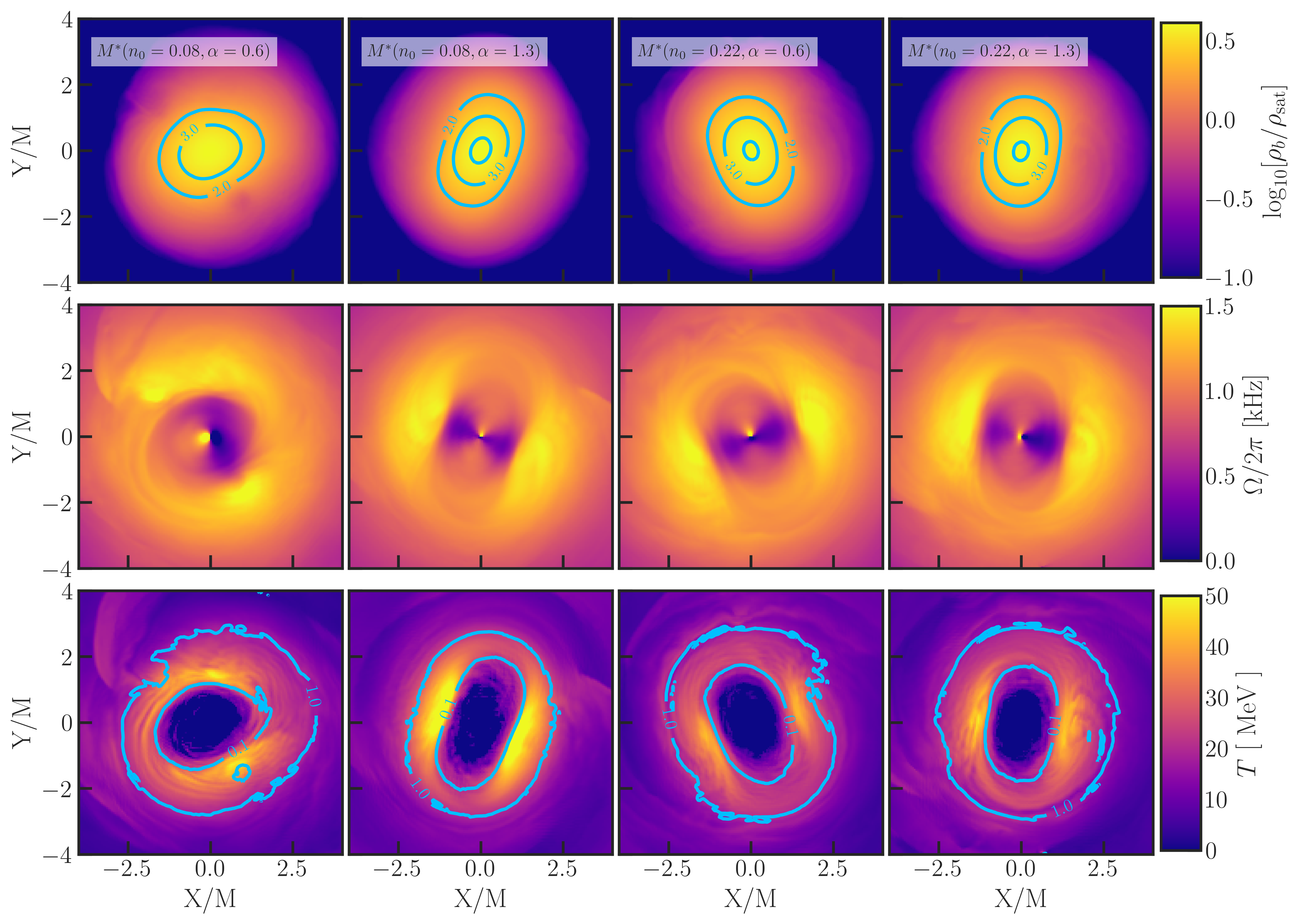} 
\caption{\label{fig:PthT_h4} Same as Fig.~\ref{fig:PthT_wff2}, but for the \Rbig EoS evolutions. These late-time snapshots correspond to $t-t_{\rm merger}=18.6$~ms. }
\end{figure*}

\subsection{Thermal evolution over time}
\label{sec:thermal_evo}
In order to assess the heating of the neutron stars during the inspiral, merger, and post-merger phases, Fig.~\ref{fig:Pth_vs_t} shows the average thermal pressure of the neutron stars, $P_{\rm th}$, relative to the zero-temperature pressure, $P_{\rm cold}$, over time. We calculate these density-weighted averages from slices taken along the equatorial ($Z=0$) plane. In order to focus on the bulk neutron star matter, these averages only include matter with rest-mass density $\rho_b > 0.1 \rns$, where $\rns=2.7\times10^{14}$g/cm$^3$ is the nuclear saturation density.
 
For all evolutions considered, we find that there is negligible pre-merger heating, with $\langle P_{\rm th}/P_{\rm cold}\rangle \lesssim 0.1$. This is consistent with previous studies, which have shown that tidal heating during the inspiral is minimal \cite[e.g.,][]{Oechslin:2006uk,Bauswein:2010dn}. Figure~\ref{fig:Pth_vs_t} shows a sharp increase in the thermal pressure at merger, as expected by the shock heating of the collision. Following merger, the average thermal pressure is a large fraction (up to $\sim 100\%$) of the cold pressure and, as a result, may be expected to play a role in the post-merger evolution. We investigate the differences in the thermal properties of the post-merger remnant  in detail in the following sections.

\begin{table*}
\centering
\begin{tabular}{ccccccccc}
\hline \hline
Cold EoS & $M^*$-parameters & $C_V(\rho_{\rm sat}, T=10$ MeV) & $\langle P_{\rm th}/P_{\rm cold} \rangle$ & $\langle P_{\rm th} \rangle$ [MeV/fm$^{-3}$] & $\langle{T}\rangle$ [MeV] \\
\hline \hline
       &  $n_0=0.08$ fm$^{-3}$, $\alpha=0.6$  &0.60 & 0.33 & 4.6 & 33.1 \\
\Rsmall &  $n_0=0.08$ fm$^{-3}$, $\alpha=1.3$  &0.50 & 0.27 & 3.8 & 36.5 \\
       &  $n_0=0.22$ fm$^{-3}$, $\alpha=0.6$  &0.71 & 0.34 & 4.0 & 26.3 \\
       &  $n_0=0.22$ fm$^{-3}$, $\alpha=1.3$  &0.74 & 0.40 & 4.7 & 27.7 \\
\hline \hline
       &  $n_0=0.08$ fm$^{-3}$, $\alpha=0.6$  &0.60 & 0.12 & 1.7 & 17.3 \\
\Rbig &  $n_0=0.08$ fm$^{-3}$, $\alpha=1.3$  &0.50 & 0.12 & 1.6 & 18.5 \\
       &  $n_0=0.22$ fm$^{-3}$, $\alpha=0.6$  &0.71 & 0.10 & 1.5 & 14.3 \\
       &  $n_0=0.22$ fm$^{-3}$, $\alpha=1.3$  &0.74 & 0.14 & 2.0 & 16.4 \\ 
\hline \hline
\end{tabular}
\caption{Average late-time properties of the remnant from each simulation. All averages are computed including matter with densities $\rho \ge \rho_{\rm sat}$. From left to right, the columns indicate the cold EoS, the thermal prescription, the specific heat at $\rns$ and $T=10$~MeV, and the density-weighted averages of the following thermal quantities: the thermal pressure relative to the cold pressure, the thermal pressure, and the temperature.}
  \label{table:thermal}
\end{table*}

\begin{figure*}[!ht]
\centering
\includegraphics[width=0.9\textwidth]{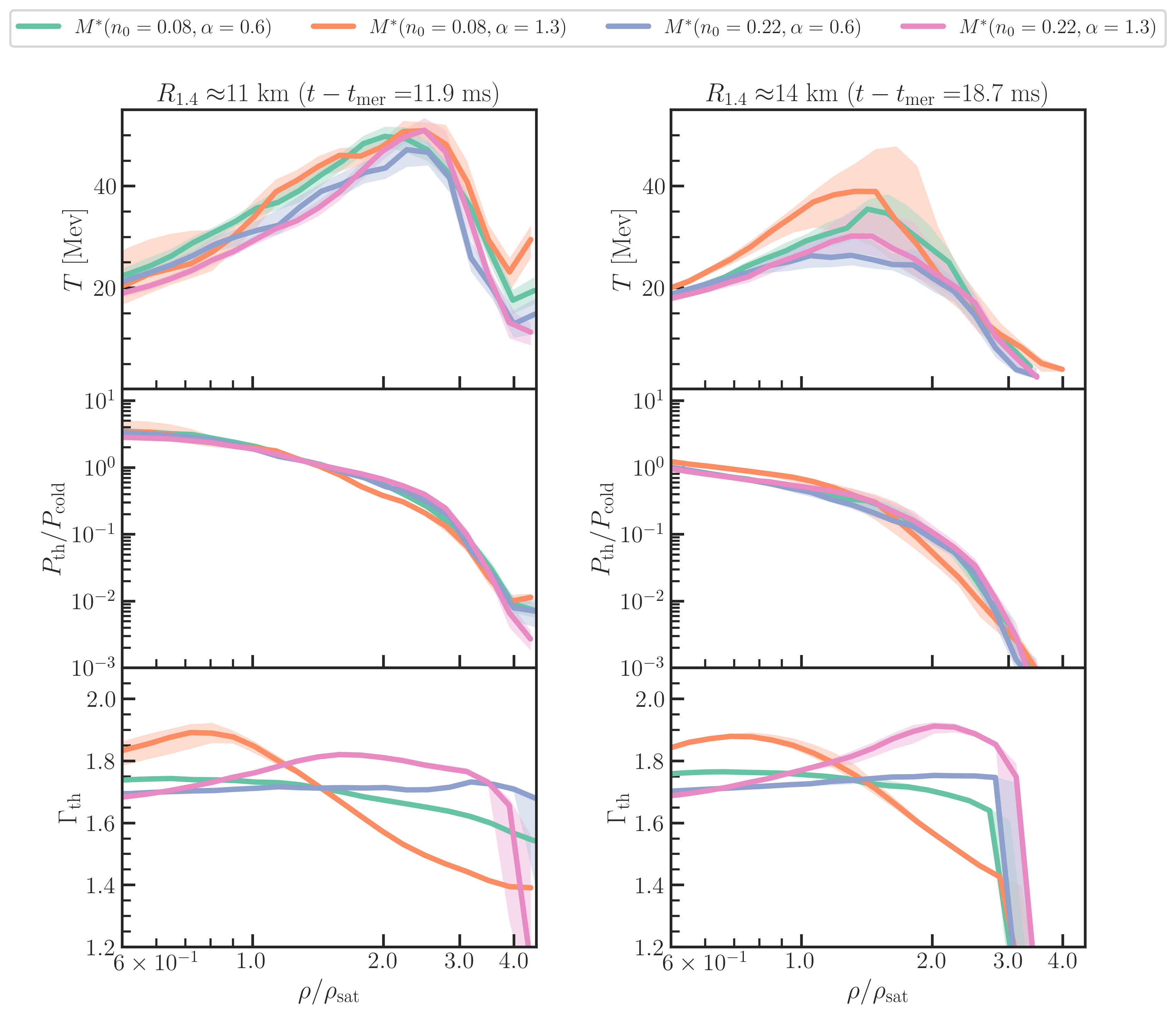}  
\caption{\label{fig:PthT_dens} Median thermal properties from the late-time remnant. These median quantities are extracted for the equatorial slices shown in Figs.~\ref{fig:PthT_wff2} and \ref{fig:PthT_h4}. From top to bottom, we show: the temperature, $T$; the thermal pressure relative to the cold pressure, $P_{\rm th}/P_{\rm cold}$; and the effective thermal index, $\Gamma_{\rm th}$. The solid line indicates the median value, calculated for bins that are spaced logarithmically in density, with the shaded regions indicating 68\% bounds. }
\end{figure*}
  
\subsection{Late-time properties of the neutron star remnants}
\label{sec:thermal}

Figures~\ref{fig:PthT_wff2} and \ref{fig:PthT_h4} show equatorial snapshots of the remnants for the evolutions with the \Rsmall EoS at 12~ms post-merger and \Rbig EoS at 18.6~ms post-merger, respectively. The top row of each figure shows the rest mass density $\rho_b$ relative to $\rho_{\rm sat}$, the middle row shows the angular velocity of the fluid within the remnant for each snapshot, and the bottom row shows the corresponding temperature. We highlight a few key features of these snapshots here, and will discuss the details in the following subsections.

In these late-time snapshots, the remnants evolved with the \Rsmall cold EoS (shown in Fig.~\ref{fig:PthT_wff2}) have become nearly axisymmetric, while the the late-time remnants for the \Rbig evolutions (Fig.~\ref{fig:PthT_h4})  show a pronounced bar-mode in the density distribution. 
This bar mode drives significant GW emission (see Sec.~\ref{sec:gw}),
which drains angular momentum from the remnant and will cause the structure to 
become more axisymmetric over time. 
As a result, this bar-mode may help to explain the late-time contraction
discussed in Sec.~\ref{sec:overview}, for this stiff cold EoS.
For all of our simulations, the remnants exhibit strong differential rotation within the remnants, with small differences depending on the choice of $M^*$-parameters (consistent with our results from Paper I).  We also find the merger resulted in significant heating, with temperatures above 50~MeV reached in most of the remnants,
and overall more heating in the \Rsmall evolutions compared to the \Rbig evolutions.

From these late-time snapshots, we calculate the average thermal properties  of the remnants and report these global summary statistics in Table~\ref{table:thermal}. Finally, we also compute the median thermal properties as a function of the density within the remnant, using bins that are spaced log-uniformly in the density. Figure~\ref{fig:PthT_dens} shows these  median values, along with their 68\% scatter,  for the temperature, the normalized thermal pressure $P_{\rm th}/P_{\rm cold}$, and the effective thermal index $\Gamma_{\rm th}$. We discuss these thermal properties in detail in the following subsections.

\subsubsection{Temperature}

We start with the temperature profiles of the remnants, which directly influence the neutrino opacity of the remnant and thus may affect the long-term cooling of the remnant and  neutrino irradiation of the disk \cite[e.g.,][and references therein]{Paschalidis:2016vmz,Radice:2020ddv,Foucart:2022bth}.

For all four thermal prescriptions with the \Rsmall EoS (shown  in Fig.~\ref{fig:PthT_wff2} across columns) , we find evidence of a high-temperature ring in the remnant,  located at $X/M\simeq1.5$ in these spatial snapshots. In contrast, the temperature of the innermost core ($X/M < 1$) varies significantly, depending on the choice of $M^*$-parameters. In particular, the evolutions with $M^*$-parameters of $n_0=0.08$~fm$^{-3}$ have hotter cores, while the evolutions with $n_0=0.22$~fm$^{-3}$ are characterized by temperatures of $\lesssim10$~MeV in the innermost core ($X/M < 1$).
 
 We found a similar behavior in Paper I, in which the temperature of the inner core of the remnants  depended sensitively on the choice of $M^*$-parameters. However, in that work, for a cold EoS with $R_{1.4}=12$~km, the inner-core temperature was correlated with $\alpha$ and was relatively insensitive to $n_0$, in contrast to what we now find in Fig.~\ref{fig:PthT_wff2}. Thus, already, this suggests that the mapping between the $M^*$-parameters and the thermal properties of the remnant is not straightforward for models with different stellar compactness.

The location of the high-temperature ring aligns closely with the maximum angular velocity within the remnant, as shown in the middle row of Fig.~\ref{fig:PthT_wff2}, and
as has been seen previous studies \cite[e.g.,][]{Hanauske:2016gia,Kastaun:2016yaf}. 
The ring-like structures in Fig.~\ref{fig:PthT_wff2} also correspond to a strong peak in the 1D density-profiles of the temperature in Fig.~\ref{fig:PthT_dens}. From these 1D profiles, we find that the maximum temperatures reach up to 40-50~MeV, depending on the choice of $M^*$-parameters, and occur at densities of $\sim2-2.5 \rho_{\rm sat}$ (or, equivalently, at $\sim0.5\rho_{\rm b,~max}$). The density-weighted average temperatures within the remnants also depend on the $M^*$-parameters, with the evolutions with $n_0=0.08$~fm$^{-3}$ leading to larger average temperatures of $\langle T \rangle$ = 33-36~MeV, compared to the $n_0=0.22$~fm$^{-3}$ evolutions, which yield $\langle T \rangle$ = 26-28~MeV (see Table~\ref{table:thermal}).

The picture changes somewhat for the evolutions performed with the \Rbig cold EoS. Perhaps as a consequence of the continued contraction for these remnants, we do not find a high-temperature ring, although we note the presence of two hot spots at $X/M\simeq1.5$ in the bottom row of Fig.~\ref{fig:PthT_h4}, that suggest such a ring may be forming. This is further apparent in the 1D density profiles in Fig.~\ref{fig:PthT_dens}, where we see a weaker and broader, but nonetheless well-defined peak in the median temperature distribution located at $\sim1.5-2 \rho_{\rm sat}$ (or, equivalently, 0.3-0.4$\rho_{\rm b,~max}$). Interestingly, for all sets of $M^*$-parameters, we find that the temperature of the innermost core is low, $\lesssim8$~MeV, for these evolutions.

Compared to the \Rsmall evolutions, the \Rbig evolutions generally have lower average temperatures in the remnant, with $\langle{T}\rangle=14-18$~MeV (see Table~\ref{table:thermal}). The more significant heating for the \Rsmall evolutions is not surprising:  given the smaller radii, the stars reach shorter separations before merging and thus experience higher-velocity impacts \cite[e.g.,][]{Bauswein:2013yna}. Nevertheless, even with this reduced heating, we find that the same trend between $n_0$ and the average remnant temperature persists with this stiffer cold EoS: i.e., evolutions with $n_0=0.08$~fm$^{-3}$ lead to larger average remnant temperatures, compared to the evolutions with $n_0=0.22$~fm$^{-3}$. 

Because $n_0$ determines the density at which the effective mass starts to decrease, the correlation between $\langle T \rangle$ and $n_0$ also implies a correlation with $M^*$. Indeed, we find that, for both cold EoSs, the average remnant temperatures are loosely correlated with the effective mass or with the specific heat at $\rns$: in general, smaller values of $C_v(\rns, T=10 \text{MeV})$ or $M^*(\rns)$ are correlated with smaller $\langle T \rangle$ (see Table~\ref{table:thermal}). However, the correlation is not exact, as seen in Table~\ref{table:thermal}, suggesting that the density-dependence of $M^*$ or $C_v$ at densities beyond $\rns$ may also play a role in the heating of the remnant.

\subsubsection{Thermal pressure}

The middle panel of Fig.~\ref{fig:PthT_dens} shows the median profiles (and $68\%$ bounds) of $P_{\rm th}/P_{\rm cold}$ for the late-time remnants.  For all of the evolutions, we find that the thermal pressure is a significant fraction of the cold pressure to high densities within the remnant. For the \Rsmall evolutions, the median thermal pressure exceeds 10\% of the cold pressure at densities of up to $3\rns$ or, equivalently, $0.7\rho_{\rm b,~max}$. For the \Rbig evolutions, $P_{\rm th}$ exceeds this threshold at densities of up to $\sim2\rho_{\rm sat}$ (or $0.5-0.6\rho_{\rm b,~max}$). For both cold EoSs, we find that the average ${P_{\rm th}/P_{\rm cold}}$ in the remnant can vary by up to 40-50\%, depending on the choice of $M^*$-parameters, but that the dependence on individual $M^*$-parameters is not straightforward (see Table~\ref{sec:thermal}). As we saw with the remnant temperatures, the \Rsmall evolutions lead to systematically higher $\langle {P_{\rm th}/P_{\rm cold}} \rangle$, as a result of the more extreme collisions.

Given the sensitivity of the thermal pressure to the choice of the $M^*$-parameters, we expect the long-term stability of the remnant and the post-merger dynamical evolution to also be sensitive to these parameters. We discuss the possible signatures of this dependence in the post-merger gravitational wave signals in Sec.~\ref{sec:gw}.

\subsubsection{Thermal index}

Finally, the bottom row of Fig.~\ref{fig:PthT_dens} shows the median thermal index, $\Gamma_{\rm th}$ within each finite-temperature remnant. In the $M^*$-framework, the parameter $\alpha$ controls the density-dependence of $\Gamma_{\rm th}$ at fixed temperature. This can be seen in Fig.~\ref{fig:gth}, where the choice of $\alpha=1.3$ leads to a steeper $\Gamma_{\rm th}$ function, while $\alpha=0.6$ leads to a shallower density-dependence.

Due to the non-uniform temperature profile within the remnants, the actual $\Gamma_{\rm th}(n,T)$ distributions within the remnant differ significantly from the constant temperature predictions of Fig.~\ref{fig:gth}. In particular, the $(n_0=0.22$~fm$^{-3}$, $\alpha=1.3)$ prescription leads to the most rapid density-variation in $\Gamma_{\rm th}$, even compared to the other $\alpha=1.3$ evolution. This has important implications
for the ejected mass, as we discuss in Sec.~\ref{sec:ejecta}.

When comparing between the evolutions with the  \Rsmall and \Rbig cold EoSs, we find small differences between the $\Gamma_{\rm th}$ distributions within the remnants but, overall, we conclude that they are qualitatively similar.

\subsection{Gravitational wave signals}
\label{sec:gw}

\begin{figure}[!ht]
\centering
\includegraphics[width=0.45\textwidth]{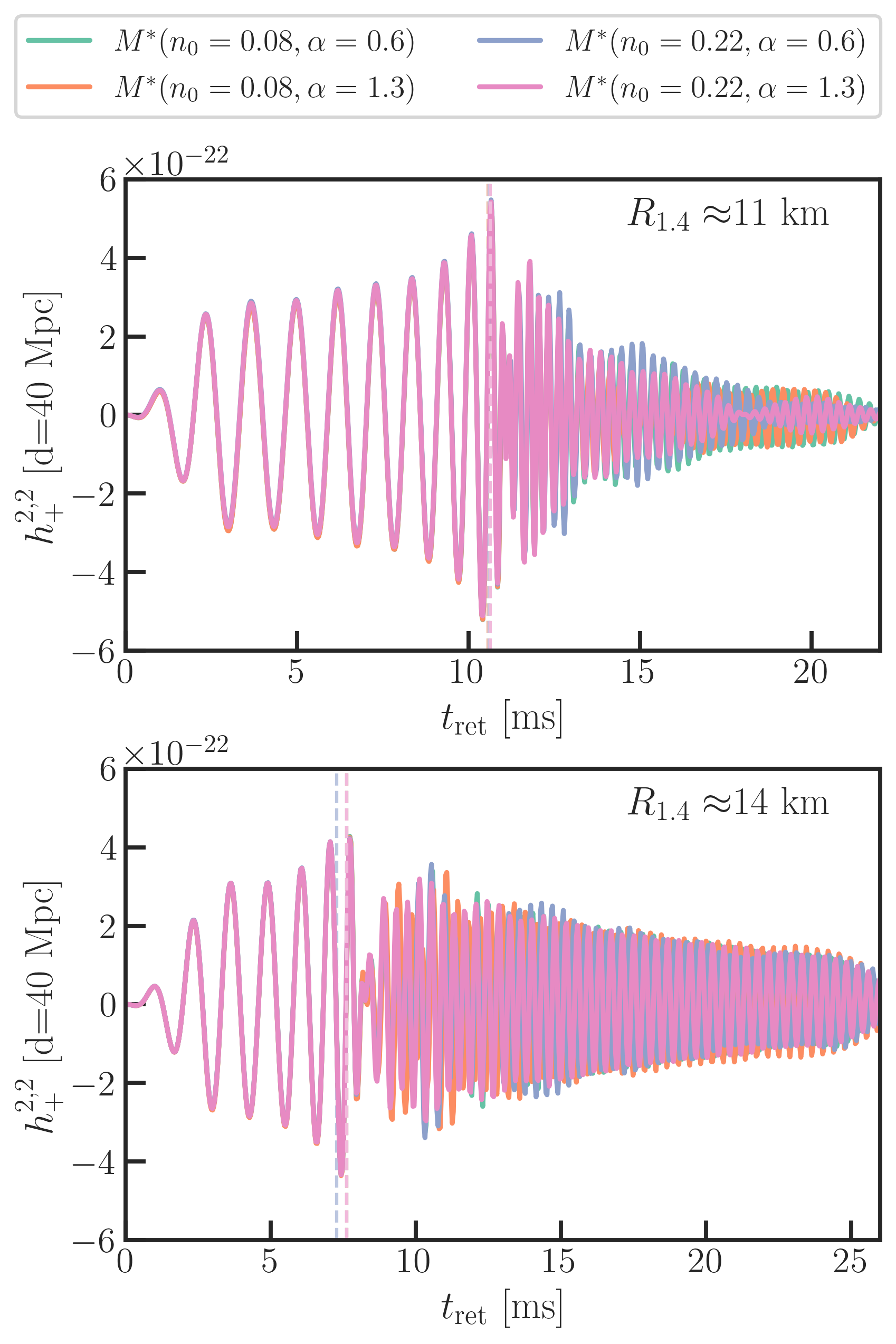} 
\caption{\label{fig:strain} Plus-polarized gravitational wave strain, for the  $\ell=m=2$ mode. The strain is plotted relative to the retarded time and is scaled to a distance of 40~Mpc. The evolutions performed with the \Rsmall cold EoS are shown in the top figure, while the \Rbig evolutions are shown in the bottom figure. For a given cold EoS, the inspiral gravitational waves are identical, but differences emerge after the merger (indicated with the vertical dashed lines), depending on the choice of $M^*$-parameters.} 
\end{figure}

We extract the GW radiation from each of our simulations using the Newman-Penrose scalar $\psi_4$, which is decomposed onto $s=-2$ spin-weighted spherical harmonics at large radii ($r\ge 300 \Ms$). We calculate the $+$ and $\times$ polarizations of the strain, $h$, via the relation $\psi_4 = \ddot{h}_{+} - i \ddot{h}_{\times}$, using the fixed-frequency integration method of \cite{Reisswig:2010di}. We show the resulting plus-polarization of the dominant $\ell=m=2$ mode of the GW strain in Fig.~\ref{fig:strain}, for each of our evolutions.

For a given cold EoS, we find negligible differences in the inspiral GWs between the different $M^*$-parameters, as expected given the negligible inspiral heating that was found prior to merger (see Fig.~\ref{fig:Pth_vs_t}). In contrast, differences in the strain emerge following the merger in both cases, with variations in the oscillations and amplitude of the post-merger GWs that depend on the choice of $M^*$-parameters.

\begin{figure}[!ht]
\centering
\includegraphics[width=0.45\textwidth]{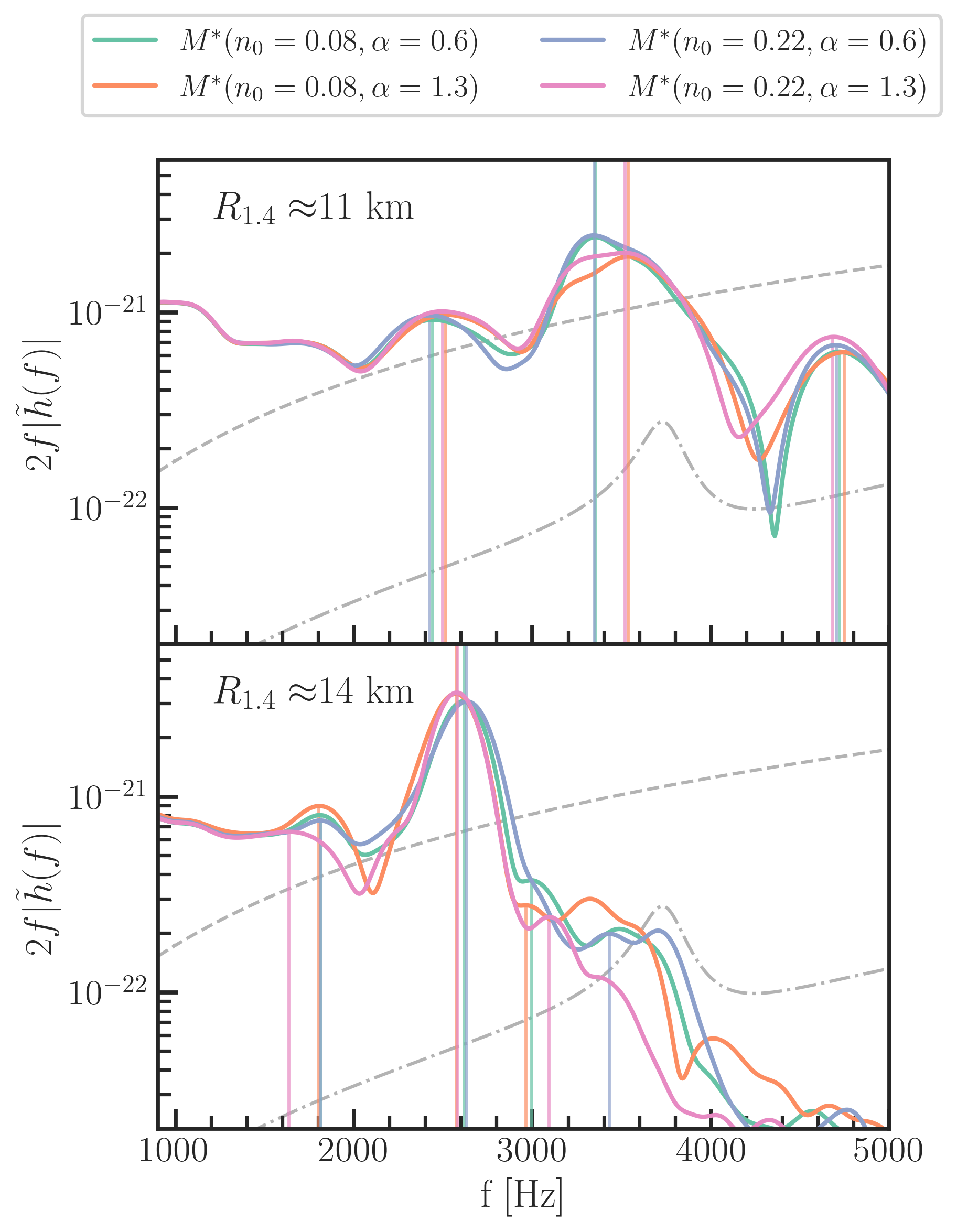} 
\caption{\label{fig:hc} Characteristic strain for a face-on merger located at 40 Mpc, for evolutions with the \Rsmall cold EoS (top) and \Rbig EoS (bottom).  The vertical lines indicate the three spectral peaks, $f_1, f_2$, and $f_3$. The dashed gray line represents the sensitivity of aLIGO at design sensitivity \cite{aLIGOsensitivity}, while the dash-dotted line indicates the proposed sensitivity for the next-generation detector Cosmic Explorer \cite{CEsensitivity}. The characteristic strain has been Welch-averaged and normalized, according to the procedure described in Appendix~\ref{sec:appendixGW}.}
\end{figure}

In order to explore these differences in the post-merger GWs in more detail, we compute the characteristic strain according to
\begin{equation}
h_c = 2 f |\tilde{h}(f)|
\end{equation}
where $f$ is the frequency and $\tilde{h}(f)$ is the Fourier transform of the strain $h(t)$. We describe the details of how we calculate and Welch-average this characteristic strain in Appendix~\ref{sec:appendixGW}, and we show the resulting spectra in Fig.~\ref{fig:hc}.

The spectra are characterized by a small number of peaks, which we refer to generically as $f_1$, $f_2$, and $f_3$. We summarize these frequencies in Table~\ref{table:fpeaks} and indicate them with vertical lines in Fig.~\ref{fig:hc}. The dominant peak, $f_2$, is typically associated with quadrupolar oscillations of the remnant \cite{Stergioulas:2011gd,Takami:2014tva,Rezzolla:2016nxn} and has been empirically correlated with particular properties of the zero-temperature EoS, such as the radius or tidal deformability of an intermediate-mass neutron star \cite[e.g.,][]{Bauswein:2012ya,Bauswein:2011tp,Takami:2014zpa,Bernuzzi:2015rla}, with corrections possible due to  the slope of the mass-radius relation \cite{Raithel:2022orm}; or with, e.g., the maximum density of neutron stars \cite{Breschi:2021xrx}.
 
In general, these studies find that the peak frequency is inversely correlated with $R_{1.4}$, and our spectra are globally consistent with this trend:   that is, for the \Rbig evolutions, we find $f_2\approx2.5$~kHz, while for the \Rsmall evolutions, $f_2\approx3.4$~kHz. However, for a given cold EoS, we find that the post-merger GWs are also mildly sensitive to the choice of $M^*$-parameters, with $f_2$ differing by up to 190~Hz (fractional difference of 5.5\%) for the \Rsmall evolutions, and by up to 60~Hz (2.3\%) for the \Rbig evolutions.  
 
\begin{table*}
\centering
\begin{tabularx}{0.85\textwidth}{@{\extracolsep{\fill}}ccccccccc}
\hline \hline
Cold EoS     &  $M^*$-parameters  & $f_1$  & $f_2$     & $f_3$    &  $f_2-f_1$   &  $f_3-f_2$    & $f_0$ \\
\hline
       &  $n_0=0.08$ fm$^{-3}$, $\alpha=0.6$  &2.44  &  3.35  & 4.72  & 0.91  &  1.37  & 1.27  \\
\Rsmall ~~ &  $n_0=0.08$ fm$^{-3}$, $\alpha=1.3$  &2.51  &  3.54  & 4.75  & 1.02  &  1.21  & 1.29  \\
       &  $n_0=0.22$ fm$^{-3}$, $\alpha=0.6$  &2.42  &  3.35  & 4.70  & 0.92  &  1.36  & 1.34  \\
       &  $n_0=0.22$ fm$^{-3}$, $\alpha=1.3$  &2.50  &  3.52  & 4.68  & 1.02  &  1.17  & 1.34  \\
\hline
 & Max. difference    & 0.09  & 0.19  & 0.07  & 0.11  & 0.20  & 0.07 \\
\hline \hline
       &  $n_0=0.08$ fm$^{-3}$, $\alpha=0.6$  &1.81  &  2.62  & 2.99  & 0.81  &  0.38  & 0.74  \\
\Rbig ~~ &  $n_0=0.08$ fm$^{-3}$, $\alpha=1.3$  &1.80  &  2.57  & 2.96  & 0.77  &  0.39  & 1.04  \\
       &  $n_0=0.22$ fm$^{-3}$, $\alpha=0.6$  &1.81  &  2.63  & 3.43  & 0.82  &  0.80  & 1.02  \\
       &  $n_0=0.22$ fm$^{-3}$, $\alpha=1.3$  &1.63  &  2.58  & 3.09  & 0.94  &  0.51  & 1.01  \\
\hline
 & Max. difference    & 0.18  & 0.06  & 0.47  & 0.17  & 0.42  & 0.30 \\
\hline \hline
\end{tabularx}
\caption{Peak frequencies of the post-merger GW spectra for each EoS. All frequencies are given in kHz. The cold EoS is given in the first column and the thermal prescription is given in the second column. The next three columns, labeled $f_{1,2,3}$, correspond to the peaks identified from the Welch-averaged spectra in Fig.~\ref{fig:hc}. These peaks are used to compute the differences  $f_2-f_1$ and  $f_3-f_2$. The last column provides the quasi-radial mode frequency, estimated from Fourier analysis of the rest-mass density evolution from Fig.~\ref{fig:rhob}. The maximum difference between each set of four simulations (governed by the same cold EoS) is given in the bottom rows. }
  \label{table:fpeaks}
\end{table*}

Although the $\lesssim200$~Hz variation in $f_2$ due to thermal effects is sub-dominant compared to the $\sim$1~kHz scatter due to uncertainties in the cold EoS, this is a strong indication that the post-merger GWs are indeed sensitive to the details of the thermal prescription. This finding is consistent with early studies that used a constant-$\Gamma_{\rm th}$ prescription  to bracket the range of uncertainty in finite-temperature effects, and showed that  varying $\Gamma_{\rm th}$ from 1.5 to 2 can change  $f_2$ by up to 390~Hz, depending on the  cold EoS \cite{Bauswein:2010dn} \cite[see also][]{Figura:2020fkj}. In a recent study, Ref. \cite{Fields:2023bhs} explored the same question with the phenomenological EoS framework of \cite{Schneider:2017tfi}, which is based on a liquid drop model with Skyrme interactions and includes density-dependent thermal effects. They found differences of up to 245~Hz (8\%) in $f_2$ depending on their choice of effective mass parameters, which is a similar -- though somewhat larger-- range of $f_2$ than found here (for further comparison with this work, see Appendix~\ref{sec:compareM}).

Within the observed scatter in $f_2$, we do not find a unique trend with the $M^*$-parameters that persists across stellar compactness. For example, for the \Rsmall cold EoS, we find that thermal prescriptions with $\alpha=1.3$ lead to larger values of $f_2$, while $\alpha=0.6$ leads to systematically smaller values of $f_2$. But the trend with $\alpha$ is reversed for the remnants evolved with the \Rbig cold EoS, suggesting a complex interplay between the $M^*$-parameters, stellar compactness, and the spectrum of post-merger GWs. We also do not find any clear trends  between $f_2$ and the average thermal properties of the remnant in Table~\ref{table:thermal}. Notably, we do not find a unique correlation between $f_2$ and the specific heat at $\rho_{\rm sat}$, as was suggested in \cite{Fields:2023bhs}, that persists across both cold EoSs and all four thermal prescriptions.

 Table~\ref{table:fpeaks} also reports the values of the secondary peaks, $f_1$ and $f_3$, along with their distances from the dominant peak $f_2$. Lastly, we include in Table~\ref{table:fpeaks} the quasi-radial oscillation frequency, $f_0$, for each remnant, which we estimate by  Fourier analysis of the rest-mass density evolution (shown in Fig.~\ref{fig:rhob}).

 There are various physical interpretations for the origin of the secondary peaks, including that they are powered by a non-linear interaction between the $m=2$ mode and  the quasi-radial oscillations, or that they are generated by the propagation of a spiral mode \cite[e.g.,][]{Stergioulas:2011gd,Takami:2014tva,Bauswein:2015yca}. We do not find a definitive correlation (or lack thereof) between $f_0$ and the $f_2-f_1$ or $f_3-f_2$ spacings, making it unclear which mechanism is operating here. We thus take an agnostic interpretation of these secondary peak frequencies in the following discussion.

In general, we find that the secondary peaks are also sensitive to the choice of $M^*$-parameters. The low-frequency secondary peak $f_1$ shifts by up to 90 (180)~Hz for the soft (stiff) cold EoS, corresponding to $\sim4$-10\% fractional variations. For the high-frequency peak $f_3$, we find shifts of up to 70 (470)~Hz, for the soft (stiff) cold EoSs. We note that the large scatter in $f_3$ for the \Rbig models is dominated by a single spectrum (see Fig.~\ref{fig:hc}) and that, in general, these high-frequency secondary peaks are not well resolved for the \Rbig models. As a result, this scatter should be interpreted cautiously. Nevertheless, we find that the secondary peaks
are also sensitive to the choice of $M^*$-parameters. As was found for $f_2$, we again find that there is no unique trend between the secondary peak frequencies and either $C_v$ or $M^*$ at $\rns$ that persists for both cold EoSs and all four thermal prescriptions.
 
 In summary, we find $\sim2-5\%$ variation in the location of $f_2$, depending on the choice of $M^*$-parameters. This dependence persists across both cold EoSs considered here, but the effect is largest for the softer (smaller radius) models. In principle, if the zero-temperature EoS were known precisely -- e.g., from a large number of tidal deformability measurements, observed across a range of masses -- then these small variations in the location of $f_2$  could be used to constrain the finite-temperature part of the EoS. This would provide insight into the properties of dense matter in a region of parameter space that is distinct from what can be probed during the inspiral.

 In practice, the large uncertainties in the cold EoS mean that, for the time being, the sensitivity of the post-merger GWs on the thermal prescription may simply  introduce an additional source of uncertainty into  inferences that constrain  properties of the cold  EoS (such as $R_{1.4}$) from $f_2$.  This uncertainty due to the finite-temperature physics -- shown here to affect $f_2$ by up to 190~Hz --  will be important to take into account in  future inferences from the post-merger GWs.

We leave a further investigation into the detectability of these thermal effects for a future work.

\subsection{Dynamical ejecta}
\label{sec:ejecta}

Finally, we turn to the properties of the  shock-heated ejecta
that is dynamically launched during and immediately
following the merger. We calculate the ejected 
mass for each of our evolutions by integrating the total rest-mass density 
outside of a sphere of radius $r=100M$, including all matter for which $-u_t >1$ and $u^r > 0$, 
according to
\begin{equation}
M_{\rm ej}(>r) = \int_{>r} \rho_b u^t \sqrt{-g} d^3x,
\end{equation}
where $u^t$ is the time-component of the four-velocity, $u^r$ is the radial velocity, and $g$ is the determinant of the metric.
We report the total amounts of ejecta thus calculated for each of our simulations in Table~\ref{table:ej}.

\begin{table}
\centering
\begin{tabular}{ccccccccc}
\hline \hline
Cold EoS     &  $M^*$-parameters      &  $t-t_{\rm merger}$ (ms)      &   $M_{\rm ej} (10^{-2} \Ms)$    \\
 \hline 
   &   $n_0=0.08, \alpha=0.6$     & 		     &   1.84     \\
\Rsmall   &   $n_0=0.08, \alpha=1.3$     &   12.1   &   3.21       \\
   &   $n_0=0.22,  \alpha=0.6$     & 		     &   1.73      \\
   &   $n_0=0.22,  \alpha=1.3$     &   		   &   1.81     \\
\hline 
   &   $n_0=0.08,  \alpha=0.6$     &    		  	&   0.41      \\
\Rbig   &   $n_0=0.08,  \alpha=1.3$     &  18.6    &   0.35     \\
   &   $n_0=0.22, \alpha=0.6$     &   		   &   0.49      \\
   &   $n_0=0.22, \alpha=1.3$     &   		   &   0.36      \\
\hline 
\end{tabular}
\caption{Total dynamical ejecta extracted at the end of each simulation,
for all unbound matter outside a sphere placed at 100$M$. From left to right, the columns
indicate:  the cold EoS, the thermal prescription, the time at which the ejecta are calculated,
and the total amount of ejecta in units of $10^{-2} \Ms$.}
  \label{table:ej}
\end{table}

 \begin{figure}[!ht]
\centering
\includegraphics[width=0.45\textwidth]{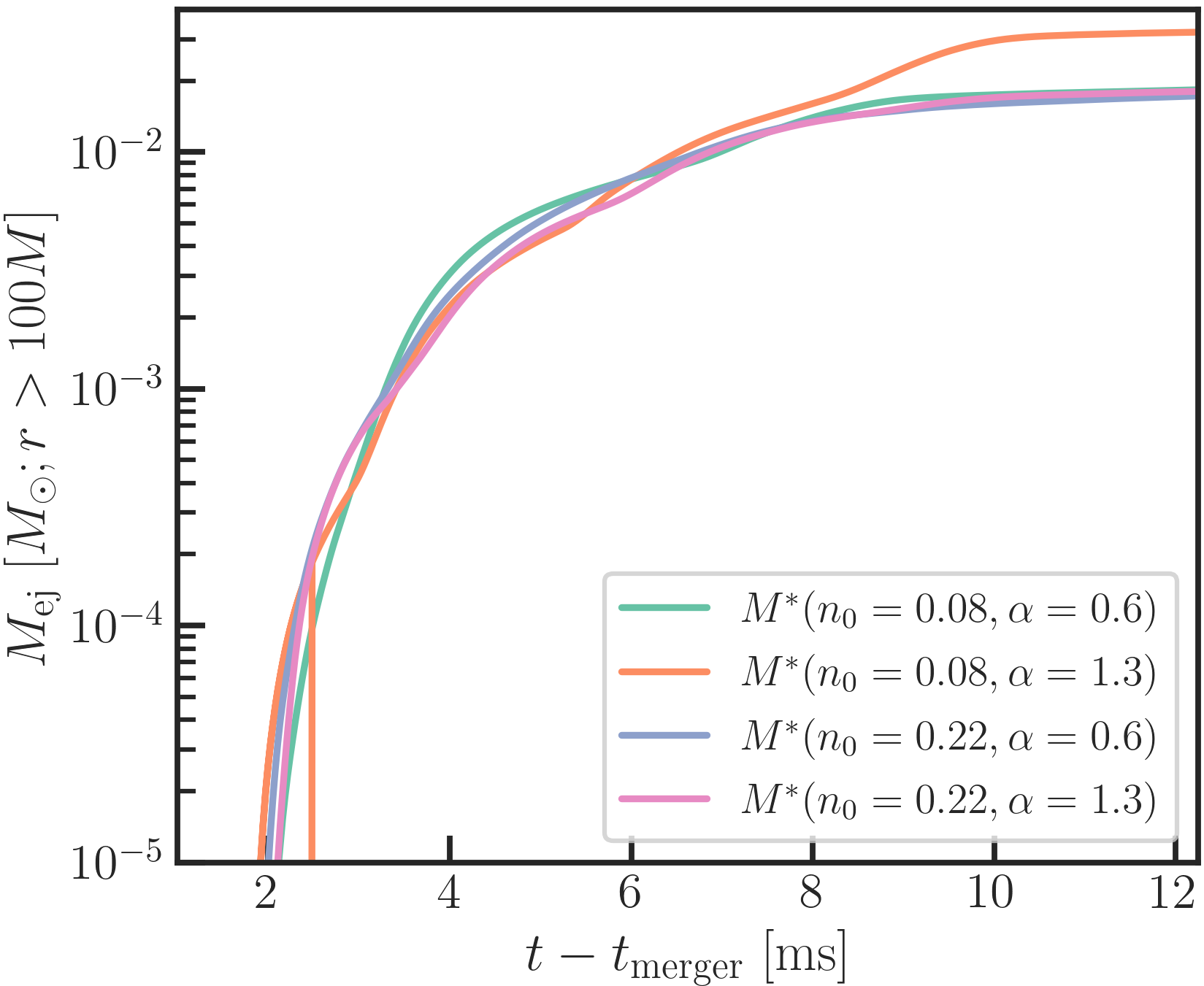} 
\includegraphics[width=0.45\textwidth]{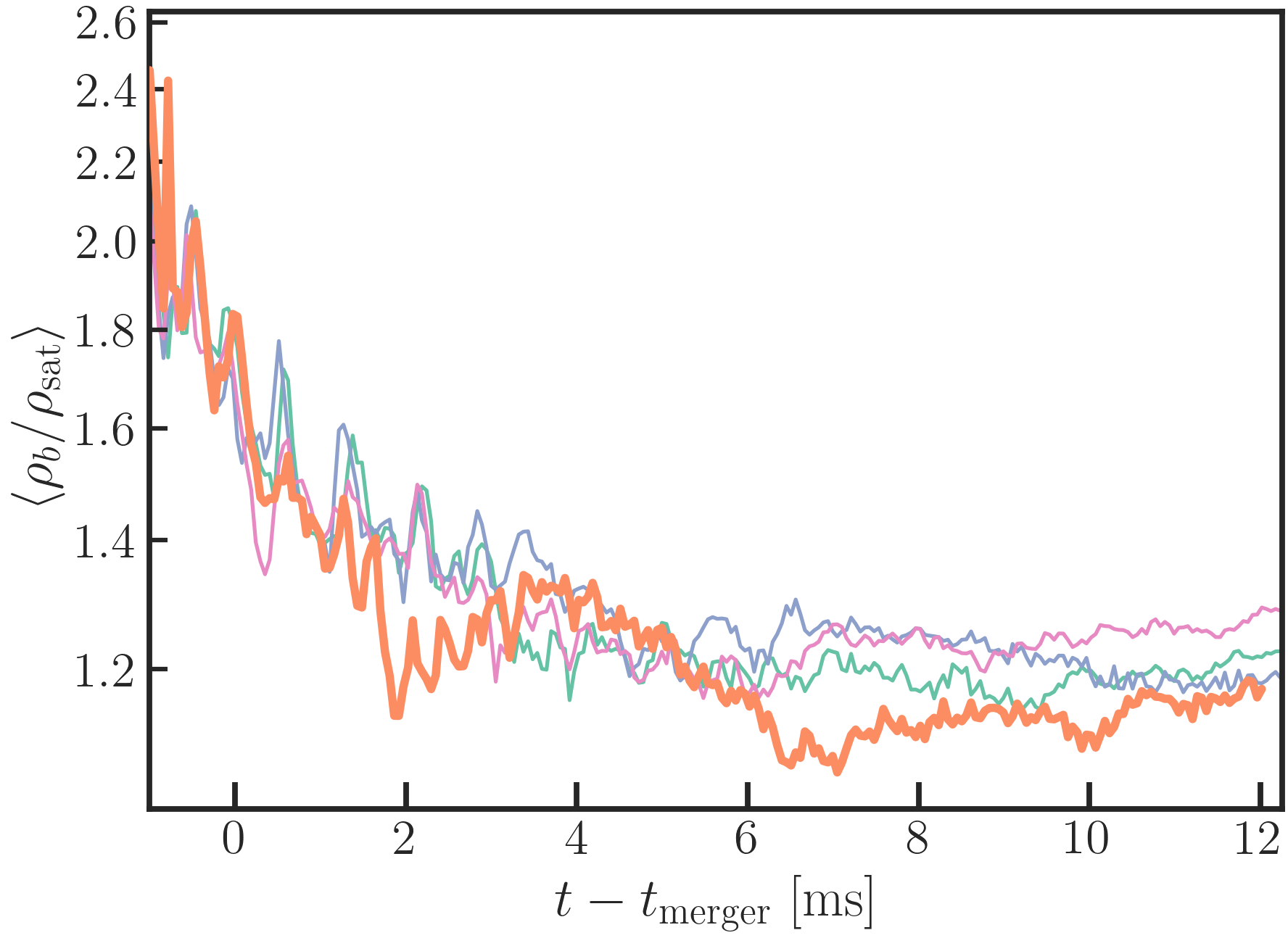} 
\caption{\label{fig:ejecta} Top: Unbound (dynamical) ejecta outside a radius of 100$M$, as a function
of time since the merger, for the \Rsmall evolutions. Bottom: Average density of the remnant as a function of
time, where the averages are computed from equatorial slices and include all matter with $\rho_b>0.1\rns$.
The $(n_0=0.08$ fm$^{-3}, \alpha=1.3)$ evolution produces a second wave of ejecta, which passes through the
detector at 100$M$ approximately 9~ms after the merger. The same evolution also features a pronounced core bounce, 
as evidenced by the bump in the average density of the remnant $\approx4$~ms after merger,
which likely launched the second wave of ejecta. }
\end{figure}

Overall, we find that the evolutions with the \Rsmall EoSs produce
 more dynamical ejecta than the evolutions with the \Rbig EoSs.
This is consistent with the overall stronger shock heating
that we find for the more compact stars, and 
with general trends found in previous studies
\cite{Bauswein:2013yna,Shibata:2019wef} 
(though see also \cite{Radice:2018pdn,Most:2021ktk}).

When comparing the differences between thermal prescriptions, we generally find a minimal
dependence on the choice of $M^*$-parameters.
For the \Rbig models, we find differences in $M_{\rm ej}$ of only $1.4\times10^{-3}~\Ms$ 
(fractional difference of $\sim$30\%), depending on the thermal prescription. This difference
 is comparable to our estimate of the numerical error in the ejecta 
 ($\sim$30\%; see Appendix~\ref{sec:appendixConv}), and thus we conclude that
 these differences are not numerically significant.
 
For the \Rsmall EoSs, we find similarly negligible differences in $M_{\rm ej}$ between three of
the four thermal prescriptions. However, we find significantly more dynamical ejecta
for the set of $M^*$-parameters
with $n_0=0.08~\text{fm}^{-3}$ and $\alpha=1.3$.
This prescription leads to an additional $\sim$1.4$\times10^{-2}~\Ms$ ejecta (80\% fractional effect) ,
compared to the other choices of $M^*$-parameters. 

We explore the anomalously high ejecta for this case in Fig.~\ref{fig:ejecta},
which compares the time evolution of $M_{\rm ej}$ for all four evolutions 
with the \Rsmall cold EoS. We find that the early evolution of all
four thermal treatments is similar, but that the $(n_0=0.08~\text{fm}^{-3}, \alpha=1.3)$ 
prescription produces a second wave of ejecta, which reaches the imaginary detector
at $r=100M$ approximately 9~ms after merger. This second wave is the source
of the excess ejecta.

We find that the second wave of ejecta is launched by a significant core bounce in
the post-merger remnant, for this thermal treatment. The core bounce is not apparent from the 
maximum density of the remnant (Fig.~\ref{fig:rhob}),
but it become evident in the \textit{average} density of the remnant, which we show in the bottom
panel of Fig.~\ref{fig:ejecta}. We compute the average density from equatorial snapshots
of the remnant, including all matter with $\rho_b > 0.1\rns$. The evolution of the 
average density of the post-merger remnant is initially similar for all four thermal treatments,
but the remnant evolved with the $(n_0=0.08~\text{fm}^{-3}, \alpha=1.3)$ thermal
prescription experiences a sharper 
contraction at 2~ms post-merger, followed by an increase in the central density 
that peaks 4~ms after the merger. The second wave of ejecta passes through the detector 
(at $r=100M$)
approximately 5~ms later, indicating characteristic speeds of $\sim0.3c$.

\begin{figure}[!ht]
\centering
\includegraphics[width=0.45\textwidth]{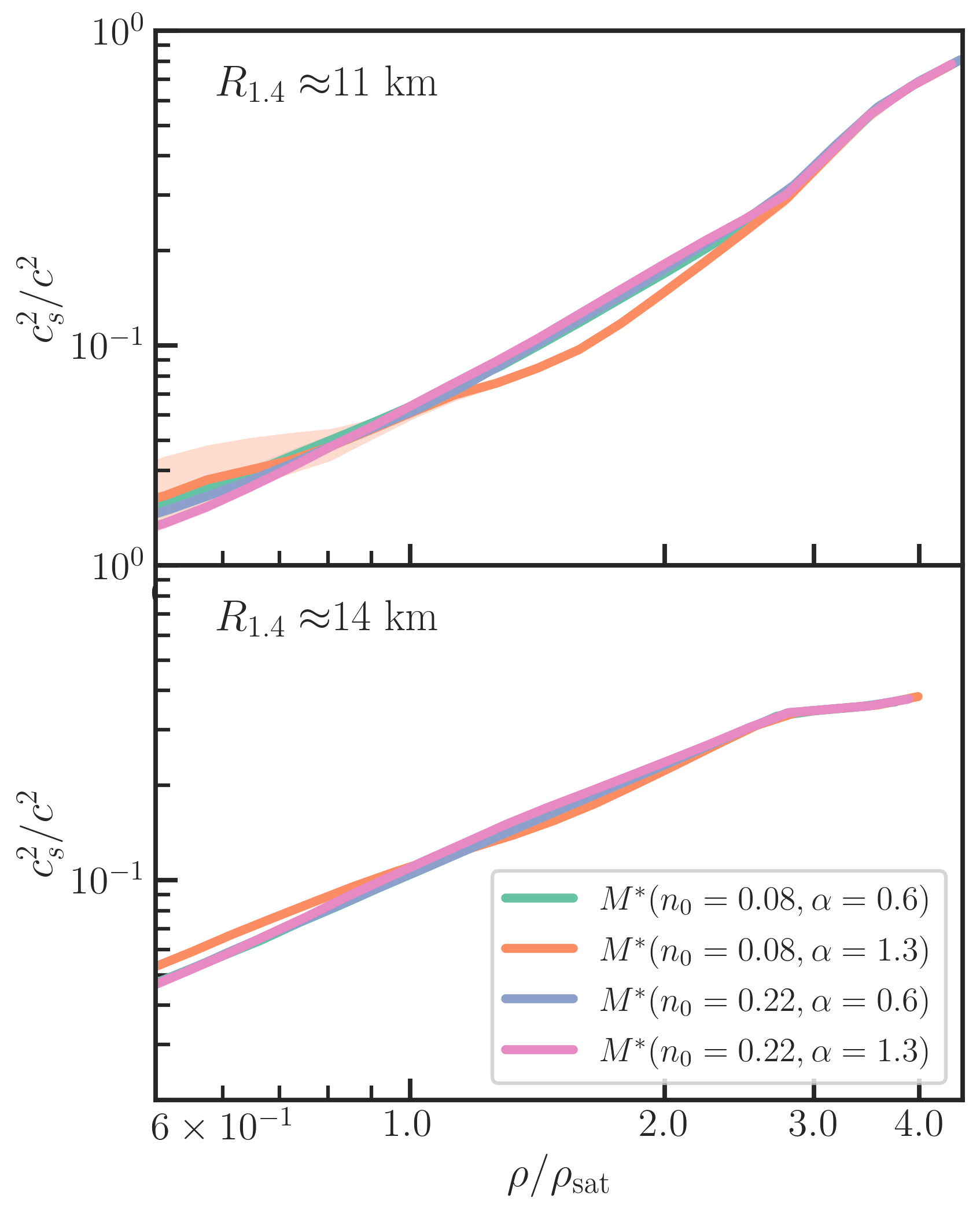} 
\caption{\label{fig:cs} Median sound speed profile within each post-merger remnant, 
with  the 68\% uncertainties shown in the shaded regions. The sound speed profiles
are similar for a given zero-temperature EoS (separated into the top and bottom figures),
with the exception that the $n_0=0.08$~fm$^{-3}$ and $\alpha=1.3$ set of parameters
leads to a reduced sound speed at intermediate densities for the \Rsmall EoS, 
and thus produces a more compressible remnant.}
\end{figure}

This remnant is susceptible to the additional core-bounce
due to its softer (i.e., lower) sound speed at intermediate densities, which we
show in Fig.~\ref{fig:cs} for the \Rsmall EoS remnants and explain as follows.
The adiabatic sound speed, $c_s$, is defined as 
\begin{equation}
\left(\frac{c_s}{c}\right)^2 = \frac{\partial P}{\partial \rho}\biggr\rvert_{S,Y_e} \left( \frac{1}{h}\right)
\end{equation}
where $h$ is the enthalpy and the derivative is evaluated at constant entropy $S$ and
electron fraction. For a polytropic EoS, the derivative term can be expanded as 
\begin{equation}
\label{eq:dPdn}
\frac{\partial P(\rho,T)}{\partial n} \biggr\rvert_{S,Y_e} =\frac{1}{\rho}\left[ \Gamma_{\rm cold} P_{\rm cold} + P_{\rm th} \left(\Gamma_{\rm th} + \rho \ln \rho  \frac{\partial \Gamma_{\rm th}}{\partial \rho} 
\right)\right] 
\end{equation}
where $\Gamma_{\rm cold}$ is the adiabatic index of the zero-temperature EoS
and $\Gamma_{\rm th}$ is the (density-dependent) effective thermal index.
As shown in the bottom row of Fig.~\ref{fig:PthT_dens}, the 
$(n_0=0.08~\text{fm}^{-3}, \alpha=1.3)$ set of parameters leads to 
the most rapidly-decreasing $\Gamma_{\rm th}$
as a function of density, within the finite-temperature remnant.
In addition, in eq.~(\ref{eq:dPdn}), the derivative term is multiplied by the
overall thermal pressure. Thus, for the \Rsmall evolutions,
which undergo significant heating, the derivative term
is weighted strongly and this choice of parameters
leads to a significant reduction in the sound speed
within the finite-temperature remnant. This is shown in
the top panel of Fig.~\ref{fig:cs}. 

In contrast, because the \Rbig collisions are so gentle, 
the thermal pressures are overall lower and thus the contributions
from the $\Gamma_{\rm th}$ terms in eq.~(\ref{eq:dPdn}) are suppressed.
As a result, we do not see significant differences in the sound speed profile of
the remnant, and the dynamical
ejecta are similar for all four thermal treatments with this stiffer cold EoS.

We note that the ejecta properties discussed in this section
may be modified with the inclusion of neutrinos into
the simulations. For example, \cite{Radice:2016dwd} showed that neglecting
neutrino cooling and heating can
cause the unbound matter to be overestimated by factors of $>2$. 
We leave the investigation of the interplay between finite-temperature 
and neutrino effects to a future study.
Nevertheless, these results provide a first indication of 
the degree to which the uncertainties solely in the thermal
part of the EoS can influence the dynamically-launched ejecta.

\section{Discussion and conclusions}
\label{sec:discussion}
In this paper, we have investigated the role of finite-temperature
effects in BNS mergers for two different cold
EoSs, corresponding to one soft zero-temperature model (predicting \Rsmall stars)
and one stiff zero-temperature model (\Rbig). All evolutions
 are performed for equal mass ($1.3+1.3\Ms$) binaries, with
 one of four different thermal prescriptions attached to each cold EoS.
 The thermal prescriptions are calculated using a phenomenological model of
 the particle effective mass $M^*$, following the framework of \cite{Raithel:2019gws}.
 
 We confirm previous results that tidal heating during
 the inspiral is small and that it, accordingly, has a negligible effect on the inspiral dynamics. In contrast, strong 
 heating during the merger leads to differences in the post-merger
 evolution, depending on the choice of $M^*$-parameters.

 For the \Rsmall evolutions, we find that the average temperature
 of the late-time, neutron star remnant is 26-37~MeV, depending
 on the choice of $M^*$-parameters, while the average
 $\langle P_{\rm th}/P_{\rm cold} \rangle$ can range from 0.27-0.40,
 indicating significant thermal support within the remnant neutron star.
For all of these models, we find a strong density-dependence
for the temperature distribution within the remnants, leading
to the formation of a high-temperature ring in the outer core. 
Together with the results from Paper I, these
findings confirm that the high-temperature ring forms generically
for any choice of $M^*$-parameters and across a range of
different cold EoSs. Inside this high-temperature ring, however, the temperature of the innermost core
depends sensitively on the $M^*$-parameters in a non-trivial way,
which may affect the long-term stability and cooling of the remnant.
 
 The evolutions with the stiffer \Rbig EoS lead to gentler collisions,
 with less significant heating. As a result, the average temperatures of the late-time remnants for this
 EoS are only 14-18~MeV, with  
 $\langle P_{\rm th}/P_{\rm cold}\rangle \approx0.1-0.14$, depending
 on the choice of $M^*$-parameters. Because this EoS is
 so stiff, the remnant object has still not become fully axisymmetric
 by the end of our simulations and we observe a significant bar mode in the remnant. 
 Nevertheless, we 
 see evidence of high-temperature hot spots forming within these remnants,
 which we expect to evolve into a ring-like structure at later times. 

We have also investigated the GW emissions for these
evolutions and conclude that the post-merger GWs
are modestly sensitive to the choice of $M^*$-parameters,
as was first indicated in the GW strain in Paper I for a single cold EoS.
We find that the shifts to the peak frequency $f_2$
are $\lesssim200$~Hz ($\sim$2-5\%) for a given cold EoS, with
the largest shifts occuring for the \Rsmall cold EoS.
The peak frequency depends most sensitively on the
parameter $\alpha$, which governs the rate of decay
for the effective mass function at high densities. However,
we find opposite trends between $f_2$ and $\alpha$ 
for the \Rsmall and \Rbig cold EoSs, suggesting
a complex interplay between the $M^*$-parameters,
the stellar compactness, and the structure
of the post-merger remnant. Notably, we do not
find a correlation between $f_2$ and the specific
heat at $\rho_{\rm sat}$ for our models,
as was recently suggested by \cite{Fields:2023bhs}, nor do we
find a correlation between $f_2$ and the value of $M^*$ at
$\rho_{\rm sat}$.
The locations of the secondary peaks of the post-merger
GW spectra are also sensitive to the choice of
$M^*$-parameters at the $\sim4-10\%$ level, but again
without a persistent correlation with either the $M^*$-parameters
of specific heat at $\rns$.

We note that, like most modern merger codes, our post-merger gravitational wave signals are not convergent \cite[e.g.,][]{Raithel:2022san}. In a recent convergence study with a similar numerical set-up to this work, we estimated that the approximate, fractional error in $f_2$ for our simulations is $\sim$0.2\% as the resolution is increased \cite{Raithel:2022san}. However, without strict convergence, a robust error estimation is difficult. Other studies have estimated the numerical error in $f_2$ as being as high as 2-8\%, due to the discrete Fourier transform \cite{Breschi:2019srl}. We note also that the dependence of $f_2$ on the $M^*$-parameters is comparable to the $\sim$100~Hz dependences of the post-merger GWs on e.g., microphysical bulk transport \cite{Most:2022yhe} or the spin of the initial neutron stars \cite{East:2019lbk}, which may complicate efforts to disentangle the $M^*$-parameters from a future detection of $f_2$. 
 
Furthermore, attributing the spectral differences
to finite-temperature effects will first require precision
knowledge of the zero-temperature EoS. This 
may be possible with the hundreds of thousands
of inspiral events that will be observed in the
era of next-generation facilities such as
Cosmic Explorer \cite{Reitze:2019iox}, Einstein Telescope \cite{Punturo:2010zz}, and NEMO \cite{Ackley:2020atn}.
We plan to further investigate the detectability
of these spectral differences in a future work.
 
 Finally, we have also compared the total amounts of dynamical ejecta 
 for these evolutions and we find that $M_{\rm ej}$ is 
 not very sensitive to the choice of $M^*$-parameters, 
 except for one extreme case, for which
the ejecta can be enhanced by $\sim$85\% simply
by changing the finite-temperature part of the EoS. We attribute
this increase in the ejecta to an additional core bounce,
which is possible for the softer cold EoS with the
$(n_0=0.08$~fm$^{-3}, \alpha=1.3)$ set of $M^*$-parameters,
due to the lower sound speed in the finite-temperature remnant for this combination of parameters.

While this work presents a first exploration
of realistic finite-temperature effects in mergers
across a range of stellar compactness, 
we note that the results presented here are limited to a single binary mass and an 
equal mass ratio.  Changing the binary parameters may further influence the strength of the thermal effects and the 
sensitivity to the $M^*$-parameters.

As we look forward to new discoveries with the upcoming
LIGO-Virgo-Kagra observation runs and 
as the community plans for the construction
of the next-generation of detectors, it is essential
to quantify the key uncertainties in our modeling of
neutron star mergers, and to understand
how these uncertainties will influence our ability
to constrain the EoS from future observations.
At present, the dependences that we
have found here between the post-merger observables 
(such as ejecta and GWs) 
and the finite-temperature part of the 
EoS will effectively introduce an additional
source of uncertainty into post-merger inferences of the cold EoS.
 In the future, however, 
as the zero-temperature EoS becomes
better constrained, these dependences
can be exploited to derive novel constraints
on the EoS at high densities and temperatures,
thus probing a fundamentally 
new region of parameter space.

\acknowledgements{CR thanks Elias Most for useful 
conversations related to this work.
CR is supported by a joint postdoctoral fellowship at the Princeton Center
for Theoretical Science, the Princeton Gravity Initiative, and as a
John N. Bahcall Fellow at the Institute for Advanced Study.  
This work
was in part supported by NSF Grant PHY-1912619 and PHY-2145421 to the
University of Arizona.  The simulations presented in this work were
carried out with the {\tt Stampede2} cluster at the Texas Advanced 
Computing Center and the {\tt Expanse} cluster at San Diego Supercomputer 
Center, under XSEDE allocation PHY190020. The simulations were also
performed, in part, with the Princeton Research Computing resources at
Princeton University, which is a consortium of groups led by the
Princeton Institute for Computational Science and Engineering
(PIC-SciE) and Office of Information Technology's Research
Computing. }

\bibliography{inspire,non_inspire}

\begin{thebibliography}{74}%
\makeatletter
\providecommand \@ifxundefined [1]{%
 \@ifx{#1\undefined}
}%
\providecommand \@ifnum [1]{%
 \ifnum #1\expandafter \@firstoftwo
 \else \expandafter \@secondoftwo
 \fi
}%
\providecommand \@ifx [1]{%
 \ifx #1\expandafter \@firstoftwo
 \else \expandafter \@secondoftwo
 \fi
}%
\providecommand \natexlab [1]{#1}%
\providecommand \enquote  [1]{``#1''}%
\providecommand \bibnamefont  [1]{#1}%
\providecommand \bibfnamefont [1]{#1}%
\providecommand \citenamefont [1]{#1}%
\providecommand \href@noop [0]{\@secondoftwo}%
\providecommand \href [0]{\begingroup \@sanitize@url \@href}%
\providecommand \@href[1]{\@@startlink{#1}\@@href}%
\providecommand \@@href[1]{\endgroup#1\@@endlink}%
\providecommand \@sanitize@url [0]{\catcode `\\12\catcode `\$12\catcode
  `\&12\catcode `\#12\catcode `\^12\catcode `\_12\catcode `\%12\relax}%
\providecommand \@@startlink[1]{}%
\providecommand \@@endlink[0]{}%
\providecommand \url  [0]{\begingroup\@sanitize@url \@url }%
\providecommand \@url [1]{\endgroup\@href {#1}{\urlprefix }}%
\providecommand \urlprefix  [0]{URL }%
\providecommand \Eprint [0]{\href }%
\providecommand \doibase [0]{http://dx.doi.org/}%
\providecommand \selectlanguage [0]{\@gobble}%
\providecommand \bibinfo  [0]{\@secondoftwo}%
\providecommand \bibfield  [0]{\@secondoftwo}%
\providecommand \translation [1]{[#1]}%
\providecommand \BibitemOpen [0]{}%
\providecommand \bibitemStop [0]{}%
\providecommand \bibitemNoStop [0]{.\EOS\space}%
\providecommand \EOS [0]{\spacefactor3000\relax}%
\providecommand \BibitemShut  [1]{\csname bibitem#1\endcsname}%
\let\auto@bib@innerbib\@empty
\bibitem [{\citenamefont {Abbott}\ \emph {et~al.}(2017)\citenamefont {Abbott}
  \emph {et~al.}}]{LIGOScientific:2017vwq}%
  \BibitemOpen
  \bibfield  {author} {\bibinfo {author} {\bibfnamefont {B.~P.}\ \bibnamefont
  {Abbott}} \emph {et~al.} (\bibinfo {collaboration} {LIGO Scientific,
  Virgo}),\ }\href {\doibase 10.1103/PhysRevLett.119.161101} {\bibfield
  {journal} {\bibinfo  {journal} {Phys. Rev. Lett.}\ }\textbf {\bibinfo
  {volume} {119}},\ \bibinfo {pages} {161101} (\bibinfo {year} {2017})},\
  \Eprint {http://arxiv.org/abs/1710.05832} {arXiv:1710.05832 [gr-qc]}
  \BibitemShut {NoStop}%
\bibitem [{\citenamefont {Abbott}\ \emph {et~al.}(2018)\citenamefont {Abbott}
  \emph {et~al.}}]{LIGOScientific:2018cki}%
  \BibitemOpen
  \bibfield  {author} {\bibinfo {author} {\bibfnamefont {B.~P.}\ \bibnamefont
  {Abbott}} \emph {et~al.} (\bibinfo {collaboration} {LIGO Scientific,
  Virgo}),\ }\href {\doibase 10.1103/PhysRevLett.121.161101} {\bibfield
  {journal} {\bibinfo  {journal} {Phys. Rev. Lett.}\ }\textbf {\bibinfo
  {volume} {121}},\ \bibinfo {pages} {161101} (\bibinfo {year} {2018})},\
  \Eprint {http://arxiv.org/abs/1805.11581} {arXiv:1805.11581 [gr-qc]}
  \BibitemShut {NoStop}%
\bibitem [{\citenamefont {Baiotti}(2019)}]{Baiotti:2019sew}%
  \BibitemOpen
  \bibfield  {author} {\bibinfo {author} {\bibfnamefont {L.}~\bibnamefont
  {Baiotti}},\ }\href {\doibase 10.1016/j.ppnp.2019.103714} {\bibfield
  {journal} {\bibinfo  {journal} {Prog. Part. Nucl. Phys.}\ }\textbf {\bibinfo
  {volume} {109}},\ \bibinfo {pages} {103714} (\bibinfo {year} {2019})},\
  \Eprint {http://arxiv.org/abs/1907.08534} {arXiv:1907.08534 [astro-ph.HE]}
  \BibitemShut {NoStop}%
\bibitem [{\citenamefont {Guerra~Chaves}\ and\ \citenamefont
  {Hinderer}(2019)}]{GuerraChaves:2019foa}%
  \BibitemOpen
  \bibfield  {author} {\bibinfo {author} {\bibfnamefont {A.}~\bibnamefont
  {Guerra~Chaves}}\ and\ \bibinfo {author} {\bibfnamefont {T.}~\bibnamefont
  {Hinderer}},\ }\href {\doibase 10.1088/1361-6471/ab45be} {\bibfield
  {journal} {\bibinfo  {journal} {J. Phys. G}\ }\textbf {\bibinfo {volume}
  {46}},\ \bibinfo {pages} {123002} (\bibinfo {year} {2019})},\ \Eprint
  {http://arxiv.org/abs/1912.01461} {arXiv:1912.01461 [nucl-th]} \BibitemShut
  {NoStop}%
\bibitem [{\citenamefont {Raithel}(2019)}]{Raithel:2019uzi}%
  \BibitemOpen
  \bibfield  {author} {\bibinfo {author} {\bibfnamefont {C.~A.}\ \bibnamefont
  {Raithel}},\ }\href {\doibase 10.1140/epja/i2019-12759-5} {\bibfield
  {journal} {\bibinfo  {journal} {Eur. Phys. J. A}\ }\textbf {\bibinfo {volume}
  {55}},\ \bibinfo {pages} {80} (\bibinfo {year} {2019})},\ \Eprint
  {http://arxiv.org/abs/1904.10002} {arXiv:1904.10002 [astro-ph.HE]}
  \BibitemShut {NoStop}%
\bibitem [{\citenamefont {Chatziioannou}(2020)}]{Chatziioannou:2020pqz}%
  \BibitemOpen
  \bibfield  {author} {\bibinfo {author} {\bibfnamefont {K.}~\bibnamefont
  {Chatziioannou}},\ }\href {\doibase 10.1007/s10714-020-02754-3} {\bibfield
  {journal} {\bibinfo  {journal} {Gen. Rel. Grav.}\ }\textbf {\bibinfo {volume}
  {52}},\ \bibinfo {pages} {109} (\bibinfo {year} {2020})},\ \Eprint
  {http://arxiv.org/abs/2006.03168} {arXiv:2006.03168 [gr-qc]} \BibitemShut
  {NoStop}%
\bibitem [{\citenamefont {Annala}\ \emph {et~al.}(2022)\citenamefont {Annala},
  \citenamefont {Gorda}, \citenamefont {Katerini}, \citenamefont {Kurkela},
  \citenamefont {N\"attil\"a}, \citenamefont {Paschalidis},\ and\ \citenamefont
  {Vuorinen}}]{Annala:2021gom}%
  \BibitemOpen
  \bibfield  {author} {\bibinfo {author} {\bibfnamefont {E.}~\bibnamefont
  {Annala}}, \bibinfo {author} {\bibfnamefont {T.}~\bibnamefont {Gorda}},
  \bibinfo {author} {\bibfnamefont {E.}~\bibnamefont {Katerini}}, \bibinfo
  {author} {\bibfnamefont {A.}~\bibnamefont {Kurkela}}, \bibinfo {author}
  {\bibfnamefont {J.}~\bibnamefont {N\"attil\"a}}, \bibinfo {author}
  {\bibfnamefont {V.}~\bibnamefont {Paschalidis}}, \ and\ \bibinfo {author}
  {\bibfnamefont {A.}~\bibnamefont {Vuorinen}},\ }\href {\doibase
  10.1103/PhysRevX.12.011058} {\bibfield  {journal} {\bibinfo  {journal} {Phys.
  Rev. X}\ }\textbf {\bibinfo {volume} {12}},\ \bibinfo {pages} {011058}
  (\bibinfo {year} {2022})},\ \Eprint {http://arxiv.org/abs/2105.05132}
  {arXiv:2105.05132 [astro-ph.HE]} \BibitemShut {NoStop}%
\bibitem [{\citenamefont {Baiotti}\ and\ \citenamefont
  {Rezzolla}(2017)}]{Baiotti:2016qnr}%
  \BibitemOpen
  \bibfield  {author} {\bibinfo {author} {\bibfnamefont {L.}~\bibnamefont
  {Baiotti}}\ and\ \bibinfo {author} {\bibfnamefont {L.}~\bibnamefont
  {Rezzolla}},\ }\href {\doibase 10.1088/1361-6633/aa67bb} {\bibfield
  {journal} {\bibinfo  {journal} {Rept. Prog. Phys.}\ }\textbf {\bibinfo
  {volume} {80}},\ \bibinfo {pages} {096901} (\bibinfo {year} {2017})},\
  \Eprint {http://arxiv.org/abs/1607.03540} {arXiv:1607.03540 [gr-qc]}
  \BibitemShut {NoStop}%
\bibitem [{\citenamefont {Paschalidis}\ and\ \citenamefont
  {Stergioulas}(2017)}]{Paschalidis:2016vmz}%
  \BibitemOpen
  \bibfield  {author} {\bibinfo {author} {\bibfnamefont {V.}~\bibnamefont
  {Paschalidis}}\ and\ \bibinfo {author} {\bibfnamefont {N.}~\bibnamefont
  {Stergioulas}},\ }\href {\doibase 10.1007/s41114-017-0008-x} {\bibfield
  {journal} {\bibinfo  {journal} {Living Rev. Rel.}\ }\textbf {\bibinfo
  {volume} {20}},\ \bibinfo {pages} {7} (\bibinfo {year} {2017})},\ \Eprint
  {http://arxiv.org/abs/1612.03050} {arXiv:1612.03050 [astro-ph.HE]}
  \BibitemShut {NoStop}%
\bibitem [{\citenamefont {Bauswein}\ \emph {et~al.}(2010)\citenamefont
  {Bauswein}, \citenamefont {Janka},\ and\ \citenamefont
  {Oechslin}}]{Bauswein:2010dn}%
  \BibitemOpen
  \bibfield  {author} {\bibinfo {author} {\bibfnamefont {A.}~\bibnamefont
  {Bauswein}}, \bibinfo {author} {\bibfnamefont {H.~T.}\ \bibnamefont {Janka}},
  \ and\ \bibinfo {author} {\bibfnamefont {R.}~\bibnamefont {Oechslin}},\
  }\href {\doibase 10.1103/PhysRevD.82.084043} {\bibfield  {journal} {\bibinfo
  {journal} {Phys. Rev. D}\ }\textbf {\bibinfo {volume} {82}},\ \bibinfo
  {pages} {084043} (\bibinfo {year} {2010})},\ \Eprint
  {http://arxiv.org/abs/1006.3315} {arXiv:1006.3315 [astro-ph.SR]} \BibitemShut
  {NoStop}%
\bibitem [{\citenamefont {Paschalidis}\ \emph {et~al.}(2012)\citenamefont
  {Paschalidis}, \citenamefont {Etienne},\ and\ \citenamefont
  {Shapiro}}]{Paschalidis:2012ff}%
  \BibitemOpen
  \bibfield  {author} {\bibinfo {author} {\bibfnamefont {V.}~\bibnamefont
  {Paschalidis}}, \bibinfo {author} {\bibfnamefont {Z.~B.}\ \bibnamefont
  {Etienne}}, \ and\ \bibinfo {author} {\bibfnamefont {S.~L.}\ \bibnamefont
  {Shapiro}},\ }\href {\doibase 10.1103/PhysRevD.86.064032} {\bibfield
  {journal} {\bibinfo  {journal} {Phys. Rev. D}\ }\textbf {\bibinfo {volume}
  {86}},\ \bibinfo {pages} {064032} (\bibinfo {year} {2012})},\ \Eprint
  {http://arxiv.org/abs/1208.5487} {arXiv:1208.5487 [astro-ph.HE]} \BibitemShut
  {NoStop}%
\bibitem [{\citenamefont {{Janka}}\ \emph {et~al.}(1993)\citenamefont
  {{Janka}}, \citenamefont {{Zwerger}},\ and\ \citenamefont
  {{Moenchmeyer}}}]{Janka1993}%
  \BibitemOpen
  \bibfield  {author} {\bibinfo {author} {\bibfnamefont {H.-T.}\ \bibnamefont
  {{Janka}}}, \bibinfo {author} {\bibfnamefont {T.}~\bibnamefont {{Zwerger}}},
  \ and\ \bibinfo {author} {\bibfnamefont {R.}~\bibnamefont {{Moenchmeyer}}},\
  }\href@noop {} {\bibfield  {journal} {\bibinfo  {journal} {\aap}\ }\textbf
  {\bibinfo {volume} {268}},\ \bibinfo {pages} {360} (\bibinfo {year}
  {1993})}\BibitemShut {NoStop}%
\bibitem [{\citenamefont {Figura}\ \emph {et~al.}(2020)\citenamefont {Figura},
  \citenamefont {Lu}, \citenamefont {Burgio}, \citenamefont {Li},\ and\
  \citenamefont {Schulze}}]{Figura:2020fkj}%
  \BibitemOpen
  \bibfield  {author} {\bibinfo {author} {\bibfnamefont {A.}~\bibnamefont
  {Figura}}, \bibinfo {author} {\bibfnamefont {J.~J.}\ \bibnamefont {Lu}},
  \bibinfo {author} {\bibfnamefont {G.~F.}\ \bibnamefont {Burgio}}, \bibinfo
  {author} {\bibfnamefont {Z.~H.}\ \bibnamefont {Li}}, \ and\ \bibinfo {author}
  {\bibfnamefont {H.~J.}\ \bibnamefont {Schulze}},\ }\href {\doibase
  10.1103/PhysRevD.102.043006} {\bibfield  {journal} {\bibinfo  {journal}
  {Phys. Rev. D}\ }\textbf {\bibinfo {volume} {102}},\ \bibinfo {pages}
  {043006} (\bibinfo {year} {2020})},\ \Eprint
  {http://arxiv.org/abs/2005.08691} {arXiv:2005.08691 [gr-qc]} \BibitemShut
  {NoStop}%
\bibitem [{\citenamefont {Constantinou}\ \emph
  {et~al.}(2015{\natexlab{a}})\citenamefont {Constantinou}, \citenamefont
  {Muccioli}, \citenamefont {Prakash},\ and\ \citenamefont
  {Lattimer}}]{Constantinou:2015mna}%
  \BibitemOpen
  \bibfield  {author} {\bibinfo {author} {\bibfnamefont {C.}~\bibnamefont
  {Constantinou}}, \bibinfo {author} {\bibfnamefont {B.}~\bibnamefont
  {Muccioli}}, \bibinfo {author} {\bibfnamefont {M.}~\bibnamefont {Prakash}}, \
  and\ \bibinfo {author} {\bibfnamefont {J.~M.}\ \bibnamefont {Lattimer}},\
  }\href {\doibase 10.1103/PhysRevC.92.025801} {\bibfield  {journal} {\bibinfo
  {journal} {Phys. Rev. C}\ }\textbf {\bibinfo {volume} {92}},\ \bibinfo
  {pages} {025801} (\bibinfo {year} {2015}{\natexlab{a}})},\ \Eprint
  {http://arxiv.org/abs/1504.03982} {arXiv:1504.03982 [astro-ph.SR]}
  \BibitemShut {NoStop}%
\bibitem [{\citenamefont {Carbone}\ and\ \citenamefont
  {Schwenk}(2019)}]{Carbone:2019pkr}%
  \BibitemOpen
  \bibfield  {author} {\bibinfo {author} {\bibfnamefont {A.}~\bibnamefont
  {Carbone}}\ and\ \bibinfo {author} {\bibfnamefont {A.}~\bibnamefont
  {Schwenk}},\ }\href {\doibase 10.1103/PhysRevC.100.025805} {\bibfield
  {journal} {\bibinfo  {journal} {Phys. Rev. C}\ }\textbf {\bibinfo {volume}
  {100}},\ \bibinfo {pages} {025805} (\bibinfo {year} {2019})},\ \Eprint
  {http://arxiv.org/abs/1904.00924} {arXiv:1904.00924 [nucl-th]} \BibitemShut
  {NoStop}%
\bibitem [{\citenamefont {Huth}\ \emph {et~al.}(2021)\citenamefont {Huth},
  \citenamefont {Wellenhofer},\ and\ \citenamefont {Schwenk}}]{Huth:2020ozf}%
  \BibitemOpen
  \bibfield  {author} {\bibinfo {author} {\bibfnamefont {S.}~\bibnamefont
  {Huth}}, \bibinfo {author} {\bibfnamefont {C.}~\bibnamefont {Wellenhofer}}, \
  and\ \bibinfo {author} {\bibfnamefont {A.}~\bibnamefont {Schwenk}},\ }\href
  {\doibase 10.1103/PhysRevC.103.025803} {\bibfield  {journal} {\bibinfo
  {journal} {Phys. Rev. C}\ }\textbf {\bibinfo {volume} {103}},\ \bibinfo
  {pages} {025803} (\bibinfo {year} {2021})},\ \Eprint
  {http://arxiv.org/abs/2009.08885} {arXiv:2009.08885 [nucl-th]} \BibitemShut
  {NoStop}%
\bibitem [{\citenamefont {Raithel}\ \emph {et~al.}(2019)\citenamefont
  {Raithel}, \citenamefont {Ozel},\ and\ \citenamefont
  {Psaltis}}]{Raithel:2019gws}%
  \BibitemOpen
  \bibfield  {author} {\bibinfo {author} {\bibfnamefont {C.~A.}\ \bibnamefont
  {Raithel}}, \bibinfo {author} {\bibfnamefont {F.}~\bibnamefont {Ozel}}, \
  and\ \bibinfo {author} {\bibfnamefont {D.}~\bibnamefont {Psaltis}},\ }\href
  {\doibase 10.3847/1538-4357/ab08ea} {\bibfield  {journal} {\bibinfo
  {journal} {Astrophys. J.}\ }\textbf {\bibinfo {volume} {875}},\ \bibinfo
  {pages} {12} (\bibinfo {year} {2019})},\ \Eprint
  {http://arxiv.org/abs/1902.10735} {arXiv:1902.10735 [astro-ph.HE]}
  \BibitemShut {NoStop}%
\bibitem [{\citenamefont {{Baym}}\ and\ \citenamefont
  {{Pethick}}()}]{Baym1991}%
  \BibitemOpen
  \bibfield  {author} {\bibinfo {author} {\bibfnamefont {G.}~\bibnamefont
  {{Baym}}}\ and\ \bibinfo {author} {\bibfnamefont {C.}~\bibnamefont
  {{Pethick}}},\ }\href@noop {} {\emph {\bibinfo {title} {Landau Fermi‐Liquid
  Theory: Concepts and Applications}}}\BibitemShut {NoStop}%
\bibitem [{\citenamefont {Constantinou}\ \emph
  {et~al.}(2015{\natexlab{b}})\citenamefont {Constantinou}, \citenamefont
  {Muccioli}, \citenamefont {Prakash},\ and\ \citenamefont
  {Lattimer}}]{Constantinou:2015zia}%
  \BibitemOpen
  \bibfield  {author} {\bibinfo {author} {\bibfnamefont {C.}~\bibnamefont
  {Constantinou}}, \bibinfo {author} {\bibfnamefont {B.}~\bibnamefont
  {Muccioli}}, \bibinfo {author} {\bibfnamefont {M.}~\bibnamefont {Prakash}}, \
  and\ \bibinfo {author} {\bibfnamefont {J.~M.}\ \bibnamefont {Lattimer}},\
  }\href {\doibase 10.1016/j.aop.2015.10.003} {\bibfield  {journal} {\bibinfo
  {journal} {Annals Phys.}\ }\textbf {\bibinfo {volume} {363}},\ \bibinfo
  {pages} {533} (\bibinfo {year} {2015}{\natexlab{b}})},\ \Eprint
  {http://arxiv.org/abs/1507.07874} {arXiv:1507.07874 [nucl-th]} \BibitemShut
  {NoStop}%
\bibitem [{\citenamefont {Raithel}\ \emph {et~al.}(2022)\citenamefont
  {Raithel}, \citenamefont {Espino},\ and\ \citenamefont
  {Paschalidis}}]{Raithel:2022nab}%
  \BibitemOpen
  \bibfield  {author} {\bibinfo {author} {\bibfnamefont {C.~A.}\ \bibnamefont
  {Raithel}}, \bibinfo {author} {\bibfnamefont {P.}~\bibnamefont {Espino}}, \
  and\ \bibinfo {author} {\bibfnamefont {V.}~\bibnamefont {Paschalidis}},\
  }\href {\doibase 10.1093/mnras/stac2450} {\bibfield  {journal} {\bibinfo
  {journal} {Mon. Not. Roy. Astron. Soc.}\ }\textbf {\bibinfo {volume} {516}},\
  \bibinfo {pages} {4792} (\bibinfo {year} {2022})},\ \Eprint
  {http://arxiv.org/abs/2206.14838} {arXiv:2206.14838 [astro-ph.HE]}
  \BibitemShut {NoStop}%
\bibitem [{\citenamefont {Raithel}\ \emph {et~al.}(2021)\citenamefont
  {Raithel}, \citenamefont {Paschalidis},\ and\ \citenamefont
  {\"Ozel}}]{Raithel:2021hye}%
  \BibitemOpen
  \bibfield  {author} {\bibinfo {author} {\bibfnamefont {C.}~\bibnamefont
  {Raithel}}, \bibinfo {author} {\bibfnamefont {V.}~\bibnamefont
  {Paschalidis}}, \ and\ \bibinfo {author} {\bibfnamefont {F.}~\bibnamefont
  {\"Ozel}},\ }\href {\doibase 10.1103/PhysRevD.104.063016} {\bibfield
  {journal} {\bibinfo  {journal} {Phys. Rev. D}\ }\textbf {\bibinfo {volume}
  {104}},\ \bibinfo {pages} {063016} (\bibinfo {year} {2021})},\ \Eprint
  {http://arxiv.org/abs/2104.07226} {arXiv:2104.07226 [astro-ph.HE]}
  \BibitemShut {NoStop}%
\bibitem [{\citenamefont {Raithel}\ and\ \citenamefont
  {Paschalidis}(2022)}]{Raithel:2022san}%
  \BibitemOpen
  \bibfield  {author} {\bibinfo {author} {\bibfnamefont {C.~A.}\ \bibnamefont
  {Raithel}}\ and\ \bibinfo {author} {\bibfnamefont {V.}~\bibnamefont
  {Paschalidis}},\ }\href {\doibase 10.1103/PhysRevD.106.023015} {\bibfield
  {journal} {\bibinfo  {journal} {Phys. Rev. D}\ }\textbf {\bibinfo {volume}
  {106}},\ \bibinfo {pages} {023015} (\bibinfo {year} {2022})},\ \Eprint
  {http://arxiv.org/abs/2204.00698} {arXiv:2204.00698 [gr-qc]} \BibitemShut
  {NoStop}%
\bibitem [{\citenamefont {Duez}\ \emph {et~al.}(2005)\citenamefont {Duez},
  \citenamefont {Liu}, \citenamefont {Shapiro},\ and\ \citenamefont
  {Stephens}}]{Duez:2005sg}%
  \BibitemOpen
  \bibfield  {author} {\bibinfo {author} {\bibfnamefont {M.~D.}\ \bibnamefont
  {Duez}}, \bibinfo {author} {\bibfnamefont {Y.~T.}\ \bibnamefont {Liu}},
  \bibinfo {author} {\bibfnamefont {S.~L.}\ \bibnamefont {Shapiro}}, \ and\
  \bibinfo {author} {\bibfnamefont {B.~C.}\ \bibnamefont {Stephens}},\ }\href
  {\doibase 10.1103/PhysRevD.72.024029} {\bibfield  {journal} {\bibinfo
  {journal} {Phys. Rev. D}\ }\textbf {\bibinfo {volume} {72}},\ \bibinfo
  {pages} {024029} (\bibinfo {year} {2005})},\ \Eprint
  {http://arxiv.org/abs/astro-ph/0503421} {arXiv:astro-ph/0503421} \BibitemShut
  {NoStop}%
\bibitem [{\citenamefont {Paschalidis}\ \emph {et~al.}(2011)\citenamefont
  {Paschalidis}, \citenamefont {Etienne}, \citenamefont {Liu},\ and\
  \citenamefont {Shapiro}}]{Paschalidis:2010dh}%
  \BibitemOpen
  \bibfield  {author} {\bibinfo {author} {\bibfnamefont {V.}~\bibnamefont
  {Paschalidis}}, \bibinfo {author} {\bibfnamefont {Z.}~\bibnamefont
  {Etienne}}, \bibinfo {author} {\bibfnamefont {Y.~T.}\ \bibnamefont {Liu}}, \
  and\ \bibinfo {author} {\bibfnamefont {S.~L.}\ \bibnamefont {Shapiro}},\
  }\href {\doibase 10.1103/PhysRevD.83.064002} {\bibfield  {journal} {\bibinfo
  {journal} {Phys. Rev. D}\ }\textbf {\bibinfo {volume} {83}},\ \bibinfo
  {pages} {064002} (\bibinfo {year} {2011})},\ \Eprint
  {http://arxiv.org/abs/1009.4932} {arXiv:1009.4932 [astro-ph.HE]} \BibitemShut
  {NoStop}%
\bibitem [{\citenamefont {Etienne}\ \emph {et~al.}(2012)\citenamefont
  {Etienne}, \citenamefont {Paschalidis}, \citenamefont {Liu},\ and\
  \citenamefont {Shapiro}}]{Etienne:2011re}%
  \BibitemOpen
  \bibfield  {author} {\bibinfo {author} {\bibfnamefont {Z.~B.}\ \bibnamefont
  {Etienne}}, \bibinfo {author} {\bibfnamefont {V.}~\bibnamefont
  {Paschalidis}}, \bibinfo {author} {\bibfnamefont {Y.~T.}\ \bibnamefont
  {Liu}}, \ and\ \bibinfo {author} {\bibfnamefont {S.~L.}\ \bibnamefont
  {Shapiro}},\ }\href {\doibase 10.1103/PhysRevD.85.024013} {\bibfield
  {journal} {\bibinfo  {journal} {Phys. Rev. D}\ }\textbf {\bibinfo {volume}
  {85}},\ \bibinfo {pages} {024013} (\bibinfo {year} {2012})},\ \Eprint
  {http://arxiv.org/abs/1110.4633} {arXiv:1110.4633 [astro-ph.HE]} \BibitemShut
  {NoStop}%
\bibitem [{\citenamefont {Etienne}\ \emph {et~al.}(2015)\citenamefont
  {Etienne}, \citenamefont {Paschalidis}, \citenamefont {Haas}, \citenamefont
  {M\"osta},\ and\ \citenamefont {Shapiro}}]{Etienne:2015cea}%
  \BibitemOpen
  \bibfield  {author} {\bibinfo {author} {\bibfnamefont {Z.~B.}\ \bibnamefont
  {Etienne}}, \bibinfo {author} {\bibfnamefont {V.}~\bibnamefont
  {Paschalidis}}, \bibinfo {author} {\bibfnamefont {R.}~\bibnamefont {Haas}},
  \bibinfo {author} {\bibfnamefont {P.}~\bibnamefont {M\"osta}}, \ and\
  \bibinfo {author} {\bibfnamefont {S.~L.}\ \bibnamefont {Shapiro}},\ }\href
  {\doibase 10.1088/0264-9381/32/17/175009} {\bibfield  {journal} {\bibinfo
  {journal} {Class. Quant. Grav.}\ }\textbf {\bibinfo {volume} {32}},\ \bibinfo
  {pages} {175009} (\bibinfo {year} {2015})},\ \Eprint
  {http://arxiv.org/abs/1501.07276} {arXiv:1501.07276 [astro-ph.HE]}
  \BibitemShut {NoStop}%
\bibitem [{\citenamefont {{Allen}}\ \emph {et~al.}(2001)\citenamefont
  {{Allen}}, \citenamefont {{Angulo}}, \citenamefont {{Foster}}, \citenamefont
  {{Lanfermann}}, \citenamefont {{Liu}}, \citenamefont {{Radke}}, \citenamefont
  {{Seidel}},\ and\ \citenamefont {{Shalf}}}]{Allen2001}%
  \BibitemOpen
  \bibfield  {author} {\bibinfo {author} {\bibfnamefont {G.}~\bibnamefont
  {{Allen}}}, \bibinfo {author} {\bibfnamefont {D.}~\bibnamefont {{Angulo}}},
  \bibinfo {author} {\bibfnamefont {I.}~\bibnamefont {{Foster}}}, \bibinfo
  {author} {\bibfnamefont {G.}~\bibnamefont {{Lanfermann}}}, \bibinfo {author}
  {\bibfnamefont {C.}~\bibnamefont {{Liu}}}, \bibinfo {author} {\bibfnamefont
  {T.}~\bibnamefont {{Radke}}}, \bibinfo {author} {\bibfnamefont
  {E.}~\bibnamefont {{Seidel}}}, \ and\ \bibinfo {author} {\bibfnamefont
  {J.}~\bibnamefont {{Shalf}}},\ }\href@noop {} {\bibfield  {journal} {\bibinfo
   {journal} {The International Journal of High Performance Computing
  Applications}\ }\textbf {\bibinfo {volume} {15}},\ \bibinfo {pages} {345}
  (\bibinfo {year} {2001})}\BibitemShut {NoStop}%
\bibitem [{\citenamefont {Schnetter}\ \emph {et~al.}(2004)\citenamefont
  {Schnetter}, \citenamefont {Hawley},\ and\ \citenamefont
  {Hawke}}]{Schnetter:2003rb}%
  \BibitemOpen
  \bibfield  {author} {\bibinfo {author} {\bibfnamefont {E.}~\bibnamefont
  {Schnetter}}, \bibinfo {author} {\bibfnamefont {S.~H.}\ \bibnamefont
  {Hawley}}, \ and\ \bibinfo {author} {\bibfnamefont {I.}~\bibnamefont
  {Hawke}},\ }\href {\doibase 10.1088/0264-9381/21/6/014} {\bibfield  {journal}
  {\bibinfo  {journal} {Class. Quant. Grav.}\ }\textbf {\bibinfo {volume}
  {21}},\ \bibinfo {pages} {1465} (\bibinfo {year} {2004})},\ \Eprint
  {http://arxiv.org/abs/gr-qc/0310042} {arXiv:gr-qc/0310042} \BibitemShut
  {NoStop}%
\bibitem [{\citenamefont {Schnetter}\ \emph {et~al.}(2006)\citenamefont
  {Schnetter}, \citenamefont {Diener}, \citenamefont {Dorband},\ and\
  \citenamefont {Tiglio}}]{Schnetter:2006pg}%
  \BibitemOpen
  \bibfield  {author} {\bibinfo {author} {\bibfnamefont {E.}~\bibnamefont
  {Schnetter}}, \bibinfo {author} {\bibfnamefont {P.}~\bibnamefont {Diener}},
  \bibinfo {author} {\bibfnamefont {E.~N.}\ \bibnamefont {Dorband}}, \ and\
  \bibinfo {author} {\bibfnamefont {M.}~\bibnamefont {Tiglio}},\ }\href
  {\doibase 10.1088/0264-9381/23/16/S14} {\bibfield  {journal} {\bibinfo
  {journal} {Class. Quant. Grav.}\ }\textbf {\bibinfo {volume} {23}},\ \bibinfo
  {pages} {S553} (\bibinfo {year} {2006})},\ \Eprint
  {http://arxiv.org/abs/gr-qc/0602104} {arXiv:gr-qc/0602104} \BibitemShut
  {NoStop}%
\bibitem [{\citenamefont {Wiringa}\ \emph {et~al.}(1988)\citenamefont
  {Wiringa}, \citenamefont {Fiks},\ and\ \citenamefont
  {Fabrocini}}]{Wiringa:1988tp}%
  \BibitemOpen
  \bibfield  {author} {\bibinfo {author} {\bibfnamefont {R.~B.}\ \bibnamefont
  {Wiringa}}, \bibinfo {author} {\bibfnamefont {V.}~\bibnamefont {Fiks}}, \
  and\ \bibinfo {author} {\bibfnamefont {A.}~\bibnamefont {Fabrocini}},\ }\href
  {\doibase 10.1103/PhysRevC.38.1010} {\bibfield  {journal} {\bibinfo
  {journal} {Phys. Rev. C}\ }\textbf {\bibinfo {volume} {38}},\ \bibinfo
  {pages} {1010} (\bibinfo {year} {1988})}\BibitemShut {NoStop}%
\bibitem [{\citenamefont {Lackey}\ \emph {et~al.}(2006)\citenamefont {Lackey},
  \citenamefont {Nayyar},\ and\ \citenamefont {Owen}}]{Lackey:2005tk}%
  \BibitemOpen
  \bibfield  {author} {\bibinfo {author} {\bibfnamefont {B.~D.}\ \bibnamefont
  {Lackey}}, \bibinfo {author} {\bibfnamefont {M.}~\bibnamefont {Nayyar}}, \
  and\ \bibinfo {author} {\bibfnamefont {B.~J.}\ \bibnamefont {Owen}},\ }\href
  {\doibase 10.1103/PhysRevD.73.024021} {\bibfield  {journal} {\bibinfo
  {journal} {Phys. Rev. D}\ }\textbf {\bibinfo {volume} {73}},\ \bibinfo
  {pages} {024021} (\bibinfo {year} {2006})},\ \Eprint
  {http://arxiv.org/abs/astro-ph/0507312} {arXiv:astro-ph/0507312} \BibitemShut
  {NoStop}%
\bibitem [{\citenamefont {O'Boyle}\ \emph {et~al.}(2020)\citenamefont
  {O'Boyle}, \citenamefont {Markakis}, \citenamefont {Stergioulas},\ and\
  \citenamefont {Read}}]{OBoyle:2020qvf}%
  \BibitemOpen
  \bibfield  {author} {\bibinfo {author} {\bibfnamefont {M.~F.}\ \bibnamefont
  {O'Boyle}}, \bibinfo {author} {\bibfnamefont {C.}~\bibnamefont {Markakis}},
  \bibinfo {author} {\bibfnamefont {N.}~\bibnamefont {Stergioulas}}, \ and\
  \bibinfo {author} {\bibfnamefont {J.~S.}\ \bibnamefont {Read}},\ }\href
  {\doibase 10.1103/PhysRevD.102.083027} {\bibfield  {journal} {\bibinfo
  {journal} {Phys. Rev. D}\ }\textbf {\bibinfo {volume} {102}},\ \bibinfo
  {pages} {083027} (\bibinfo {year} {2020})},\ \Eprint
  {http://arxiv.org/abs/2008.03342} {arXiv:2008.03342 [astro-ph.HE]}
  \BibitemShut {NoStop}%
\bibitem [{\citenamefont {Most}\ \emph {et~al.}(2022)\citenamefont {Most},
  \citenamefont {Haber}, \citenamefont {Harris}, \citenamefont {Zhang},
  \citenamefont {Alford},\ and\ \citenamefont {Noronha}}]{Most:2022yhe}%
  \BibitemOpen
  \bibfield  {author} {\bibinfo {author} {\bibfnamefont {E.~R.}\ \bibnamefont
  {Most}}, \bibinfo {author} {\bibfnamefont {A.}~\bibnamefont {Haber}},
  \bibinfo {author} {\bibfnamefont {S.~P.}\ \bibnamefont {Harris}}, \bibinfo
  {author} {\bibfnamefont {Z.}~\bibnamefont {Zhang}}, \bibinfo {author}
  {\bibfnamefont {M.~G.}\ \bibnamefont {Alford}}, \ and\ \bibinfo {author}
  {\bibfnamefont {J.}~\bibnamefont {Noronha}},\ }\href@noop {} {\  (\bibinfo
  {year} {2022})},\ \Eprint {http://arxiv.org/abs/2207.00442} {arXiv:2207.00442
  [astro-ph.HE]} \BibitemShut {NoStop}%
\bibitem [{\citenamefont {Li}\ \emph {et~al.}(2021)\citenamefont {Li},
  \citenamefont {Cai}, \citenamefont {Xie},\ and\ \citenamefont
  {Zhang}}]{Li:2021thg}%
  \BibitemOpen
  \bibfield  {author} {\bibinfo {author} {\bibfnamefont {B.-A.}\ \bibnamefont
  {Li}}, \bibinfo {author} {\bibfnamefont {B.-J.}\ \bibnamefont {Cai}},
  \bibinfo {author} {\bibfnamefont {W.-J.}\ \bibnamefont {Xie}}, \ and\
  \bibinfo {author} {\bibfnamefont {N.-B.}\ \bibnamefont {Zhang}},\ }\href
  {\doibase 10.3390/universe7060182} {\bibfield  {journal} {\bibinfo  {journal}
  {Universe}\ }\textbf {\bibinfo {volume} {7}},\ \bibinfo {pages} {182}
  (\bibinfo {year} {2021})},\ \Eprint {http://arxiv.org/abs/2105.04629}
  {arXiv:2105.04629 [nucl-th]} \BibitemShut {NoStop}%
\bibitem [{\citenamefont {Raithel}\ and\ \citenamefont
  {Ozel}(2019)}]{Raithel:2019ejc}%
  \BibitemOpen
  \bibfield  {author} {\bibinfo {author} {\bibfnamefont {C.~A.}\ \bibnamefont
  {Raithel}}\ and\ \bibinfo {author} {\bibfnamefont {F.}~\bibnamefont {Ozel}},\
  }\href {\doibase 10.3847/1538-4357/ab48e6} {\  (\bibinfo {year} {2019}),\
  10.3847/1538-4357/ab48e6},\ \Eprint {http://arxiv.org/abs/1908.00018}
  {arXiv:1908.00018 [astro-ph.HE]} \BibitemShut {NoStop}%
\bibitem [{\citenamefont {Most}\ and\ \citenamefont
  {Raithel}(2021)}]{Most:2021ktk}%
  \BibitemOpen
  \bibfield  {author} {\bibinfo {author} {\bibfnamefont {E.~R.}\ \bibnamefont
  {Most}}\ and\ \bibinfo {author} {\bibfnamefont {C.~A.}\ \bibnamefont
  {Raithel}},\ }\href {\doibase 10.1103/PhysRevD.104.124012} {\bibfield
  {journal} {\bibinfo  {journal} {Phys. Rev. D}\ }\textbf {\bibinfo {volume}
  {104}},\ \bibinfo {pages} {124012} (\bibinfo {year} {2021})},\ \Eprint
  {http://arxiv.org/abs/2107.06804} {arXiv:2107.06804 [astro-ph.HE]}
  \BibitemShut {NoStop}%
\bibitem [{\citenamefont {{Gourgoulhon}}\ \emph {et~al.}()\citenamefont
  {{Gourgoulhon}}, \citenamefont {{Grandclement}}, \citenamefont {{Marck}},
  \citenamefont {{Novak}},\ and\ \citenamefont {{Taniguchi}}}]{Lorene}%
  \BibitemOpen
  \bibfield  {author} {\bibinfo {author} {\bibfnamefont {E.}~\bibnamefont
  {{Gourgoulhon}}}, \bibinfo {author} {\bibfnamefont {P.}~\bibnamefont
  {{Grandclement}}}, \bibinfo {author} {\bibfnamefont {J.-A.}\ \bibnamefont
  {{Marck}}}, \bibinfo {author} {\bibfnamefont {J.}~\bibnamefont {{Novak}}}, \
  and\ \bibinfo {author} {\bibfnamefont {K.}~\bibnamefont {{Taniguchi}}},\
  }\href@noop {} {\enquote {\bibinfo {title} {Lorene},}\ }\bibinfo
  {howpublished} {\url{https://lorene.obspm.fr/}}\BibitemShut {NoStop}%
\bibitem [{\citenamefont {Oechslin}\ \emph {et~al.}(2007)\citenamefont
  {Oechslin}, \citenamefont {Janka},\ and\ \citenamefont
  {Marek}}]{Oechslin:2006uk}%
  \BibitemOpen
  \bibfield  {author} {\bibinfo {author} {\bibfnamefont {R.}~\bibnamefont
  {Oechslin}}, \bibinfo {author} {\bibfnamefont {H.~T.}\ \bibnamefont {Janka}},
  \ and\ \bibinfo {author} {\bibfnamefont {A.}~\bibnamefont {Marek}},\ }\href
  {\doibase 10.1051/0004-6361:20066682} {\bibfield  {journal} {\bibinfo
  {journal} {Astron. Astrophys.}\ }\textbf {\bibinfo {volume} {467}},\ \bibinfo
  {pages} {395} (\bibinfo {year} {2007})},\ \Eprint
  {http://arxiv.org/abs/astro-ph/0611047} {arXiv:astro-ph/0611047} \BibitemShut
  {NoStop}%
\bibitem [{\citenamefont {Radice}\ \emph {et~al.}(2020)\citenamefont {Radice},
  \citenamefont {Bernuzzi},\ and\ \citenamefont {Perego}}]{Radice:2020ddv}%
  \BibitemOpen
  \bibfield  {author} {\bibinfo {author} {\bibfnamefont {D.}~\bibnamefont
  {Radice}}, \bibinfo {author} {\bibfnamefont {S.}~\bibnamefont {Bernuzzi}}, \
  and\ \bibinfo {author} {\bibfnamefont {A.}~\bibnamefont {Perego}},\ }\href
  {\doibase 10.1146/annurev-nucl-013120-114541} {\bibfield  {journal} {\bibinfo
   {journal} {Ann. Rev. Nucl. Part. Sci.}\ }\textbf {\bibinfo {volume} {70}},\
  \bibinfo {pages} {95} (\bibinfo {year} {2020})},\ \Eprint
  {http://arxiv.org/abs/2002.03863} {arXiv:2002.03863 [astro-ph.HE]}
  \BibitemShut {NoStop}%
\bibitem [{\citenamefont {Foucart}(2022)}]{Foucart:2022bth}%
  \BibitemOpen
  \bibfield  {author} {\bibinfo {author} {\bibfnamefont {F.}~\bibnamefont
  {Foucart}},\ }\href {\doibase 10.1007/s41115-023-00016-y} {\  (\bibinfo
  {year} {2022}),\ 10.1007/s41115-023-00016-y},\ \Eprint
  {http://arxiv.org/abs/2209.02538} {arXiv:2209.02538 [astro-ph.HE]}
  \BibitemShut {NoStop}%
\bibitem [{\citenamefont {Hanauske}\ \emph {et~al.}(2017)\citenamefont
  {Hanauske}, \citenamefont {Takami}, \citenamefont {Bovard}, \citenamefont
  {Rezzolla}, \citenamefont {Font}, \citenamefont {Galeazzi},\ and\
  \citenamefont {St\"ocker}}]{Hanauske:2016gia}%
  \BibitemOpen
  \bibfield  {author} {\bibinfo {author} {\bibfnamefont {M.}~\bibnamefont
  {Hanauske}}, \bibinfo {author} {\bibfnamefont {K.}~\bibnamefont {Takami}},
  \bibinfo {author} {\bibfnamefont {L.}~\bibnamefont {Bovard}}, \bibinfo
  {author} {\bibfnamefont {L.}~\bibnamefont {Rezzolla}}, \bibinfo {author}
  {\bibfnamefont {J.~A.}\ \bibnamefont {Font}}, \bibinfo {author}
  {\bibfnamefont {F.}~\bibnamefont {Galeazzi}}, \ and\ \bibinfo {author}
  {\bibfnamefont {H.}~\bibnamefont {St\"ocker}},\ }\href {\doibase
  10.1103/PhysRevD.96.043004} {\bibfield  {journal} {\bibinfo  {journal} {Phys.
  Rev. D}\ }\textbf {\bibinfo {volume} {96}},\ \bibinfo {pages} {043004}
  (\bibinfo {year} {2017})},\ \Eprint {http://arxiv.org/abs/1611.07152}
  {arXiv:1611.07152 [gr-qc]} \BibitemShut {NoStop}%
\bibitem [{\citenamefont {Kastaun}\ \emph {et~al.}(2016)\citenamefont
  {Kastaun}, \citenamefont {Ciolfi},\ and\ \citenamefont
  {Giacomazzo}}]{Kastaun:2016yaf}%
  \BibitemOpen
  \bibfield  {author} {\bibinfo {author} {\bibfnamefont {W.}~\bibnamefont
  {Kastaun}}, \bibinfo {author} {\bibfnamefont {R.}~\bibnamefont {Ciolfi}}, \
  and\ \bibinfo {author} {\bibfnamefont {B.}~\bibnamefont {Giacomazzo}},\
  }\href {\doibase 10.1103/PhysRevD.94.044060} {\bibfield  {journal} {\bibinfo
  {journal} {Phys. Rev. D}\ }\textbf {\bibinfo {volume} {94}},\ \bibinfo
  {pages} {044060} (\bibinfo {year} {2016})},\ \Eprint
  {http://arxiv.org/abs/1607.02186} {arXiv:1607.02186 [astro-ph.HE]}
  \BibitemShut {NoStop}%
\bibitem [{\citenamefont {Bauswein}\ \emph {et~al.}(2013)\citenamefont
  {Bauswein}, \citenamefont {Goriely},\ and\ \citenamefont
  {Janka}}]{Bauswein:2013yna}%
  \BibitemOpen
  \bibfield  {author} {\bibinfo {author} {\bibfnamefont {A.}~\bibnamefont
  {Bauswein}}, \bibinfo {author} {\bibfnamefont {S.}~\bibnamefont {Goriely}}, \
  and\ \bibinfo {author} {\bibfnamefont {H.~T.}\ \bibnamefont {Janka}},\ }\href
  {\doibase 10.1088/0004-637X/773/1/78} {\bibfield  {journal} {\bibinfo
  {journal} {Astrophys. J.}\ }\textbf {\bibinfo {volume} {773}},\ \bibinfo
  {pages} {78} (\bibinfo {year} {2013})},\ \Eprint
  {http://arxiv.org/abs/1302.6530} {arXiv:1302.6530 [astro-ph.SR]} \BibitemShut
  {NoStop}%
\bibitem [{\citenamefont {Reisswig}\ and\ \citenamefont
  {Pollney}(2011)}]{Reisswig:2010di}%
  \BibitemOpen
  \bibfield  {author} {\bibinfo {author} {\bibfnamefont {C.}~\bibnamefont
  {Reisswig}}\ and\ \bibinfo {author} {\bibfnamefont {D.}~\bibnamefont
  {Pollney}},\ }\href {\doibase 10.1088/0264-9381/28/19/195015} {\bibfield
  {journal} {\bibinfo  {journal} {Class. Quant. Grav.}\ }\textbf {\bibinfo
  {volume} {28}},\ \bibinfo {pages} {195015} (\bibinfo {year} {2011})},\
  \Eprint {http://arxiv.org/abs/1006.1632} {arXiv:1006.1632 [gr-qc]}
  \BibitemShut {NoStop}%
\bibitem [{aLI()}]{aLIGOsensitivity}%
  \BibitemOpen
  \href@noop {} {\enquote {\bibinfo {title} {aligo design sensitivity},}\
  }\bibinfo {howpublished}
  {\url{https://dcc.ligo.org/public/0149/T1800044/005/aLIGODesign.txt}},\
  \bibinfo {note} {accessed: 2023-04-03}\BibitemShut {NoStop}%
\bibitem [{CEs()}]{CEsensitivity}%
  \BibitemOpen
  \href@noop {} {\enquote {\bibinfo {title} {Cosmic explorer sensitivity
  curves},}\ }\bibinfo {howpublished}
  {\url{}https://dcc.ligo.org/LIGO-P1600143/public},\ \bibinfo {note}
  {accessed: 2023-04-03}\BibitemShut {NoStop}%
\bibitem [{\citenamefont {Stergioulas}\ \emph {et~al.}(2011)\citenamefont
  {Stergioulas}, \citenamefont {Bauswein}, \citenamefont {Zagkouris},\ and\
  \citenamefont {Janka}}]{Stergioulas:2011gd}%
  \BibitemOpen
  \bibfield  {author} {\bibinfo {author} {\bibfnamefont {N.}~\bibnamefont
  {Stergioulas}}, \bibinfo {author} {\bibfnamefont {A.}~\bibnamefont
  {Bauswein}}, \bibinfo {author} {\bibfnamefont {K.}~\bibnamefont {Zagkouris}},
  \ and\ \bibinfo {author} {\bibfnamefont {H.-T.}\ \bibnamefont {Janka}},\
  }\href {\doibase 10.1111/j.1365-2966.2011.19493.x} {\bibfield  {journal}
  {\bibinfo  {journal} {Mon. Not. Roy. Astron. Soc.}\ }\textbf {\bibinfo
  {volume} {418}},\ \bibinfo {pages} {427} (\bibinfo {year} {2011})},\ \Eprint
  {http://arxiv.org/abs/1105.0368} {arXiv:1105.0368 [gr-qc]} \BibitemShut
  {NoStop}%
\bibitem [{\citenamefont {Takami}\ \emph {et~al.}(2015)\citenamefont {Takami},
  \citenamefont {Rezzolla},\ and\ \citenamefont {Baiotti}}]{Takami:2014tva}%
  \BibitemOpen
  \bibfield  {author} {\bibinfo {author} {\bibfnamefont {K.}~\bibnamefont
  {Takami}}, \bibinfo {author} {\bibfnamefont {L.}~\bibnamefont {Rezzolla}}, \
  and\ \bibinfo {author} {\bibfnamefont {L.}~\bibnamefont {Baiotti}},\ }\href
  {\doibase 10.1103/PhysRevD.91.064001} {\bibfield  {journal} {\bibinfo
  {journal} {Phys. Rev. D}\ }\textbf {\bibinfo {volume} {91}},\ \bibinfo
  {pages} {064001} (\bibinfo {year} {2015})},\ \Eprint
  {http://arxiv.org/abs/1412.3240} {arXiv:1412.3240 [gr-qc]} \BibitemShut
  {NoStop}%
\bibitem [{\citenamefont {Rezzolla}\ and\ \citenamefont
  {Takami}(2016)}]{Rezzolla:2016nxn}%
  \BibitemOpen
  \bibfield  {author} {\bibinfo {author} {\bibfnamefont {L.}~\bibnamefont
  {Rezzolla}}\ and\ \bibinfo {author} {\bibfnamefont {K.}~\bibnamefont
  {Takami}},\ }\href {\doibase 10.1103/PhysRevD.93.124051} {\bibfield
  {journal} {\bibinfo  {journal} {Phys. Rev. D}\ }\textbf {\bibinfo {volume}
  {93}},\ \bibinfo {pages} {124051} (\bibinfo {year} {2016})},\ \Eprint
  {http://arxiv.org/abs/1604.00246} {arXiv:1604.00246 [gr-qc]} \BibitemShut
  {NoStop}%
\bibitem [{\citenamefont {Bauswein}\ \emph {et~al.}(2012)\citenamefont
  {Bauswein}, \citenamefont {Janka}, \citenamefont {Hebeler},\ and\
  \citenamefont {Schwenk}}]{Bauswein:2012ya}%
  \BibitemOpen
  \bibfield  {author} {\bibinfo {author} {\bibfnamefont {A.}~\bibnamefont
  {Bauswein}}, \bibinfo {author} {\bibfnamefont {H.~T.}\ \bibnamefont {Janka}},
  \bibinfo {author} {\bibfnamefont {K.}~\bibnamefont {Hebeler}}, \ and\
  \bibinfo {author} {\bibfnamefont {A.}~\bibnamefont {Schwenk}},\ }\href
  {\doibase 10.1103/PhysRevD.86.063001} {\bibfield  {journal} {\bibinfo
  {journal} {Phys. Rev. D}\ }\textbf {\bibinfo {volume} {86}},\ \bibinfo
  {pages} {063001} (\bibinfo {year} {2012})},\ \Eprint
  {http://arxiv.org/abs/1204.1888} {arXiv:1204.1888 [astro-ph.SR]} \BibitemShut
  {NoStop}%
\bibitem [{\citenamefont {Bauswein}\ and\ \citenamefont
  {Janka}(2012)}]{Bauswein:2011tp}%
  \BibitemOpen
  \bibfield  {author} {\bibinfo {author} {\bibfnamefont {A.}~\bibnamefont
  {Bauswein}}\ and\ \bibinfo {author} {\bibfnamefont {H.~T.}\ \bibnamefont
  {Janka}},\ }\href {\doibase 10.1103/PhysRevLett.108.011101} {\bibfield
  {journal} {\bibinfo  {journal} {Phys. Rev. Lett.}\ }\textbf {\bibinfo
  {volume} {108}},\ \bibinfo {pages} {011101} (\bibinfo {year} {2012})},\
  \Eprint {http://arxiv.org/abs/1106.1616} {arXiv:1106.1616 [astro-ph.SR]}
  \BibitemShut {NoStop}%
\bibitem [{\citenamefont {Takami}\ \emph {et~al.}(2014)\citenamefont {Takami},
  \citenamefont {Rezzolla},\ and\ \citenamefont {Baiotti}}]{Takami:2014zpa}%
  \BibitemOpen
  \bibfield  {author} {\bibinfo {author} {\bibfnamefont {K.}~\bibnamefont
  {Takami}}, \bibinfo {author} {\bibfnamefont {L.}~\bibnamefont {Rezzolla}}, \
  and\ \bibinfo {author} {\bibfnamefont {L.}~\bibnamefont {Baiotti}},\ }\href
  {\doibase 10.1103/PhysRevLett.113.091104} {\bibfield  {journal} {\bibinfo
  {journal} {Phys. Rev. Lett.}\ }\textbf {\bibinfo {volume} {113}},\ \bibinfo
  {pages} {091104} (\bibinfo {year} {2014})},\ \Eprint
  {http://arxiv.org/abs/1403.5672} {arXiv:1403.5672 [gr-qc]} \BibitemShut
  {NoStop}%
\bibitem [{\citenamefont {Bernuzzi}\ \emph {et~al.}(2015)\citenamefont
  {Bernuzzi}, \citenamefont {Dietrich},\ and\ \citenamefont
  {Nagar}}]{Bernuzzi:2015rla}%
  \BibitemOpen
  \bibfield  {author} {\bibinfo {author} {\bibfnamefont {S.}~\bibnamefont
  {Bernuzzi}}, \bibinfo {author} {\bibfnamefont {T.}~\bibnamefont {Dietrich}},
  \ and\ \bibinfo {author} {\bibfnamefont {A.}~\bibnamefont {Nagar}},\ }\href
  {\doibase 10.1103/PhysRevLett.115.091101} {\bibfield  {journal} {\bibinfo
  {journal} {Phys. Rev. Lett.}\ }\textbf {\bibinfo {volume} {115}},\ \bibinfo
  {pages} {091101} (\bibinfo {year} {2015})},\ \Eprint
  {http://arxiv.org/abs/1504.01764} {arXiv:1504.01764 [gr-qc]} \BibitemShut
  {NoStop}%
\bibitem [{\citenamefont {Raithel}\ and\ \citenamefont
  {Most}(2022)}]{Raithel:2022orm}%
  \BibitemOpen
  \bibfield  {author} {\bibinfo {author} {\bibfnamefont {C.~A.}\ \bibnamefont
  {Raithel}}\ and\ \bibinfo {author} {\bibfnamefont {E.~R.}\ \bibnamefont
  {Most}},\ }\href {\doibase 10.3847/2041-8213/ac7c75} {\bibfield  {journal}
  {\bibinfo  {journal} {Astrophys. J. Lett.}\ }\textbf {\bibinfo {volume}
  {933}},\ \bibinfo {pages} {L39} (\bibinfo {year} {2022})},\ \Eprint
  {http://arxiv.org/abs/2201.03594} {arXiv:2201.03594 [astro-ph.HE]}
  \BibitemShut {NoStop}%
\bibitem [{\citenamefont {Breschi}\ \emph {et~al.}(2022)\citenamefont
  {Breschi}, \citenamefont {Bernuzzi}, \citenamefont {Godzieba}, \citenamefont
  {Perego},\ and\ \citenamefont {Radice}}]{Breschi:2021xrx}%
  \BibitemOpen
  \bibfield  {author} {\bibinfo {author} {\bibfnamefont {M.}~\bibnamefont
  {Breschi}}, \bibinfo {author} {\bibfnamefont {S.}~\bibnamefont {Bernuzzi}},
  \bibinfo {author} {\bibfnamefont {D.}~\bibnamefont {Godzieba}}, \bibinfo
  {author} {\bibfnamefont {A.}~\bibnamefont {Perego}}, \ and\ \bibinfo {author}
  {\bibfnamefont {D.}~\bibnamefont {Radice}},\ }\href {\doibase
  10.1103/PhysRevLett.128.161102} {\bibfield  {journal} {\bibinfo  {journal}
  {Phys. Rev. Lett.}\ }\textbf {\bibinfo {volume} {128}},\ \bibinfo {pages}
  {161102} (\bibinfo {year} {2022})},\ \Eprint
  {http://arxiv.org/abs/2110.06957} {arXiv:2110.06957 [gr-qc]} \BibitemShut
  {NoStop}%
\bibitem [{\citenamefont {Fields}\ \emph {et~al.}(2023)\citenamefont {Fields},
  \citenamefont {Prakash}, \citenamefont {Breschi}, \citenamefont {Radice},
  \citenamefont {Bernuzzi},\ and\ \citenamefont {Schneider}}]{Fields:2023bhs}%
  \BibitemOpen
  \bibfield  {author} {\bibinfo {author} {\bibfnamefont {J.}~\bibnamefont
  {Fields}}, \bibinfo {author} {\bibfnamefont {A.}~\bibnamefont {Prakash}},
  \bibinfo {author} {\bibfnamefont {M.}~\bibnamefont {Breschi}}, \bibinfo
  {author} {\bibfnamefont {D.}~\bibnamefont {Radice}}, \bibinfo {author}
  {\bibfnamefont {S.}~\bibnamefont {Bernuzzi}}, \ and\ \bibinfo {author}
  {\bibfnamefont {A.~d.~S.}\ \bibnamefont {Schneider}},\ }\href@noop {} {\
  (\bibinfo {year} {2023})},\ \Eprint {http://arxiv.org/abs/2302.11359}
  {arXiv:2302.11359 [astro-ph.HE]} \BibitemShut {NoStop}%
\bibitem [{\citenamefont {Schneider}\ \emph {et~al.}(2017)\citenamefont
  {Schneider}, \citenamefont {Roberts},\ and\ \citenamefont
  {Ott}}]{Schneider:2017tfi}%
  \BibitemOpen
  \bibfield  {author} {\bibinfo {author} {\bibfnamefont {A.~S.}\ \bibnamefont
  {Schneider}}, \bibinfo {author} {\bibfnamefont {L.~F.}\ \bibnamefont
  {Roberts}}, \ and\ \bibinfo {author} {\bibfnamefont {C.~D.}\ \bibnamefont
  {Ott}},\ }\href {\doibase 10.1103/PhysRevC.96.065802} {\bibfield  {journal}
  {\bibinfo  {journal} {Phys. Rev. C}\ }\textbf {\bibinfo {volume} {96}},\
  \bibinfo {pages} {065802} (\bibinfo {year} {2017})},\ \Eprint
  {http://arxiv.org/abs/1707.01527} {arXiv:1707.01527 [astro-ph.HE]}
  \BibitemShut {NoStop}%
\bibitem [{\citenamefont {Bauswein}\ and\ \citenamefont
  {Stergioulas}(2015)}]{Bauswein:2015yca}%
  \BibitemOpen
  \bibfield  {author} {\bibinfo {author} {\bibfnamefont {A.}~\bibnamefont
  {Bauswein}}\ and\ \bibinfo {author} {\bibfnamefont {N.}~\bibnamefont
  {Stergioulas}},\ }\href {\doibase 10.1103/PhysRevD.91.124056} {\bibfield
  {journal} {\bibinfo  {journal} {Phys. Rev. D}\ }\textbf {\bibinfo {volume}
  {91}},\ \bibinfo {pages} {124056} (\bibinfo {year} {2015})},\ \Eprint
  {http://arxiv.org/abs/1502.03176} {arXiv:1502.03176 [astro-ph.SR]}
  \BibitemShut {NoStop}%
\bibitem [{\citenamefont {Shibata}\ and\ \citenamefont
  {Hotokezaka}(2019)}]{Shibata:2019wef}%
  \BibitemOpen
  \bibfield  {author} {\bibinfo {author} {\bibfnamefont {M.}~\bibnamefont
  {Shibata}}\ and\ \bibinfo {author} {\bibfnamefont {K.}~\bibnamefont
  {Hotokezaka}},\ }\href {\doibase 10.1146/annurev-nucl-101918-023625}
  {\bibfield  {journal} {\bibinfo  {journal} {Ann. Rev. Nucl. Part. Sci.}\
  }\textbf {\bibinfo {volume} {69}},\ \bibinfo {pages} {41} (\bibinfo {year}
  {2019})},\ \Eprint {http://arxiv.org/abs/1908.02350} {arXiv:1908.02350
  [astro-ph.HE]} \BibitemShut {NoStop}%
\bibitem [{\citenamefont {Radice}\ \emph {et~al.}(2018)\citenamefont {Radice},
  \citenamefont {Perego}, \citenamefont {Hotokezaka}, \citenamefont {Fromm},
  \citenamefont {Bernuzzi},\ and\ \citenamefont {Roberts}}]{Radice:2018pdn}%
  \BibitemOpen
  \bibfield  {author} {\bibinfo {author} {\bibfnamefont {D.}~\bibnamefont
  {Radice}}, \bibinfo {author} {\bibfnamefont {A.}~\bibnamefont {Perego}},
  \bibinfo {author} {\bibfnamefont {K.}~\bibnamefont {Hotokezaka}}, \bibinfo
  {author} {\bibfnamefont {S.~A.}\ \bibnamefont {Fromm}}, \bibinfo {author}
  {\bibfnamefont {S.}~\bibnamefont {Bernuzzi}}, \ and\ \bibinfo {author}
  {\bibfnamefont {L.~F.}\ \bibnamefont {Roberts}},\ }\href {\doibase
  10.3847/1538-4357/aaf054} {\bibfield  {journal} {\bibinfo  {journal}
  {Astrophys. J.}\ }\textbf {\bibinfo {volume} {869}},\ \bibinfo {pages} {130}
  (\bibinfo {year} {2018})},\ \Eprint {http://arxiv.org/abs/1809.11161}
  {arXiv:1809.11161 [astro-ph.HE]} \BibitemShut {NoStop}%
\bibitem [{\citenamefont {Radice}\ \emph {et~al.}(2016)\citenamefont {Radice},
  \citenamefont {Galeazzi}, \citenamefont {Lippuner}, \citenamefont {Roberts},
  \citenamefont {Ott},\ and\ \citenamefont {Rezzolla}}]{Radice:2016dwd}%
  \BibitemOpen
  \bibfield  {author} {\bibinfo {author} {\bibfnamefont {D.}~\bibnamefont
  {Radice}}, \bibinfo {author} {\bibfnamefont {F.}~\bibnamefont {Galeazzi}},
  \bibinfo {author} {\bibfnamefont {J.}~\bibnamefont {Lippuner}}, \bibinfo
  {author} {\bibfnamefont {L.~F.}\ \bibnamefont {Roberts}}, \bibinfo {author}
  {\bibfnamefont {C.~D.}\ \bibnamefont {Ott}}, \ and\ \bibinfo {author}
  {\bibfnamefont {L.}~\bibnamefont {Rezzolla}},\ }\href {\doibase
  10.1093/mnras/stw1227} {\bibfield  {journal} {\bibinfo  {journal} {Mon. Not.
  Roy. Astron. Soc.}\ }\textbf {\bibinfo {volume} {460}},\ \bibinfo {pages}
  {3255} (\bibinfo {year} {2016})},\ \Eprint {http://arxiv.org/abs/1601.02426}
  {arXiv:1601.02426 [astro-ph.HE]} \BibitemShut {NoStop}%
\bibitem [{\citenamefont {Breschi}\ \emph {et~al.}(2019)\citenamefont
  {Breschi}, \citenamefont {Bernuzzi}, \citenamefont {Zappa}, \citenamefont
  {Agathos}, \citenamefont {Perego}, \citenamefont {Radice},\ and\
  \citenamefont {Nagar}}]{Breschi:2019srl}%
  \BibitemOpen
  \bibfield  {author} {\bibinfo {author} {\bibfnamefont {M.}~\bibnamefont
  {Breschi}}, \bibinfo {author} {\bibfnamefont {S.}~\bibnamefont {Bernuzzi}},
  \bibinfo {author} {\bibfnamefont {F.}~\bibnamefont {Zappa}}, \bibinfo
  {author} {\bibfnamefont {M.}~\bibnamefont {Agathos}}, \bibinfo {author}
  {\bibfnamefont {A.}~\bibnamefont {Perego}}, \bibinfo {author} {\bibfnamefont
  {D.}~\bibnamefont {Radice}}, \ and\ \bibinfo {author} {\bibfnamefont
  {A.}~\bibnamefont {Nagar}},\ }\href {\doibase 10.1103/PhysRevD.100.104029}
  {\bibfield  {journal} {\bibinfo  {journal} {Phys. Rev. D}\ }\textbf {\bibinfo
  {volume} {100}},\ \bibinfo {pages} {104029} (\bibinfo {year} {2019})},\
  \Eprint {http://arxiv.org/abs/1908.11418} {arXiv:1908.11418 [gr-qc]}
  \BibitemShut {NoStop}%
\bibitem [{\citenamefont {East}\ \emph {et~al.}(2019)\citenamefont {East},
  \citenamefont {Paschalidis}, \citenamefont {Pretorius},\ and\ \citenamefont
  {Tsokaros}}]{East:2019lbk}%
  \BibitemOpen
  \bibfield  {author} {\bibinfo {author} {\bibfnamefont {W.~E.}\ \bibnamefont
  {East}}, \bibinfo {author} {\bibfnamefont {V.}~\bibnamefont {Paschalidis}},
  \bibinfo {author} {\bibfnamefont {F.}~\bibnamefont {Pretorius}}, \ and\
  \bibinfo {author} {\bibfnamefont {A.}~\bibnamefont {Tsokaros}},\ }\href
  {\doibase 10.1103/PhysRevD.100.124042} {\bibfield  {journal} {\bibinfo
  {journal} {Phys. Rev. D}\ }\textbf {\bibinfo {volume} {100}},\ \bibinfo
  {pages} {124042} (\bibinfo {year} {2019})},\ \Eprint
  {http://arxiv.org/abs/1906.05288} {arXiv:1906.05288 [astro-ph.HE]}
  \BibitemShut {NoStop}%
\bibitem [{\citenamefont {Reitze}\ \emph {et~al.}(2019)\citenamefont {Reitze}
  \emph {et~al.}}]{Reitze:2019iox}%
  \BibitemOpen
  \bibfield  {author} {\bibinfo {author} {\bibfnamefont {D.}~\bibnamefont
  {Reitze}} \emph {et~al.},\ }\href@noop {} {\bibfield  {journal} {\bibinfo
  {journal} {Bull. Am. Astron. Soc.}\ }\textbf {\bibinfo {volume} {51}},\
  \bibinfo {pages} {035} (\bibinfo {year} {2019})},\ \Eprint
  {http://arxiv.org/abs/1907.04833} {arXiv:1907.04833 [astro-ph.IM]}
  \BibitemShut {NoStop}%
\bibitem [{\citenamefont {Punturo}\ \emph {et~al.}(2010)\citenamefont {Punturo}
  \emph {et~al.}}]{Punturo:2010zz}%
  \BibitemOpen
  \bibfield  {author} {\bibinfo {author} {\bibfnamefont {M.}~\bibnamefont
  {Punturo}} \emph {et~al.},\ }\href {\doibase 10.1088/0264-9381/27/19/194002}
  {\bibfield  {journal} {\bibinfo  {journal} {Class. Quant. Grav.}\ }\textbf
  {\bibinfo {volume} {27}},\ \bibinfo {pages} {194002} (\bibinfo {year}
  {2010})}\BibitemShut {NoStop}%
\bibitem [{\citenamefont {Ackley}\ \emph {et~al.}(2020)\citenamefont {Ackley}
  \emph {et~al.}}]{Ackley:2020atn}%
  \BibitemOpen
  \bibfield  {author} {\bibinfo {author} {\bibfnamefont {K.}~\bibnamefont
  {Ackley}} \emph {et~al.},\ }\href {\doibase 10.1017/pasa.2020.39} {\bibfield
  {journal} {\bibinfo  {journal} {Publ. Astron. Soc. Austral.}\ }\textbf
  {\bibinfo {volume} {37}},\ \bibinfo {pages} {e047} (\bibinfo {year}
  {2020})},\ \Eprint {http://arxiv.org/abs/2007.03128} {arXiv:2007.03128
  [astro-ph.HE]} \BibitemShut {NoStop}%
\bibitem [{\citenamefont {Hempel}\ and\ \citenamefont
  {Schaffner-Bielich}(2010)}]{Hempel:2009mc}%
  \BibitemOpen
  \bibfield  {author} {\bibinfo {author} {\bibfnamefont {M.}~\bibnamefont
  {Hempel}}\ and\ \bibinfo {author} {\bibfnamefont {J.}~\bibnamefont
  {Schaffner-Bielich}},\ }\href {\doibase 10.1016/j.nuclphysa.2010.02.010}
  {\bibfield  {journal} {\bibinfo  {journal} {Nucl. Phys. A}\ }\textbf
  {\bibinfo {volume} {837}},\ \bibinfo {pages} {210} (\bibinfo {year}
  {2010})},\ \Eprint {http://arxiv.org/abs/0911.4073} {arXiv:0911.4073
  [nucl-th]} \BibitemShut {NoStop}%
\bibitem [{\citenamefont {Steiner}\ \emph {et~al.}(2013)\citenamefont
  {Steiner}, \citenamefont {Hempel},\ and\ \citenamefont
  {Fischer}}]{Steiner:2012rk}%
  \BibitemOpen
  \bibfield  {author} {\bibinfo {author} {\bibfnamefont {A.~W.}\ \bibnamefont
  {Steiner}}, \bibinfo {author} {\bibfnamefont {M.}~\bibnamefont {Hempel}}, \
  and\ \bibinfo {author} {\bibfnamefont {T.}~\bibnamefont {Fischer}},\ }\href
  {\doibase 10.1088/0004-637X/774/1/17} {\bibfield  {journal} {\bibinfo
  {journal} {Astrophys. J.}\ }\textbf {\bibinfo {volume} {774}},\ \bibinfo
  {pages} {17} (\bibinfo {year} {2013})},\ \Eprint
  {http://arxiv.org/abs/1207.2184} {arXiv:1207.2184 [astro-ph.SR]} \BibitemShut
  {NoStop}%
\bibitem [{\citenamefont {Shen}\ \emph {et~al.}(2011)\citenamefont {Shen},
  \citenamefont {Horowitz},\ and\ \citenamefont {Teige}}]{Shen:2011kr}%
  \BibitemOpen
  \bibfield  {author} {\bibinfo {author} {\bibfnamefont {G.}~\bibnamefont
  {Shen}}, \bibinfo {author} {\bibfnamefont {C.~J.}\ \bibnamefont {Horowitz}},
  \ and\ \bibinfo {author} {\bibfnamefont {S.}~\bibnamefont {Teige}},\ }\href
  {\doibase 10.1103/PhysRevC.83.035802} {\bibfield  {journal} {\bibinfo
  {journal} {Phys. Rev. C}\ }\textbf {\bibinfo {volume} {83}},\ \bibinfo
  {pages} {035802} (\bibinfo {year} {2011})},\ \Eprint
  {http://arxiv.org/abs/1101.3715} {arXiv:1101.3715 [astro-ph.SR]} \BibitemShut
  {NoStop}%
\bibitem [{\citenamefont {Shen}\ \emph {et~al.}(1998)\citenamefont {Shen},
  \citenamefont {Toki}, \citenamefont {Oyamatsu},\ and\ \citenamefont
  {Sumiyoshi}}]{Shen:1998gq}%
  \BibitemOpen
  \bibfield  {author} {\bibinfo {author} {\bibfnamefont {H.}~\bibnamefont
  {Shen}}, \bibinfo {author} {\bibfnamefont {H.}~\bibnamefont {Toki}}, \bibinfo
  {author} {\bibfnamefont {K.}~\bibnamefont {Oyamatsu}}, \ and\ \bibinfo
  {author} {\bibfnamefont {K.}~\bibnamefont {Sumiyoshi}},\ }\href {\doibase
  10.1016/S0375-9474(98)00236-X} {\bibfield  {journal} {\bibinfo  {journal}
  {Nucl. Phys. A}\ }\textbf {\bibinfo {volume} {637}},\ \bibinfo {pages} {435}
  (\bibinfo {year} {1998})},\ \Eprint {http://arxiv.org/abs/nucl-th/9805035}
  {arXiv:nucl-th/9805035} \BibitemShut {NoStop}%
\bibitem [{\citenamefont {Andersen}\ \emph {et~al.}(2021)\citenamefont
  {Andersen}, \citenamefont {Zha}, \citenamefont {da~Silva~Schneider},
  \citenamefont {Betranhandy}, \citenamefont {Couch},\ and\ \citenamefont
  {O'Connor}}]{Andersen:2021vzo}%
  \BibitemOpen
  \bibfield  {author} {\bibinfo {author} {\bibfnamefont {O.~E.}\ \bibnamefont
  {Andersen}}, \bibinfo {author} {\bibfnamefont {S.}~\bibnamefont {Zha}},
  \bibinfo {author} {\bibfnamefont {A.}~\bibnamefont {da~Silva~Schneider}},
  \bibinfo {author} {\bibfnamefont {A.}~\bibnamefont {Betranhandy}}, \bibinfo
  {author} {\bibfnamefont {S.~M.}\ \bibnamefont {Couch}}, \ and\ \bibinfo
  {author} {\bibfnamefont {E.~P.}\ \bibnamefont {O'Connor}},\ }\href {\doibase
  10.3847/1538-4357/ac294c} {\bibfield  {journal} {\bibinfo  {journal}
  {Astrophys. J.}\ }\textbf {\bibinfo {volume} {923}},\ \bibinfo {pages} {201}
  (\bibinfo {year} {2021})},\ \Eprint {http://arxiv.org/abs/2106.09734}
  {arXiv:2106.09734 [astro-ph.HE]} \BibitemShut {NoStop}%
\bibitem [{\citenamefont {Schneider}\ \emph {et~al.}(2019)\citenamefont
  {Schneider}, \citenamefont {Roberts}, \citenamefont {Ott},\ and\
  \citenamefont {O'connor}}]{Schneider:2019shi}%
  \BibitemOpen
  \bibfield  {author} {\bibinfo {author} {\bibfnamefont {A.~S.}\ \bibnamefont
  {Schneider}}, \bibinfo {author} {\bibfnamefont {L.~F.}\ \bibnamefont
  {Roberts}}, \bibinfo {author} {\bibfnamefont {C.~D.}\ \bibnamefont {Ott}}, \
  and\ \bibinfo {author} {\bibfnamefont {E.}~\bibnamefont {O'connor}},\ }\href
  {\doibase 10.1103/PhysRevC.100.055802} {\bibfield  {journal} {\bibinfo
  {journal} {Phys. Rev. C}\ }\textbf {\bibinfo {volume} {100}},\ \bibinfo
  {pages} {055802} (\bibinfo {year} {2019})},\ \Eprint
  {http://arxiv.org/abs/1906.02009} {arXiv:1906.02009 [astro-ph.HE]}
  \BibitemShut {NoStop}%
\bibitem [{\citenamefont {Radice}\ \emph {et~al.}(2022)\citenamefont {Radice},
  \citenamefont {Bernuzzi}, \citenamefont {Perego},\ and\ \citenamefont
  {Haas}}]{Radice:2021jtw}%
  \BibitemOpen
  \bibfield  {author} {\bibinfo {author} {\bibfnamefont {D.}~\bibnamefont
  {Radice}}, \bibinfo {author} {\bibfnamefont {S.}~\bibnamefont {Bernuzzi}},
  \bibinfo {author} {\bibfnamefont {A.}~\bibnamefont {Perego}}, \ and\ \bibinfo
  {author} {\bibfnamefont {R.}~\bibnamefont {Haas}},\ }\href {\doibase
  10.1093/mnras/stac589} {\bibfield  {journal} {\bibinfo  {journal} {Mon. Not.
  Roy. Astron. Soc.}\ }\textbf {\bibinfo {volume} {512}},\ \bibinfo {pages}
  {1499} (\bibinfo {year} {2022})},\ \Eprint {http://arxiv.org/abs/2111.14858}
  {arXiv:2111.14858 [astro-ph.HE]} \BibitemShut {NoStop}%
\bibitem [{\citenamefont {Engvik}\ \emph {et~al.}(1994)\citenamefont {Engvik},
  \citenamefont {Hjorth-Jensen}, \citenamefont {Osnes}, \citenamefont {Bao},\
  and\ \citenamefont {Ostgaard}}]{Engvik:1994tj}%
  \BibitemOpen
  \bibfield  {author} {\bibinfo {author} {\bibfnamefont {L.}~\bibnamefont
  {Engvik}}, \bibinfo {author} {\bibfnamefont {M.}~\bibnamefont
  {Hjorth-Jensen}}, \bibinfo {author} {\bibfnamefont {E.}~\bibnamefont
  {Osnes}}, \bibinfo {author} {\bibfnamefont {G.}~\bibnamefont {Bao}}, \ and\
  \bibinfo {author} {\bibfnamefont {E.}~\bibnamefont {Ostgaard}},\ }\href
  {\doibase 10.1103/PhysRevLett.73.2650} {\bibfield  {journal} {\bibinfo
  {journal} {Phys. Rev. Lett.}\ }\textbf {\bibinfo {volume} {73}},\ \bibinfo
  {pages} {2650} (\bibinfo {year} {1994})},\ \Eprint
  {http://arxiv.org/abs/nucl-th/9406028} {arXiv:nucl-th/9406028} \BibitemShut
  {NoStop}%
\end{thebibliography}%
\bibliographystyle{apsrev4-1}

\appendix
\section{Generalized piecewise polytropic fits}
\label{sec:appendixGPP}
In this appendix, we report our numerical fits for the
generalized piecewise polytrope (GPP) parametrization of the cold EoSs
used in this work. The procedure for generating these fits is exactly
as described in Appendix A of \cite{Raithel:2022san}. In particular, the
low-density portion of the EoS is described by a GPP representation
of SLy, from a high-accuracy version of Table II of \cite{OBoyle:2020qvf} 
[M. O'Boyle, priv. comm.] For the high-density EoS,
we follow \cite{OBoyle:2020qvf} in using a three-segment parametrization, which
are divided at the fiducial densities $\rho_{1}=10^{14.87}$~g/cm$^3$ and
$\rho_{2}=10^{14.99}$~g/cm$^3$. 

In this parametrization, the pressure along a given segment is given by
\begin{equation}
P(\rho) = K_i \rho^{\Gamma_i} + \Lambda_i, \quad	\rho_{i-1} < \rho \le \rho_i,
\end{equation}
where the polytropic coefficient, $K_i$, is determined by requiring differentiability,
\begin{equation}
\label{eq:Ki}
K_i = K_{i-1} \left( \frac{\Gamma_{i-1}}{\Gamma_i} \right) \rho_{i-1}^{\Gamma_{i-1}-\Gamma_i},
\end{equation}
and the parameter $\Lambda_i$ is imposed to ensure continuity in the pressure,
such that
\begin{equation}
\label{eq:Lambdai}
\Lambda_i = \Lambda{i-1} + \left( 1- \frac{\Gamma_{i-1}}{\Gamma_i} \right) K_{i-1} \rho_{i-1}^{\Gamma_{i-1}} .
\end{equation}

For a three-segment parametrization, there are four free parameters: 
${K_1, \Gamma_1, \Gamma_2, \Gamma_3}$. From these parameters and the low-density
EoS, all other $K_{i}$ and $\Lambda_i$ in eqs.~(\ref{eq:Ki}-\ref{eq:Lambdai}) 
are uniquely determined.

We perform a Markov Chain Monte Carlo simulation to fit for the parameters that
minimize the difference between the tabulated version of the wff2 
\cite{Wiringa:1988tp} and H4 \cite{Lackey:2005tk}  EoSs
and their GPP representations, as in \cite{Raithel:2022san}. We report the resulting
best-fit coefficients in Table~\ref{table:GPP}.

  \begin{table}
  \centering
\begin{tabular}{ccccccc }
\hline 
EoS  &  $R_{1.4}$ [km]  & $\rho_0$ [g/cm$^3$] &  $\log_{10} K_1$  &   $\Gamma_1$   &  $\Gamma_2$ &  $\Gamma_3$   \\
\hline \hline 
wff2 & 11.10 & 1.309 $\times 10^{14}$ &  -35.443 &  3.316 &  4.122 &  3.200   \\
H4 & 13.99 &2.931 $\times 10^{14}$ & -23.110 &  2.502 &  1.511 &  2.366  \\
\hline
\end{tabular}
\caption{\label{table:GPP} Best-fit parameters for generalized piecewise polytropic
representations of the wff2 and H4 EoSs. 
$R_{1.4}$ indicates the radius of a 1.4~$\Ms$ neutron star predicted by each EoS.
The parameter $\rho_0$ is the density at which the high-density
parametrization intersects the crust EoS, which is taken to be a GPP
 representation of SLy (see text). The remaining four columns
 provide the four free parameters that are determined via our GPP
 fitting procedure. }
 \end{table}

\section{Parametrization of the particle effective mass}
\label{sec:compareM}
The thermal prescription developed in \cite{Raithel:2019gws}
is based on a Fermi-Liquid theory based approach, in
which the high-density thermal pressure and energy
depend explicitly on the particle effective mass function, $M^*(n)$.
A two-parameter approximation of 
this effective mass function was introduced in  \cite{Raithel:2019gws}, according to
\begin{equation}
\label{eq:Meff}
M^*(n) = \left\{ (m c^2)^{-2} + \left[ mc^2 \left( \frac{n}{n_0 }\right)^{-\alpha} \right]^{-2} \right\}^{-1/2},
\end{equation}
where $n$ is the baryon number density, $mc^2$ is the vacuum rest-mass of the particle
(which the effective mass asymptotes to at low densities),  $n_0$ is the density at which
the effective mass function starts to deviate away from the vacuum value,
and $\alpha$ governs the rate of decay. For the low-density
vacuum mass, we use the energy per baryon of $^{56}$Fe of $mc^2=930.6$~MeV.
The phenomenological approximation in eq.~\ref{eq:Meff}
neglects the composition and temperature
dependences of $M^*$, which were determined to be small
for neutron star conditions in \cite{Raithel:2019gws}. Instead,
it assumes that the neutron and proton
effective masses are comparable, such that $M^*_p \approx
M^*_n \approx M^*$.  For additional details, see  \cite{Raithel:2019gws,Raithel:2021hye}.

\begin{figure}[!ht]
\centering
\includegraphics[width=0.45\textwidth]{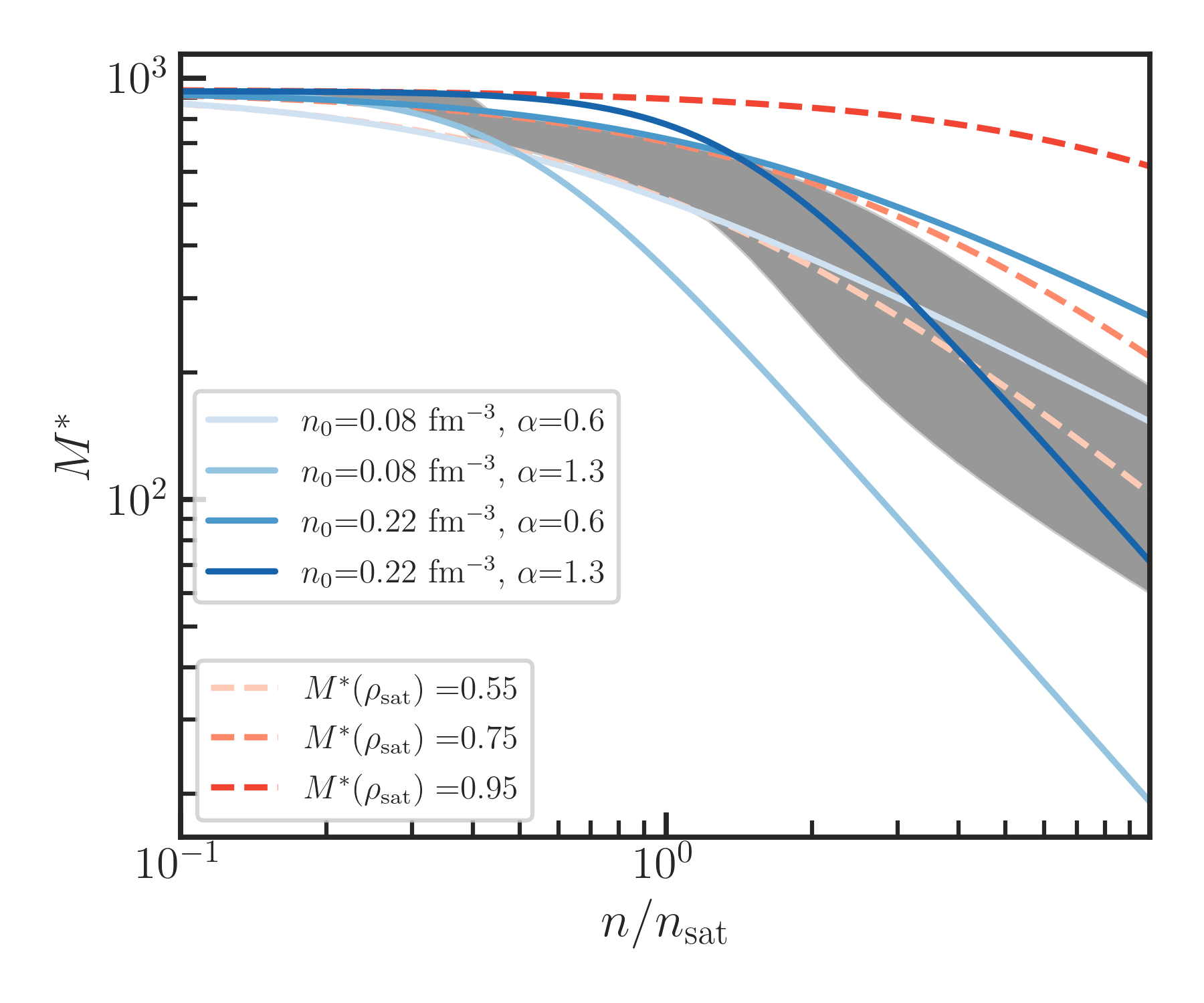} 
\caption{\label{fig:Mstar} Particle effective mass, $M^*$, as a function of the density
for symmetric nuclear matter.
The gray shaded band corresponds to the range of values spanned by a sample of tabulated,
finite-temperature EoSs. The blue lines correspond to the set of 
parameters used in this work following eq.~\ref{eq:Meff}; while the red lines indicate the models
 used in \cite{Fields:2023bhs} (eq.~\ref{eq:Mstar_SRO}). See text for details.}
\end{figure}

Figure~\ref{fig:Mstar} shows the effective mass function 
from eq.~\ref{eq:Meff} for the four sets of parameters considered in this work.
These parameters were chosen to extremally bound the range
of best fit values for a set of finite-temperature EoS tables
that are commonly used in merger simulations \cite{Raithel:2019gws},
such that the resulting thermal prescriptions would
bracket the effective thermal indices for such models 
(see Fig.~\ref{fig:gth}). We show the range of effective
masses spanned by these tabulated models
(for symmetric nuclear matter at $T=10$~MeV)
in the gray-shaded band in Fig.~\ref{fig:Mstar}.
The tabulated EoS models include the DD2, TMA, TM1, FSG
models calculated within the statistical framework of
\cite{Hempel:2009mc} (and references therein), SFHo and SFHx
\cite{Steiner:2012rk}, NL3 and FSU \cite{Shen:2011kr}, and STOS
\cite{Shen:1998gq}.

We compare these functional forms to another representation
of the particle effective mass function that was recently
explored in the context of merger simulations
in  \cite{Fields:2023bhs}.
The EoS framework used in \cite{Fields:2023bhs} is based on a
phenomenological liquid drop model
with Skyrme interactions from Schneider, Roberts, and Ott (SRO)
\cite{Schneider:2017tfi}. In this
approach, the nucleon effective mass is linearly related to the
nucleon densities, according to
\begin{equation}
\label{eq:Mstar_SRO}
\frac{\hbar^2}{2 M^*_n} = \frac{\hbar^2}{2 m_n}  + \alpha_1 n_n + \alpha_2 n_p
\end{equation}
where the subscript $n$ indicates a neutron, $p$ a proton, and
$\alpha_i$ are Skyrme parameters. There are two free parameters, 
so the function is uniquely defined by specifying: (1)
 the effective mass at saturation $M^*_n(n_{\rm sat})$, which is
 taken to be 0.55, 0.75, or 0.95 in \cite{Fields:2023bhs}, 
and (2) the neutron-proton mass difference for pure neutron matter
$\Delta$ which is taken to be $0.1 m_n$, following \cite{Andersen:2021vzo}.
From these fixed values, one can solve eq.~\ref{eq:Mstar_SRO}
for the full density-dependence of $M^*(n)$, which we extract
and plot in Fig.~\ref{fig:Mstar}. In this way, we
can compare the range of $M^*$-functions considered in
our work with the SRO approach. 

We note that two of our effective mass functions,
which have $\alpha=0.6$, are similar to the SRO
mass functions with $M^*(\rho_{\rm sat})$=0.55 and 0.75
used in \cite{Fields:2023bhs}. Our $M^*$-model
with $n_0=0.08$~fm$^{-3}$ and $\alpha=1.3$ is the most
extreme model in this space; but we note that this choice
of parameters is necessary to fully bracket the  
range of $\Gamma_{\rm th}$ in Fig.~\ref{fig:gth}.

Finally, we comment briefly on the differences in the post-merger
GWs in our work and \cite{Fields:2023bhs}, which
use these two different representations for the particle effective
mass function. We note that the SRO framework used in \cite{Fields:2023bhs}
differs additionally  beyond the $M^*$-representation; notably,
in that the EoS framework is not assumed to be separable into a
cold and thermal component (as in eq.~\ref{eq:Pth}).
As a result, the changes to $M^*(\rns)$ in that work lead to a small, but non-zero
change to the zero-temperature part of the EoS as well,
which will affect e.g. the neutron star radius (see e.g., \cite{Schneider:2019shi}).

As discussed in Sec.~\ref{sec:gw}, \cite{Fields:2023bhs} find that $f_2$
varies by up to 245~Hz, for their full range of $M^*(\rho_{\rm sat})$=0.55 to 0.95.
If we instead focus on the $M^*$-functions from \cite{Fields:2023bhs} that are the most
 similar to those used in this work (i.e., $M^*(\rho_{\rm sat})$=0.55 and 0.75),
 then their simulations show that $f_2$ varies by only 73~Hz.
 For the comparable set of $M^*$-parameters used in our work (corresponding
 to the $\alpha=0.6$ models), we find negligible differences
 in $f_2$ of $\lesssim10$~Hz for both cold EoSs. 
 
 In addition to the differences in the EoS framework discussed above,
 we note that the numerical scheme used in \cite{Fields:2023bhs} also differs
 from the code used in this work. Most notably, the evolutions in \cite{Fields:2023bhs}
 include an M1 neutrino transport scheme \cite{Radice:2021jtw},
which may account for some of the differences in the resulting
values of $f_2$.

\section{Details on the gravitational wave analysis}
\label{sec:appendixGW}
In this appendix, we briefly describe our procedure for calculating the 
characteristic strain, which is defined as
\begin{equation}
h_c = 2 f |\tilde{h}(f)|
\end{equation}
where $f$ is the frequency and $\tilde{h}(f)$ is the Fourier transform 
of the strain $h(t) \equiv h_{+}(t) - i h_{\times}(t)$. 

\begin{figure}[!ht]
\centering
\includegraphics[width=0.45\textwidth]{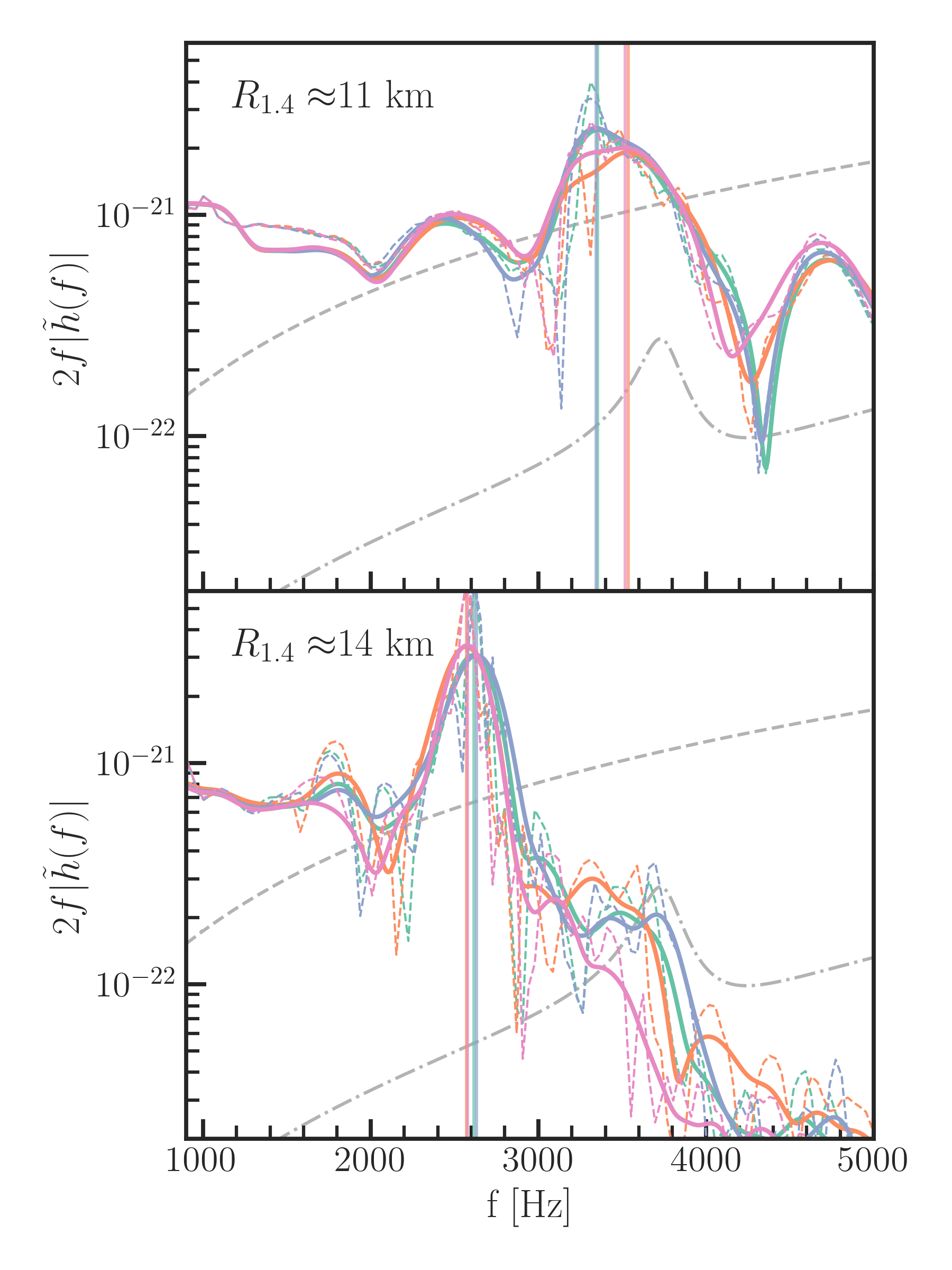} 
\caption{\label{fig:hc_fft} Same as Fig.~\ref{fig:hc}, but also including the characteristic strain
calculated from the raw (un-averaged) Fourier Transform of $\tilde{h}(f)$ in dashed lines. The solid
lines indicate the characteristic strain after the Welch-averaging and renormalization procedure described
in the text.}
\end{figure}

To compute $\tilde{h}(f)$, we first apply a Tukey window with
 a shape parameter of 0.25 to the two polarizations of the time-domain strain,
 $h_{+}(t)$ and $h_{\times}(t)$. From this windowed strain,
 we compute the Fourier transform to get an
initial spectrum, shown in dashed lines in Fig.~\ref{fig:hc_fft}. 

This initial (un-averaged) spectrum is noisy,
making individual spectral features hard to identify. 
In order to reduce the noise,
we also compute a Welch-averaged spectrum, using 6 overlapping segments
that are $\sim5$~ms each in length. Each segment is windowed
with a Hann window and zero-padded to contain a total of 4,096 points. 
After computing the Welch-averaged spectrum, we normalize it to
ensure the same total power as the raw (un-averaged) spectrum
between 1 and 5 kHz, to compensate for the loss of power due to
 Welch-averaging of a non-stationary signal.
The normalized, Welch-averaged spectra are shown with solid
lines in Fig.~\ref{fig:hc_fft}. We confirm that the spectral peaks
identified from the Welch-averaged spectra approximately agree
with what we pick out by eye from the un-averaged spectra.
The Welch-averaged spectra are used for all of the analyses
in Sec.~\ref{sec:gw}.

\section{Ejecta convergence}
\label{sec:appendixConv}

In this appendix, we provide an estimate of the numerical errors for the dynamical ejecta.
We base this estimate off of a previous set of simulations \cite{Raithel:2022san},
 which follow the same numerical set-up to the simulations described in this work, 
except that those simulations utilized a different cold EoS and a different set of $M^*$-parameters.
In particular, that work used a generalized piecewise polytropic parametrization of ENG \cite{Engvik:1994tj}
for the cold EoS, which predicts neutron star radii of $R_{1.4}=11.95$~km; while for the $M^*$-parameters,
$n_0$=0.12~fm$^{-3}$ and $\alpha=0.8$ were used. Both of these choices fall in the middle of the range
of values explored in this work, making this a convenient reference. In addition, we
note that this previous work used slightly larger total mass 
($M_{\rm ADM}=2.76$ in that work; cf. $M_{\rm ADM}=2.6$ in the present study), but
we expect the fractional error estimates made below to approximately hold across these different masses.

We report the total dynamical ejecta calculated outside a sphere at $r=100M$ 
(as in \ref{sec:ejecta}) in Table~\ref{table:converg}. The ejecta is reported for
three different resolutions.  We find approximately second-order self-convergence,
as the resolution is increased. We take advantage of this convergence to 
calculate the Richardson extrapolation of the dynamical ejecta, using the low and high resolutions,
and we find that $M_{\rm ej}\approx1.6\times 10^{-2}\Ms$ for a simulation at infinite resolution.
We take the difference between this Richardson extrapolation and $M_{\rm ej}$ extracted
from our highest-resolution evolution as an estimate of the error in $M_{\rm ej}$.
We find this error to be $\sim5\times10^{-3}~\Ms$, which is a $\sim30$\% fractional error.

\begin{table}
\centering
\begin{tabular}{ccc}
\hline \hline
Resolution  & dx (m) &   $M_{\rm ej} [10^{-2} M_{\odot}]$   \\ 
 \hline 
 Low         & 195      &   2.85   \\
 Medium   & 156.25  & 2.43   \\ 
 High       	&   125    & 2.15    \\
\hline
\end{tabular}
\caption{Dynamical ejecta extracted from simulations run at three different 
resolutions from \cite{Raithel:2022san}. From left to right: the resolution label, the grid
resolution at the innermost refinement level, and the total dynamical ejecta calculated outside
a sphere located at $r=100M$.}
  \label{table:converg}
\end{table}

\end{document}